\documentclass[a4paper,11pt]{article}
\pdfoutput=1
\usepackage{jcappub}
\usepackage[table,svgnames,dvipsnames]{xcolor}

\usepackage{orcidlink}

\usepackage{hyperref}
\usepackage{orcidlink}
\usepackage{hanging}
\usepackage{aas_macros}
\usepackage{amsmath}
\usepackage{amssymb}
\usepackage[utf8]{inputenc}

\usepackage{comment}
\usepackage{graphicx}
\usepackage{bbold}
\usepackage[normalem]{ulem}
\usepackage{gensymb}

\usepackage{lineno}

\newcommand{\Mpc}{\rm{Mpc}}
\newcommand{\hMpc}{h^{-1}\ \Mpc}

\usepackage[nameinlink,noabbrev]{cleveref}
\crefname{equation}{Eq.}{Eqs.}
\crefname{section}{Section}{Sections}
\crefname{figure}{Figure}{Figures}
\crefname{table}{Table}{Tables}
\crefname{appendix}{Appendix}{Appendices}
\Crefname{figure}{Figure}{Figures}
\Crefname{equation}{Equation}{Equations}
\Crefname{section}{Section}{Sections}
\Crefname{table}{Table}{Tables}

\title{Extensive analysis of reconstruction algorithms for DESI 2024 baryon acoustic oscillations}

\emailAdd{xinyi.chen@yale.edu}
\emailAdd{zhejied@sjtu.edu.cn}
\emailAdd{epaillas@arizona.edu}

\author[1]{{X.~Chen}\orcidlink{0000-0003-3456-0957},}
\author[2]{{Z.~Ding}\orcidlink{0000-0002-3369-3718},}
\author[3,4,5]{{E.~Paillas}\orcidlink{0000-0002-4637-2868},}
\author[6]{{S.~Nadathur}\orcidlink{0000-0001-9070-3102},}
\author[7]{{H.~Seo}\orcidlink{0000-0002-6588-3508},}
\author[8]{{S.~Chen}\orcidlink{0000-0002-5762-6405},}
\author[1]{{N.~Padmanabhan},}
\author[9,10]{{M.~White}\orcidlink{0000-0001-9912-5070},}
\author[11]{{A.~de~Mattia}\orcidlink{0000-0003-0920-2947},}
\author[12]{{P.~McDonald}\orcidlink{0000-0001-8346-8394},}
\author[13,14,15]{{A.~J.~Ross}\orcidlink{0000-0002-7522-9083},}
\author[16]{{A.~Variu}\orcidlink{0000-0001-8615-602X},}
\author[17,18]{{A.~Carnero Rosell}\orcidlink{0000-0003-3044-5150},}
\author[12,10]{{B.~Hadzhiyska}\orcidlink{0000-0002-2312-3121},}
\author[19]{{M.~M.~S~Hanif}\orcidlink{0009-0006-2583-5006},}
\author[16]{{D.~Forero-Sánchez}\orcidlink{0000-0001-5957-332X},}
\author[20]{{S.~Ahlen}\orcidlink{0000-0001-6098-7247},}
\author[19]{{O.~Alves},}
\author[21,19]{{U.~Andrade}\orcidlink{0000-0002-4118-8236},}
\author[22]{{S.~BenZvi}\orcidlink{0000-0001-5537-4710},}
\author[23]{{D.~Bianchi}\orcidlink{0000-0001-9712-0006},}
\author[24]{{D.~Brooks},}
\author[12]{{E.~Chaussidon}\orcidlink{0000-0001-8996-4874},}
\author[12]{{T.~Claybaugh},}
\author[25]{{A.~de la Macorra}\orcidlink{0000-0002-1769-1640},}
\author[26,27]{{Biprateep~Dey}\orcidlink{0000-0002-5665-7912},}
\author[28,29]{{K.~Fanning}\orcidlink{0000-0003-2371-3356},}
\author[12,10]{{S.~Ferraro}\orcidlink{0000-0003-4992-7854},}
\author[24,30]{{A.~Font-Ribera}\orcidlink{0000-0002-3033-7312},}
\author[31,32]{{J.~E.~Forero-Romero}\orcidlink{0000-0002-2890-3725},}
\author[33,34,55]{{C.~Garcia-Quintero}\orcidlink{0000-0003-1481-4294},}
\author[35,6,36]{{E.~Gaztañaga},}
\author[12]{{S.~Gontcho A Gontcho}\orcidlink{0000-0003-3142-233X},}
\author[37]{{G.~Gutierrez},}
\author[38]{{C.~Hahn}\orcidlink{0000-0003-1197-0902},}
\author[13,39,15]{{K.~Honscheid}\orcidlink{0000-0002-6550-2023},}
\author[40]{{S.~Juneau},}
\author[41]{{R.~Kehoe},}
\author[42]{{D.~Kirkby}\orcidlink{0000-0002-8828-5463},}
\author[12]{{T.~Kisner}\orcidlink{0000-0003-3510-7134},}
\author[12]{{A.~Kremin}\orcidlink{0000-0001-6356-7424},}
\author[12]{{M.~E.~Levi}\orcidlink{0000-0003-1887-1018},}
\author[40]{{A.~Meisner}\orcidlink{0000-0002-1125-7384},}
\author[43]{{J.~Mena-Fern\'andez}\orcidlink{0000-0001-9497-7266},}
\author[44,30]{{R.~Miquel},}
\author[45]{{J.~Moustakas}\orcidlink{0000-0002-2733-4559},}
\author[25]{{A.~Muñoz-Gutiérrez},}
\author[1]{{F.~Nikakhtar},}
\author[11,12]{{N.~Palanque-Delabrouille}\orcidlink{0000-0003-3188-784X},}
\author[3,46,5]{{W.~J.~Percival}\orcidlink{0000-0002-0644-5727},}
\author[47]{{F.~Prada}\orcidlink{0000-0001-7145-8674},}
\author[48]{{I.~P\'erez-R\`afols}\orcidlink{0000-0001-6979-0125},}
\author[33]{{M.~Rashkovetskyi}\orcidlink{0000-0001-7144-2349},}
\author[49]{{G.~Rossi},}
\author[50,51]{{R.~Ruggeri}\orcidlink{0000-0002-0394-0896},}
\author[52]{{E.~Sanchez}\orcidlink{0000-0002-9646-8198},}
\author[53]{{C.~Saulder}\orcidlink{0000-0002-0408-5633},}
\author[12]{{D.~Schlegel},}
\author[54,19]{{M.~Schubnell},}
\author[55]{{A.~Smith}\orcidlink{0000-0002-3712-6892},}
\author[40]{{D.~Sprayberry},}
\author[19]{{G.~Tarl\'{e}}\orcidlink{0000-0003-1704-0781},}
\author[7]{{D.~Valcin}\orcidlink{0000-0003-0129-0620},}
\author[25]{{M.~Vargas-Maga\~na}\orcidlink{0000-0003-3841-1836},}
\author[40]{{B.~A.~Weaver},}
\author[29]{{S.~Yuan}\orcidlink{0000-0002-5992-7586},}
\author[12]{{R.~Zhou}\orcidlink{0000-0001-5381-4372}}

\affiliation[1]{Physics Department, Yale University, P.O. Box 208120, New Haven, CT 06511, USA}
\affiliation[2]{Department of Astronomy, School of Physics and Astronomy, Shanghai Jiao Tong University, Shanghai 200240, China}
\affiliation[3]{Department of Physics and Astronomy, University of Waterloo, 200 University Ave W, Waterloo, ON N2L 3G1, Canada}
\affiliation[4]{Steward Observatory, University of Arizona, 933 N, Cherry Ave, Tucson, AZ 85721, USA}
\affiliation[5]{Waterloo Centre for Astrophysics, University of Waterloo, 200 University Ave W, Waterloo, ON N2L 3G1, Canada}
\affiliation[6]{Institute of Cosmology and Gravitation, University of Portsmouth, Dennis Sciama Building, Portsmouth, PO1 3FX, UK}
\affiliation[7]{Department of Physics \& Astronomy, Ohio University, 139 University Terrace, Athens, OH 45701, USA}
\affiliation[8]{Institute for Advanced Study, 1 Einstein Drive, Princeton, NJ 08540, USA}
\affiliation[9]{Department of Physics, University of California, Berkeley, 366 LeConte Hall MC 7300, Berkeley, CA 94720-7300, USA}
\affiliation[10]{University of California, Berkeley, 110 Sproul Hall \#5800 Berkeley, CA 94720, USA}
\affiliation[11]{IRFU, CEA, Universit\'{e} Paris-Saclay, F-91191 Gif-sur-Yvette, France}
\affiliation[12]{Lawrence Berkeley National Laboratory, 1 Cyclotron Road, Berkeley, CA 94720, USA}
\affiliation[13]{Center for Cosmology and AstroParticle Physics, The Ohio State University, 191 West Woodruff Avenue, Columbus, OH 43210, USA}
\affiliation[14]{Department of Astronomy, The Ohio State University, 4055 McPherson Laboratory, 140 W 18th Avenue, Columbus, OH 43210, USA}
\affiliation[15]{The Ohio State University, Columbus, 43210 OH, USA}
\affiliation[16]{Institute of Physics, Laboratory of Astrophysics, \'{E}cole Polytechnique F\'{e}d\'{e}rale de Lausanne (EPFL), Observatoire de Sauverny, Chemin Pegasi 51, CH-1290 Versoix, Switzerland}
\affiliation[17]{Departamento de Astrof\'{\i}sica, Universidad de La Laguna (ULL), E-38206, La Laguna, Tenerife, Spain}
\affiliation[18]{Instituto de Astrof\'{\i}sica de Canarias, C/ V\'{\i}a L\'{a}ctea, s/n, E-38205 La Laguna, Tenerife, Spain}
\affiliation[19]{University of Michigan, 500 S. State Street, Ann Arbor, MI 48109, USA}
\affiliation[20]{Physics Dept., Boston University, 590 Commonwealth Avenue, Boston, MA 02215, USA}
\affiliation[21]{Leinweber Center for Theoretical Physics, University of Michigan, 450 Church Street, Ann Arbor, Michigan 48109-1040, USA}
\affiliation[22]{Department of Physics \& Astronomy, University of Rochester, 206 Bausch and Lomb Hall, P.O. Box 270171, Rochester, NY 14627-0171, USA}
\affiliation[23]{Dipartimento di Fisica ``Aldo Pontremoli'', Universit\`a degli Studi di Milano, Via Celoria 16, I-20133 Milano, Italy}
\affiliation[24]{Department of Physics \& Astronomy, University College London, Gower Street, London, WC1E 6BT, UK}
\affiliation[25]{Instituto de F\'{\i}sica, Universidad Nacional Aut\'{o}noma de M\'{e}xico,  Circuito de la Investigaci\'{o}n Cient\'{\i}fica, Ciudad Universitaria, Cd. de M\'{e}xico  C.~P.~04510,  M\'{e}xico}
\affiliation[26]{Department of Astronomy \& Astrophysics, University of Toronto, Toronto, ON M5S 3H4, Canada}
\affiliation[27]{Department of Physics \& Astronomy and Pittsburgh Particle Physics, Astrophysics, and Cosmology Center (PITT PACC), University of Pittsburgh, 3941 O'Hara Street, Pittsburgh, PA 15260, USA}
\affiliation[28]{Kavli Institute for Particle Astrophysics and Cosmology, Stanford University, Menlo Park, CA 94305, USA}
\affiliation[29]{SLAC National Accelerator Laboratory, Menlo Park, CA 94305, USA}
\affiliation[30]{Institut de F\'{i}sica d’Altes Energies (IFAE), The Barcelona Institute of Science and Technology, Edifici Cn, Campus UAB, 08193, Bellaterra (Barcelona), Spain}
\affiliation[31]{Departamento de F\'isica, Universidad de los Andes, Cra. 1 No. 18A-10, Edificio Ip, CP 111711, Bogot\'a, Colombia}
\affiliation[32]{Observatorio Astron\'omico, Universidad de los Andes, Cra. 1 No. 18A-10, Edificio H, CP 111711 Bogot\'a, Colombia}
\affiliation[33]{Center for Astrophysics $|$ Harvard \& Smithsonian, 60 Garden Street, Cambridge, MA 02138, USA}
\affiliation[34]{Department of Physics, The University of Texas at Dallas, 800 W. Campbell Rd., Richardson, TX 75080, USA}
\affiliation[35]{Institut d'Estudis Espacials de Catalunya (IEEC), c/ Esteve Terradas 1, Edifici RDIT, Campus PMT-UPC, 08860 Castelldefels, Spain}
\affiliation[36]{Institute of Space Sciences, ICE-CSIC, Campus UAB, Carrer de Can Magrans s/n, 08913 Bellaterra, Barcelona, Spain}
\affiliation[37]{Fermi National Accelerator Laboratory, PO Box 500, Batavia, IL 60510, USA}
\affiliation[38]{Department of Astrophysical Sciences, Princeton University, Princeton NJ 08544, USA}
\affiliation[39]{Department of Physics, The Ohio State University, 191 West Woodruff Avenue, Columbus, OH 43210, USA}
\affiliation[40]{NSF NOIRLab, 950 N. Cherry Ave., Tucson, AZ 85719, USA}
\affiliation[41]{Department of Physics, Southern Methodist University, 3215 Daniel Avenue, Dallas, TX 75275, USA}
\affiliation[42]{Department of Physics and Astronomy, University of California, Irvine, 92697, USA}
\affiliation[43]{Laboratoire de Physique Subatomique et de Cosmologie, 53 Avenue des Martyrs, 38000 Grenoble, France}
\affiliation[44]{Instituci\'{o} Catalana de Recerca i Estudis Avan\c{c}ats, Passeig de Llu\'{\i}s Companys, 23, 08010 Barcelona, Spain}
\affiliation[45]{Department of Physics and Astronomy, Siena College, 515 Loudon Road, Loudonville, NY 12211, USA}
\affiliation[46]{Perimeter Institute for Theoretical Physics, 31 Caroline St. North, Waterloo, ON N2L 2Y5, Canada}
\affiliation[47]{Instituto de Astrof\'{i}sica de Andaluc\'{i}a (CSIC), Glorieta de la Astronom\'{i}a, s/n, E-18008 Granada, Spain}
\affiliation[48]{Departament de F\'isica, EEBE, Universitat Polit\`ecnica de Catalunya, c/Eduard Maristany 10, 08930 Barcelona, Spain}
\affiliation[49]{Department of Physics and Astronomy, Sejong University, 209 Neungdong-ro, Gwangjin-gu, Seoul 05006, Republic of Korea}
\affiliation[50]{Centre for Astrophysics \& Supercomputing, Swinburne University of Technology, P.O. Box 218, Hawthorn, VIC 3122, Australia}
\affiliation[51]{School of Mathematics and Physics, University of Queensland, Brisbane, QLD 4072, Australia}
\affiliation[52]{CIEMAT, Avenida Complutense 40, E-28040 Madrid, Spain}
\affiliation[53]{Max Planck Institute for Extraterrestrial Physics, Gie\ss enbachstra\ss e 1, 85748 Garching, Germany}
\affiliation[54]{Department of Physics, University of Michigan, 450 Church Street, Ann Arbor, MI 48109, USA}
\affiliation[55]{NASA Einstein Fellow}

\abstract{Reconstruction of the baryon acoustic oscillation (BAO) signal has been a standard procedure in BAO analyses over the past decade and has helped to improve the BAO parameter precision by a factor of $\sim$2 on average. The Dark Energy Spectroscopic Instrument (DESI) BAO analysis for the first year (DR1) data uses the ``standard'' reconstruction framework, in which the displacement field is estimated from the observed density field by solving the linearized continuity equation in redshift space, and galaxy and random positions are shifted in order to partially remove non-linearities.   
There are several approaches to 
solving for the displacement field in real survey data,
including the multigrid (MG), iterative Fast Fourier Transform (iFFT), and iterative Fast Fourier Transform particle (iFFTP) algorithms. In this work, we analyze these algorithms and compare them with various metrics including two-point statistics and the displacement itself using realistic DESI mocks. We focus on three representative DESI samples, the emission line galaxies (ELG), quasars (QSO), and the bright galaxy sample (BGS), which cover the extreme redshifts and number densities, and potential wide-angle effects. We conclude that the MG and iFFT algorithms agree within 0.4\% in post-reconstruction power spectrum on BAO scales with the \textbf{RecSym} convention, which does not remove large-scale redshift space distortions (RSDs), in all three tracers. The \textbf{RecSym} convention appears to be less sensitive to displacement errors than the \textbf{RecIso} convention, which attempts to remove large-scale RSDs.
However, iFFTP deviates from the first two; thus, we recommend against using iFFTP without further development. In addition, we provide the optimal settings for reconstruction for five years of DESI observation.
The analyses presented in this work pave the way for DESI DR1 analysis as well as future BAO analyses.}

\begin{document}
\maketitle
\flushbottom

\section{Introduction}
\label{sec:intro}
Sound waves in the baryon-photon fluid of the early universe leave an imprint in the matter clustering with a characteristic length scale that can be observed at low redshift with large galaxy surveys. This characteristic scale can be precisely measured by cosmic microwave background (CMB) experiments.
Using this scale as a standard ruler, the baryon acoustic oscillation (BAO) technique is one of the most established approaches to measuring the expansion history of the Universe and to constrain the acceleration of the cosmic expansion. The BAO feature was first detected by the Sloan Digital Sky Survey (SDSS \cite{Eisenstein05}) and 2dF Galaxy Redshift Survey (2dFGRS \cite{Cole05}). After that and over the last two decades, BAO parameter methodology has been developed \cite[e.g.][]{Padmanabhan08,Padmanabhan09} and the parameters have been measured via spectroscopic surveys of galaxies and quasars, such as 6dFGRS \cite{Beutler11}, SDSS-III Baryon Oscillation Spectroscopic Survey (BOSS \cite{Anderson14,Anderson14b,Ross15,Beutler17}), WiggleZ \cite{Kazin14}, and SDSS-IV extended Baryon Oscillation Spectroscopic Survey (eBOSS \cite{Ata18,Neveux20,Gil-Marin20,Raichoor21,Hou21,Bautista21}). The measurements from eBOSS have achieved 1\% precision in the BAO parameters at low redshift ($z<1$). There have also been measurements at higher redshift ($2<z<4$) from Lyman-$\alpha$ forests from spectroscopic surveys in BOSS and eBOSS \cite[e.g.][]{duMas20}. Measurements of BAO have also been obtained from imaging surveys, such as the dark energy survey (DES \cite{Abbott22}). 

These measurements of the observed size of the BAO combined with measurements of the physical scale of BAO from CMB or big-bang nucleosynthesis (BBN) experiments allow us to constrain the Hubble parameter $H(z)$ and in turn the nature of the accelerated expansion -- whether it takes the simplest form of the cosmological constant $\Lambda$, or evolves over time. Before Stage-IV surveys, the measurements have been consistent with a constant dark energy. More precise measurements are needed to distinguish different dark energy models, and that leads to the development of Stage-IV spectroscopic surveys, which aim to measure the BAO at sub-percent precision, among other science goals.

The Dark Energy Spectroscopic Instrument (DESI \cite{DESI13,DESI16,DESI16b,DESI19,DESI2022.KP1.Instr,FocalPlane.Silber.2023,Corrector.Miller.2023,Spectro.Pipeline.Guy.2023,SurveyOps.Schlafly.2023,DESI2023a.KP1.SV,DESI2023b.KP1.EDR,DESI2024.I.DR1,KP3,KP4,DESI2024.IV.KP6,DESI2024.V.KP5,KP7,DESI2024.VII.KP7B}), a Stage-IV spectroscopic survey, is an ongoing five-year survey of 14,000 deg$^2$ of the sky, the largest spectroscopic experiment to date optimized to measure BAO (as well as growth rate). Before the DESI DR1 results, the measurement using the early DESI data \cite{DESI2023a.KP1.SV,DESI2023b.KP1.EDR}
analyzing the luminous red galaxy (LRG) sample had already reached 1\% level precision \cite{Moon23}. Earlier this year, DESI just released an average of 0.5\% measurement of BAO parameters from its DR1 galaxy and quasar data \cite{KP4}. The cosmological analysis using DESI DR1 BAO measurements tentatively suggests a time-evolving dark energy model \cite{KP7}. More data and even more precise measurements are needed to confirm this finding.

One of the most significant sources of degradation of the BAO measurement precision is the gravitational nonlinear evolution. Bulk flows (on scales $\sim$20 ${\rm Mpc}/h$) and the gravitational collapse of large clusters in the late-time universe broaden and shift the BAO peak in the two-point correlation function and equivalently suppress the oscillations in the power spectrum, thus reducing the precision of the BAO parameter measurement \cite[][]{Eisenstein07b}. Mitigating such degradation can greatly improve the measurement. A procedure attempting to undo the effect of gravity to restore the BAO peak, known as BAO reconstruction, was first proposed in 2007 by \cite{Eisenstein07}, and has been applied in BAO analyses in the past decade. This technique takes the observed galaxy density field to estimate the displacement field from the initial distribution using a linearized form of the continuity equation. It uses this displacement field to undo the smearing of the BAO feature, and thereby improves the accuracy of the BAO technique.
We refer to this approach as the ``standard'' reconstruction method, to distinguish it from techniques that were proposed later (an incomplete list of new algorithms include \cite[e.g.][]{Schmittfull17,Hada18,Chen24,Seo22,vonHausegger22,Levy21,Nikakhtar22,Chen23,Shallue23,Parker25}). Reconstruction using the standard method has become a default procedure in BAO analysis since its first application in galaxy redshift surveys in 2012 by \cite{Padmanabhan12} and has helped to increase the BAO measurement precision by a factor of $\sim$2 on average from the past BAO measurements \cite[e.g.][]{Padmanabhan12,Xu13,Anderson14,Gil-Marin20}. Reconstruction is also assumed in the forecasts of future BAO surveys.

Real surveys have the changing line-of-sight direction, and complicated survey geometries and redshift distributions, etc, so it is not straightforward to apply the standard reconstruction algorithm. The main problem an algorithm needs to handle is the changing line-of-sight direction. The linearized redshift space continuity equation is more difficult to solve when lines of sight are non-parallel. Before DESI, two algorithms attempting to handle this problem had been employed in data. One is focused on solving the linearized redshift space continuity equation for the displacement in configuration space, using the parallel GMRES algorithm \cite[e.g.][]{Padmanabhan12}. This technique was used in SDSS and BOSS. The other technique attempts to solve the same equation in Fourier space. It iteratively 
removes redshift space distortions (RSDs) by moving galaxies, thus iteratively estimating the linear, real space density and consequently the displacement with the use of Fast Fourier Transforms 
(hereafter we refer to this algorithm as ``iFFTP''). This algorithm was inspired by the iFFT algorithm \cite{Burden15} discussed below. This technique was employed in the analyses of eBOSS \cite{Bautista18,Bautista21}.

There are two other algorithms available that attempt to solve the linearized redshift space continuity equation with changing lines of sight: (1) a multigrid (MG) method,
which also solves the differential equation in configuration space, but with an algorithm that converges faster and a code that requires no external libraries, and (2) another iterative Fast Fourier Transform technique  (hereafter ``iFFT'', \cite{Burden15}), which is the original version of the iterative FFT technique. The difference between iFFT and iFFTP is that iFFT does not move galaxies to remove RSDs at every iteration, but removes RSDs by updating potentials iteratively. Thus, the state of the art remains with two campaigns of algorithms, solving the linearized redshift space continuity equation in configuration space or in Fourier space. 
We note that these algorithms all attempt to solve the same differential equation.

In this paper, we conduct a comprehensive analysis of three algorithms mentioned above, MG, iFFT, and iFFTP, with a focus on the first two, in support of the DESI DR1 analysis. We focus on testing the consistency of these methods in the context of the accuracy of the BAO method. This is one of a series of supporting papers for the DESI 2024 BAO analysis. This study will also serve for the applications of some nonstandard reconstruction algorithms that will be applied in future DESI analysis and beyond. 
The implementation of these three algorithms are available in DESI's Github for reconstruction \textsc{Pyrecon}.\footnote{\url{https://github.com/cosmodesi/pyrecon}, developed by A.\ de Mattia. The MG code is a re-adaptation of the code at \url{https://github.com/martinjameswhite/recon_code} by M.\ White. The iFFT code is a new implementation of the algorithm. The iFFTP code is a re-implementation of the code at \url{https://github.com/julianbautista/eboss_clustering/blob/master/python/recon.py} by J.\ Bautista.} The analysis in this paper tests the algorithms as implemented in these codes. 

In addition to analyzing the three reconstruction algorithms, we also present the optimal settings for reconstruction before choosing the optimal smoothing scales, which is the most important input parameter to the standard reconstruction algorithm and is presented in a companion paper \cite{recon1}. 

This paper is structured as follows. In \cref{sec:review_recon} we review the procedure of the standard reconstruction algorithm. \cref{sec:cutsky_algorithms} details the three reconstruction algorithms for solving the linearized redshift space continuity equation with changing lines of sight, which are the focus of this paper. We introduce DESI mocks used in this study in \cref{sec:mocks}. The results of optimal setup for conducting reconstruction is presented in \cref{sec:optimal_setup}. In \cref{sec:results_algorithm_comparison}, we present results on algorithm comparison with various metrics. We discuss in \cref{sec:discussion} and conclude in \cref{sec:conclusion}.

\section{Review of standard reconstruction}\label{sec:review_recon}
Although we introduce the problem as solving the linearized continuity equation, in this section, we temporarily use the language of Lagrangian perturbation theory (LPT) and Zel'dovich approximation to connect to the majority of the literature on standard reconstruction. We will compare the two in the next section. 

As established in \cite{Eisenstein07b}, the degradation of the acoustic feature is due to motions of matter on relatively large scales relative to the initial positions. It is thus helpful to move the final overdensities back to the initial location. In standard reconstruction, this is achieved by estimating the displacements and moving galaxies as well as a set of randoms back to their initial positions, thus retaining the late-time amplitude of fluctuations, while removing the nonlinear motion of the overdensities. To estimate the displacement, one can consult LPT. In LPT, the displacement maps the initial Lagrangian space positions $\boldsymbol{q}$ to the late-time, Eulerian space positions $\boldsymbol{x}(\boldsymbol{q},t)$:
\begin{equation}
    \boldsymbol{x}(\boldsymbol{q},t)=\boldsymbol{q}+\boldsymbol{\Psi}(\boldsymbol{q},t).
\end{equation}
The displacement $\boldsymbol{\Psi}(\boldsymbol{q},t)$ is obtained by solving the Euler-Poisson system of equations. In the following, we will omit the time dependence and focus on $\boldsymbol{\Psi}(\boldsymbol{q})$. The first-order solution is the well-known Zel'dovich approximation \cite{Zeldovich70}, denoted as $\boldsymbol{\Psi}^{(1)}(\boldsymbol{q})$. The standard reconstruction estimates the Zel'dovich displacement using the measured late-time density field. It then moves particles back by this displacement to form an estimate of the linear density field. 
Hence, there are two main parts to the standard reconstruction framework:
\begin{enumerate}
    \item Estimate the displacement field in redshift space $\boldsymbol{\mathcal{S}_s}$\footnote{We use symbols like $\boldsymbol{\mathcal{S}}$ and $\boldsymbol{\mathcal{S}_s}$ for the estimated displacements and $\boldsymbol{\Psi}$, $\boldsymbol{\Psi}^{(1)}$ etc. for the true displacement solutions. }
    \item Transform the displacement vector field into the $\delta_{\rm recon}$ scalar density field
\end{enumerate}

Below, we detail the steps involved in the standard reconstruction algorithm.
\begin{enumerate}
    \item Distribute observed galaxies on a grid to form a density field $\delta_s(\boldsymbol{k})$ (subscript $s$ denotes redshift space) and smooth the density field to remove small-scale clustering that is dominated by noise and nonlinearity and therefore would break the first-order approximation. We smooth the small-scale power with a Gaussian smoothing kernel $\Sigma(k)=\exp[-k^2R^2/2]$, where $R$ is the smoothing scale. This smoothing scale is an input parameter to the algorithm and is determined empirically for different types of tracers. For DESI DR1 analysis, the choices of smoothing scales range from 15 to 30 Mpc/$h$ \cite{recon1,KP4}.
    
    \item Estimate the large-scale galaxy bias $b$ and the linear growth rate $f$. In simulations at the same time snapshot, $f\approx \Omega_{\rm m}^{0.55}$, where $\Omega_{\rm m}$ is taken to be the matter density at the redshift of this snapshot. In realistic mocks built from same time snapshot simulations, the growth rate needs to be adjusted to account for the displacement differences in the varying redshift. 
    
    \item Prepare a set of randoms, which has a number density at least the level of data. The randoms are used to estimate the three-dimentional survey footprint and the redshift distribution of targets for clustering measurements. For reconstruction, random particles will also be moved later to conserve large-scale power (i.e. retain a similar amplitude of the power spectrum to the un-reconstructed field on large scales). Without doing so, the large-scale power will be lost when only moving galaxies back to the original positions.
    \item Compute the displacement using the observed density by solving the linearized continuity equation in redshift space. In the literature, the standard reconstruction has been more often studied and modeled by Zel'dovich approximation, and the target displacement is an estimate of the Zel'dovich displacement.

    Using the Zel'dovich approximation in real space $\nabla\cdot \boldsymbol{\Psi}^{(1)}(\boldsymbol{q})=-\delta_L(\boldsymbol{q})$, where $\delta_L(\boldsymbol{q})$ is the linear density, and assuming the displacement is irrotational and hence has a scalar potential $\nabla \phi=\boldsymbol{\Psi}^{(1)}$, we can obtain in Fourier space $\tilde{\boldsymbol{\Psi}}^{(1)}(\boldsymbol{k})=i\boldsymbol{k}\tilde{\delta}_{L}(\boldsymbol{k})/k^2$ (where we use the $\tilde{}$ notation to explicitly indicate Fourier-transformed quantities).
    In redshift space, we can obtain the following estimate of the displacement in the plane-parallel approximation
\begin{equation}\label{eq:displacement}
    \tilde{\boldsymbol{\mathcal{S}}}(\boldsymbol{k})=\frac{i\boldsymbol{k}}{k^2}\frac{\tilde{\delta}_s(\boldsymbol{k})\Sigma(k)}{b(1+\beta\mu^2)}.
\end{equation}
Here, $b$ is the linear galaxy bias, $\beta=f/b$, where $f$ is the growth rate, and $\mu$ is the cosine of the angle to the line of sight, $\mu=k_z/k$ in the plane-parallel approximation, taking $\boldsymbol{\hat{z}}$ to be the line-of-sight direction. We divide out the bias $b$ and the large-scale redshift space distortion (RSD) factor, the Kaiser term $(1+\beta\mu^2)$ \cite{Kaiser87}, to approximate the real space density. Small-scale redshift space distortions, the finger-of-God effects, are not corrected in this formalism, although they are suppressed with the smoothing kernel $\Sigma(k)$. 

Note that we use the observed late-time density field in Eulerian space in this step to approximate the linear density field in Lagrangian space, which is what actually appears in Zel'dovich approximation, i.e. $\delta_L(\boldsymbol{q})$
is in Lagrangian space. 
With a sufficiently large smoothing scale, \cref{eq:displacement} is a good approximation to the Zel'dovich displacement, $\tilde{\boldsymbol{\mathcal{S}}}(\boldsymbol{k})\approx \tilde{\boldsymbol{\Psi}}^{(1)}(\boldsymbol{k})$.
The effects of approximating the Lagrangian density $\delta(\boldsymbol{q})$ by the Eulerian density $\delta(\boldsymbol{x})$ are accounted for in the modeling of the post-reconstruction two-point statistics as detailed in our companion paper on theoretical systematics \cite{KP4s2-Chen}. 

The redshift space and real space displacements in configuration space are related by 
\begin{equation}\label{eq:S_s}
\boldsymbol{\mathcal{S}}_s(\boldsymbol{q})=\boldsymbol{\mathcal{S}}(\boldsymbol{q})+f[\boldsymbol{\mathcal{S}}(\boldsymbol{q})\cdot \boldsymbol{\hat{r}}]\boldsymbol{\hat{r}},
\end{equation}
where $\boldsymbol{\hat{r}}$ is the line-of-sight direction.
The displacement in the line-of-sight direction has an extra factor of $1+f$. 
In the plane-parallel approximation, the redshift space displacement is $\boldsymbol{\mathcal{S}}_s(\boldsymbol{q})=\boldsymbol{\mathcal{S}}(\boldsymbol{q})+f[\boldsymbol{\mathcal{S}}(\boldsymbol{q})\cdot \boldsymbol{\hat{z}}]\boldsymbol{\hat{z}}$. When the line of sight is fixed, taking the derivative along the line of sight in Fourier space we will obtain $k_z$, which can be related to $k$ and Kaiser approximation simply via $\mu$. This point will be elaborated on in the next section.
\item Move the galaxies by the displacement $\boldsymbol{\mathcal{S}}_s$ and form a ``displaced'' density field, $\delta_{\rm displaced, gal}$.
\item The randoms are moved by an amount depending on different reconstruction conventions \cite{White15,Chen19}: 
\begin{itemize}
    \item \textbf{RecSym} (if RSDs are not removed, i.e. symmetric treatment of galaxies and randoms): move the randoms by $\boldsymbol{\mathcal{S}}_s$.
    
    \item \textbf{RecIso} (if RSDs are removed, i.e. isotropic reconstruction): move the randoms by $\boldsymbol{\mathcal{S}}$.
\end{itemize}
We then form the ``shifted'' density field $\delta_{\rm shifted,ran}$.
\item The reconstructed density field is then $\delta_{\rm recon}=\delta_{\rm displaced,gal}-\delta_{\rm shifted,ran}$. This is the density field whose two-point statistics is then used for BAO measurements.
\end{enumerate}

In real surveys, where lines of sight are not in one direction, the calculation of the displacement (step 4) is more complicated. The algorithms reviewed in the next section tackle this problem. The procedure of moving particles and randoms after obtaining the displacement stays the same.

\section{Reconstruction algorithms for changing lines of sight}\label{sec:cutsky_algorithms}
In this paper, we make an important nomenclature change. While up to this point, the standard reconstruction has been more often associated with estimating the Zel'dovich displacement using Zel'dovich approximation equation in the literature, we refer to the problem as solving the linearized continuity equation in redshift space and its solution as the linear or first-order solution to the continuity equation. We use the Zel'dovich notation in the last section to connect to the theory modeling the standard reconstruction, but we would like to focus on using the continuity equation in this paper. The linearized continuity equation in redshift space\footnote{Note that this equation in \cite{Burden15} is missing a 1/$b$ in the redshift space part.} reads \cite{Nusser94}
\begin{equation}\label{eq:poisson}
    \nabla \cdot \boldsymbol{\Psi}^{(1)} + \frac{f}{b}\nabla \cdot (\boldsymbol{\Psi}^{(1)} \cdot \boldsymbol{\hat{r}})\boldsymbol{\hat{r}}=-\frac{\delta_s}{b}.
\end{equation}
The second term on the LHS accounts for redshift space distortions.

An important difference, as noted in step 4 in the last section, is that the density in Zel'dovich approximation is in Lagrangian space, the derivatives are w.r.t.\ Lagrangian coordinates, and the resulting displacement is also in Lagrangian coordinates (in estimating the displacement, we approximate the Lagrangian density by the Eulerian density). On the contrary, the density, derivatives, and the displacement ($\boldsymbol{\Psi}^{(1)}(\boldsymbol{x})$ to be explicit) in the continuity equation are in Eulerian coordinates. Therefore, there are these subtle differences, although the two coordinates can be transformed to one another beyond the linear order \cite[e.g.][]{McCullagh12}. 
At linear order, $\boldsymbol{x}=\boldsymbol{q}$, so the solution to the linearized continuity equation is the same as Zel'dovich displacement on large scales, i.e. $\boldsymbol{\Psi}^{(1)}(\boldsymbol{x})=\boldsymbol{\Psi}^{(1)}(\boldsymbol{q})$. Hence, we use the same displacement symbol. Because the two solutions are very close, all the theory modeling of the standard reconstruction is based on perturbation theory and Zel'dovich approximation. 

The difference is that Zel'dovich approximation describes linear motions, where particles move in straight lines. In the process of reconstruction, we are not solving for such dynamics; we are simply solving for a displacement. So it is simpler to refer to the equation as the (linearized) continuity equation and to the solution as the linear or first-order solution. 
As mentioned above, on large scales, this solution is the same as the Zel'dovich displacement. 
The gravitational nonlinear evolution that damps the BAO signal is on large scales, so it is largely due to the Zel'dovich displacement. 
In an attempt to undo the effect of gravity, standard reconstruction estimates a displacement to move objects back. The approximate (due to starting from the Eulerian density) Zel'dovich displacement is straightforward to obtain, and the community has adopted the approximate Zel'dovich displacement thus far. However, what displacement is the best to move objects back for measuring BAO purposes is still a question under exploration\footnote{The best displacement to move objects back for general purposes might be less unclear. The full displacement might serve for other physics problems. The question is then how to estimate the full displacement, which is beyond the scope of the current paper.}. Considering the above, we drop the Zel'dovich notation from now on, and only come back to it in the Discussion section.

Further on \cref{eq:poisson}, similar to the Lagrangian displacement, at the first order, the displacement in the continuity equation, $\boldsymbol{\Psi}^{(1)}$, is also irrotational, so we can take $\nabla \phi=\boldsymbol{\Psi}^{(1)}$, where $\phi$ is the scalar potential of the displacement. \cref{eq:poisson} then becomes
\begin{equation}\label{eq:poisson_phi}
    \nabla^2\phi+\frac{f}{b}\nabla\cdot(\nabla \phi \cdot \boldsymbol{\hat{r}} )\boldsymbol{\hat{r}}=-\frac{\delta_s}{b}.
\end{equation}
It is this equation that the following methods try to solve in various ways. In the plane-parallel approximation, i.e. $\boldsymbol{\hat{r}}=\boldsymbol{\hat{z}}$, taking the derivative in Fourier space twice, we will arrive at 
\begin{equation}
    \tilde{\phi}(\boldsymbol{k})=\frac{\tilde{\delta_s}(\boldsymbol{k})}{k^2b(1+\beta\mu^2)}.
\end{equation}
Taking the gradient of the above gives back the solution for the displacement in the plane-parallel approximation in the Zel'dovich context, \cref{eq:displacement}. This shows that the solution to the linearized continuity equation is the same as the Zel'dovich displacement on large scales in the plane-parallel case.

The complication of applying reconstruction to survey data is in solving the displacement when the plane-parallel approximation is not valid, i.e. lines of sight cannot be taken to be all in the same direction. In preparation for DESI analyses, we closely compare three available algorithms that attempt to solve the problem: multigrid (MG), iterative FFT (iFFT), and iterative FFT particle (iFFTP). These algorithms are presented in order in the subsections below. The MG algorithm has not been applied to survey data in the past\footnote{Contemporaneously with the DESI DR1 BAO analysis, the MG code is also used for velocity reconstruction in the kinematic Sunyaev-Zel'dovich studies in \cite{kSZ1,kSZ2,kSZ3}.}, except that it was used in a BOSS$\times$Planck analysis \cite{Chen22} (although a different algorithm to solve the differential equation in configuration space other than MG has been applied to SDSS and BOSS before \cite[e.g. GMRES algorithm][]{Padmanabhan12}). The iFFT algorithm has also not been applied to survey data before. The iFFTP algorithm was used in the eBOSS analyses \cite{Bautista18,Bautista21}. Thus, this paper analyzes two new codes and one previously applied. All three methods attempt to solve the linearized continuity equation in redshift space.

\subsection{Multigrid (MG)}

The MG method is a general numerical analysis algorithm to solve differential equations \cite{multigrid}. Traditional single grid solvers, such as the Jacobi method or the Gauss-Siedel method, reduce the high-frequency errors rapidly, but are slow at reducing the low frequency errors. The MG algorithm accelerates the convergence by transferring between coarse and fine grids through implementing a solver as the smoother, such that the low-frequency components can be interpolated and become high-frequency ones whose errors can be efficiently reduced by the solver. For our application of the MG method in reconstruction, we use the damped Jacobi method as the smoother (and in the final step, the solver). The Gauss-Siedel method is in principle also applicable, but it is harder to parallelize than the Jacobi method.

In this MG algorithm, we use the most simple ``V-cycle'' multigrid. We conduct a few iterations of damped Jacobi on the original fine grid, move to the first-level coarse grid, and perform a few iterations of damped Jacobi at the coarser grid. We keep moving to the next-level coarse grid until reaching the coarsest grid (we reach the coarsest grid when the grid size becomes odd and has divided by 3, 5, or 7 once)\footnote{We can construct MG schemes without the power of 2 restriction, but here we use this simplest algorithm. Developing the code to allow for more flexible prime factorization would make it more efficient.}. Then we move back to the original fine grid one level at a time, and we perform linear interpolation at each level. We perform a few more iterations of damped Jacobi after reaching the original fine grid. This transfer between fine and coarse grids can be repeated a few times (each time is one V-cycle) until convergence. 
Besides moving the low-frequency part to a coarse grid to expedite convergence, when the grid becomes coarser, its size also becomes smaller, and the corresponding linear equation is then faster to solve.

We use the Jacobi algorithm to solve for $\phi$ from \cref{eq:poisson_phi} with a linear equation of the form $\boldsymbol{A}\phi=\delta_s$, with periodic boundary conditions. The derivation of matrix $\boldsymbol{A}$ on a grid is detailed in \cref{appx:algebra}. The MG algorithm solves an updated equation with the residual associated with the last iteration of Jacobi at the last-level coarse grid, and the residual becomes the source term: $\boldsymbol{A}^{\prime}\Delta \phi^{\prime}=\boldsymbol{r}^{\prime}$, where $\boldsymbol{r}^{\prime}$ is the restricted residual (smaller size than that of the residual $\boldsymbol{r}$) and residual $\boldsymbol{r}=\delta_s - \boldsymbol{A}\phi_{\rm Jacobi}$. The $\boldsymbol{A}^{\prime}$ matrix is also restricted to be of a smaller size than $\boldsymbol{A}$. After solving $\Delta \phi^{\prime}$ with the coarse grid, we have a new potential $\phi^\prime_{\rm Jacobi,new}=\phi^{\prime}_{\rm Jacobi}+\Delta \phi^{\prime}$, where $\phi^{\prime}_{\rm Jacobi}$ is the restricted solution potential at the last-level coarse grid, a new residual $\boldsymbol{r}^\prime_{\rm new}=\delta_s - \boldsymbol{A}\phi^\prime_{\rm Jacobi,new}$, and a new equation $\boldsymbol{A}^{\prime\prime}\Delta \phi^{\prime\prime}=\boldsymbol{r}^{\prime\prime}$ for the next level coarse grid.
Here, the double primes denote restricted matrices. Moving from the coarsest grid back to the original fine grid (prolongation) is achieved via linear interpolation with periodic boundary conditions.
We have three input parameters as encoded in $\textsc{Pyrecon}$: the damping factor for Jacobi, the number of iterations for damped Jacobi, and the number of iterations for the V-cycle. Our default setting is Jacobi damping factor equals to 0.4, number of iterations of Jacobi equals to 5, and number of iterations of V-cycle equals to 6; increasing the number of iterations does not result in material changes in the power spectrum as tested in \cref{appx:vary_MG_parameters}. 
Figure~\ref{fig:MG_varyparameters} shows the small changes with varying parameters. 
We use the default settings throughout this study. 

Once we obtain the potential from multigrid, the (real space) displacement is computed by finite differences on the grid, e.g. \cref{eq:dev_phi},
\begin{equation}
    \boldsymbol{\mathcal{S}}_{x,\rm MG}=\nabla_x\phi\approx\frac{\phi_{i+1,j,k}-\phi_{i-1,j,k}}{2h},
\end{equation}
where $h$ is the grid resolution. The displacement in the $\boldsymbol{\hat{y}}$ and $\boldsymbol{\hat{z}}$ directions are similar. Finally, the redshift space displacement is obtained by following \cref{eq:S_s}.

\subsection{Iterative FFT (iFFT)}
The iFFT reconstruction algorithm solves the partial differential equation of the potential $\phi$ in Fourier space utilizing the Fast Fourier Transform technique. However, solving in Fourier space is not simply Fourier transforming \cref{eq:poisson_phi} and solving it as is, because unlike the full displacement, the radial component of the displacement ($(\nabla \phi \cdot \boldsymbol{\hat{r}} )\boldsymbol{\hat{r}}$ in the LHS of \cref{eq:poisson_phi}) is not irrotational. This term entirely comes from RSDs. If there were no RSDs, the potential can be obtained by simply taking the inverse Laplacian of the real space density. So the goal of the iFFT algorithm is to iteratively estimate the linear, real space density field. Considering the linearized continuity equation in the presence of RSDs in \cref{eq:poisson_phi}, the real space version reduces to
\begin{equation}\label{eq:phi_realspace_delta}
    \nabla^2\phi=-\frac{\delta_{\rm real}}{b}.
\end{equation}
The iFFT algorithm iteratively estimates $\delta_{\rm real}$, in other words, iteratively removes RSDs. 
Starting with a potential, every iteration updates the real space density, which is then used to estimate the next iteration potential. In the end, the final real space density is used to calculate the displacement.  

The iterations to update the real space density using a potential are through an equation that describes the relation between real space and redshift space densities. Substituting \cref{eq:poisson_phi} into \cref{eq:phi_realspace_delta} and writing in the $n$-th iteration, one can obtain 
\begin{equation}\label{eq:delta_real_n}
\frac{\delta_{\rm real,n}}{b}=\frac{\delta_{s}}{b}+\frac{f}{b}\nabla\cdot(\nabla\phi_{\rm n}\cdot \boldsymbol{\hat{r}})\boldsymbol{\hat{r}}.
\end{equation}
The next iteration potential is obtained by taking the inverse Laplacian of the current iteration real space density following \cref{eq:phi_realspace_delta}. This is easier to achieve in Fourier space via 
\begin{equation}\label{eq:delta_to_phi}
    \tilde{\phi}_{\rm n+1}(\boldsymbol{k})=\frac{\tilde{\delta}_{\rm real,n}(\boldsymbol{k})}{bk^2}.
\end{equation}
In \cref{eq:delta_real_n}, because it is hard to project onto the line of sight in Fourier space, the $\nabla\cdot (\nabla \phi_{\rm est,n} \cdot \boldsymbol{\hat{r}} )\boldsymbol{\hat{r}}$ part is computed in configuration space (with the divergence and gradient taking place in Fourier space). The estimated $\delta_{\rm real,n}$ is then Fourier transformed to Fourier space, and we take the inverse Laplacian in Fourier space to obtain $\tilde{\phi}_{\rm n+1}$ via \cref{eq:delta_to_phi}. This potential is then Fourier transformed back to estimate the next iteration real space density via \cref{eq:delta_real_n}. This process of Fourier transform to compute the potential and Fourier transform back to calculate the real space density are iterated multiple times to approximate the answer. 
This is what the ``iterative'' FFT refers to.

In the plane-parallel approximation, the $n$-th real space density is related to the starting real space density by
\begin{equation}
    \frac{\delta_{\rm real,n}}{b}=\frac{\delta_{\rm real}}{b}+(-1)^{n+1}\left(\frac{f}{b}\mu^2\right)^n\left(\frac{\delta_{\rm real}}{b}-\frac{\delta_{\rm real,0}}{b}\right).
\end{equation}
Here, $\delta_{\rm real,0}$ is the starting real space density. 
The starting potential can be obtained using the starting real space density via \cref{eq:delta_to_phi}.
We can then start the iteration. In the original paper \cite{Burden15}, the starting real space density is taken to be the redshift space density, i.e. $\delta_{\rm real,0}=\delta_{s}$, which is related to the actual real space density through Kaiser approximation \cite{Kaiser87}, $\tilde{\delta}_{s}=(1+\beta\mu^2)\tilde{\delta}_{\rm real}$. In the plane-parallel approximation, this starting density gives the the following $n$-th real space density in relation to the actual real space density\footnote{This general formula in the original algorithm paper \cite{Burden15} (Eq.\ 28) presents a ``$+$'' sign in the square brackets, which should be a ``$-$'' sign. }
\begin{equation}\label{eq:iFFT_general_form_org}
    \frac{\delta_{\rm real,n}}{b}=\frac{\delta_{\rm real}}{b}\left[1-\left(-\frac{f}{b}\right)^{n+1}\mu^{2(n+1)}\right].
\end{equation}
When $n=0$, we get back the Kaiser approximation.
We can see that the iteration converges when $f/b<1$. For a sample with $f/b\geq 1$, this method does not work. However, for all relevant galaxy samples in DESI, we have $f/b<1$. We also note that in the plane-parallel approximation, this general form shows that the convergence is slower along the line of sight. The general formula suggests that more iterations leads to a better approximation to the real space density, but the code does not set a convergence criterion, so the user can decide when to stop the iterations. 

In \textsc{Pyrecon}, iFFT assumes a different starting density than the original paper does to expedite the convergence. The starting real space density here is taken to be $\delta_{\rm real,0}=(\beta/(1+\beta))\delta_{s}$, where $\beta=f/b$. This $(\beta/(1+\beta))$ factor 
is found to lead to faster convergence, which can be understood using the following argument in the plane-parallel case.
This starting density leads to a different form of the $n$-th real space density:
\begin{equation}\label{eq:iFFT_general_form_adp}
    \frac{\delta_{\rm real,n}}{b}=\frac{\delta_{\rm real}}{b}\left[1-(-\beta)^{n+1}\mu^{2(n+1)}\frac{\beta-\frac{1}{\beta\mu^2}}{\beta+1}\right].
\end{equation}
Comparing to the earlier form, we see that when $\beta<1$, the convergence is faster than before.

With the estimated real space density $\delta_{\rm real,n}$, we can obtain the displacement in Fourier space via the real space continuity equation in Fourier space
\begin{equation}\label{eq:iFFT_disp}
    \boldsymbol{\tilde{\mathcal{S}}}_{\rm iFFT}(\boldsymbol{k})=\frac{i\boldsymbol{k}\tilde{\delta}_{\rm real,n}(\boldsymbol{k})}{bk^2}.
\end{equation}
We then inverse Fourier transform $\boldsymbol{\tilde{\mathcal{S}}}_{\rm iFFT}(\boldsymbol{k})$ to obtain the displacement in configuration space $\boldsymbol{\mathcal{S}}_{\rm iFFT}(\boldsymbol{q})$ and move galaxies and randoms.

\subsection{Iterative FFT particle (iFFTP)}
The iFFTP algorithm is a variation of the original iFFT algorithm. It has the same goal of estimating the real space density by iteratively removing RSDs, but the RSDs are removed by moving galaxies at each iteration. At each iteration, a displacement (purely due to gravitational nonlinear evolution) is calculated using the $n$-th estimated real space density, 
\begin{equation}\label{eq:iFFTP_disp}
    \tilde{\boldsymbol{\mathcal{S}}}_{\rm iFFTP, disp, n}(\boldsymbol{k})=\frac{i\boldsymbol{k}\tilde{\delta}_{\rm iFFTP,real, n}}{bk^2},
\end{equation}
and the galaxies are shifted to remove the RSDs by the linear Kaiser amount
\begin{equation}\label{eq:iFFTP_rsd}
    \boldsymbol{\mathcal{S}}_{\rm iFFTP, RSD, n}=f [\boldsymbol{\mathcal{S}}_{\rm iFFTP, disp, n}\cdot \hat{\boldsymbol{r}}]\hat{\boldsymbol{r}}.
\end{equation}
We note that the galaxies are only moved by $\boldsymbol{\mathcal{S}}_{\rm iFFTP, RSD, n}$ to remove RSDs at each iteration. The real-space density field $\delta_{\rm iFFTP,real,n+1}$ at the next iteration is then estimated by placing the shifted galaxies on a grid via the cloud-in-cell (CIC) particle distribution scheme, instead of by \cref{eq:delta_real_n} as in the iFFT method. The density here $\delta_{\rm iFFTP,real, n}$ is expected to be close to the iFFT real space density $\delta_{\rm real,n}$ obtained via \cref{eq:delta_real_n}, but it is achieved by updating galaxy positions, rather than updating the potential, both of which attempt to remove RSDs. To start the iteration, the observed redshift space density is used to approximate the real-space density for the initial iteration, and a factor of $\beta/(1+\beta)$ is similarly applied to the redshift space density, as for iFFT.

In this version of the algorithm, only the galaxies are moved during the process of removing RSDs, and they are not matched by the randoms (the randoms are still moved to form the shifted field but after the iterations for removing RSDs). We note that this procedure can give rise to numerical artefacts in the density estimation if galaxies near the survey boundary get shifted outside the range covered by the randoms.
After iterations, the galaxies are then at real space positions. The displacement is calculated by \cref{eq:iFFTP_disp} again with the final estimate of the real space density, and galaxies are moved by this amount to remove gravitational nonlinear evolution. The full amount gives the $\boldsymbol{\mathcal{S}}_{\rm s}(\boldsymbol{k})$ in \cref{sec:review_recon}, i.e. $\boldsymbol{\mathcal{S}}_{\rm iFFTP, full}=\boldsymbol{\mathcal{S}}_{\rm iFFTP, RSD}+\boldsymbol{\mathcal{S}}_{\rm iFFTP, disp}$, where $\boldsymbol{\mathcal{S}}_{\rm iFFTP, RSD}$ is the total movement over all iterations.
The randoms are moved only after the iteration process, when undoing the gravity, following either the \textbf{RecSym} or \textbf{RecIso} convention. iFFTP was the method applied in the eBOSS analysis \cite{Bautista18,Bautista21}.

In the application of the three algorithms in this study, the survey boundaries are treated the same way, i.e. outside the survey boundary, the density is taken to be the mean density. 

\section{DESI mocks and representative samples}\label{sec:mocks}

For this study, we use a suite of N-body simulations, $\textsc{AbacusSummit}$ \cite{Maksimova21}, that were designed to meet the simulation requirements of DESI. The $\textsc{AbacusSummit}$ simulation suite uses the $\textsc{Abacus}$ N-body code \cite{Garrison21} and produces accurate nonlinear structure. The suite includes density field snapshots and dark matter halo fields. For DESI specific analyses, we make use of 25 realizations of the base cosmology, the Planck 2018 $\Lambda$CDM cosmology: $\Omega_ch^2=0.1200$, $\Omega_bh^2=0.02237$, $\sigma_8=0.811355$, $n_s=0.9649$, $h=0.6736$, $w_0=-1$, and $w_a=0$ \cite{Planck18}. Each realization contains $6912^3$ CDM particles in a (2 $h^{-1}$Gpc)$^3$ volume. The halos are identified with the $\textsc{CompaSO}$ halo finder \cite{Hadzhiyska21}, and after that the halo catalog is cleaned to removed over-deblended halos \cite{Sownak22}. To obtain galaxy catalogs for DESI tracers, we apply the $\textsc{AbacusHOD}$ code \cite{Yuan22} to populate galaxies in the halos, following HOD models described in \cite{Alam20} and matching DESI clustering. The cubic box galaxy catalogs are then used to generate the ``cutsky'' version of the galaxy catalogs, where boxes are replicated to map DESI tracer volumes, which are typically larger than (2 $h^{-1}$Gpc)$^3$.

There have been two generations of mocks for the DESI DR1 analysis: {\tt Abacus-1} and {\tt Abacus-2}, as denoted in \cite{KP4}. {\tt Abacus-1} mocks are tuned to match early DESI data clustering, while {\tt Abacus-2} mocks match the clustering of the final DESI early data release \cite{earlydata_release}, which include corrections for all systematics. This current paper uses {\tt Abacus-1} to test the reconstruction on the Emission Line Galaxies and the Quasars and {\tt Abacus-2} for the Bright Galaxy Sample. Our analysis focuses on the North Galactic Cap (NGC) of DESI Y5 final footprint.

The random catalogs are generated to match the survey geometry \cite{Myers23}. 
Although not included in this study, the random catalogs used in the DR1 BAO analysis also add the potential fiber assignments for each random particle to determine whether a fiber could reach its angular positions \cite{KP3}.

In this study, we select three representative samples of DESI mocks to conduct our analysis. These are
\begin{itemize}
\item \textbf{The Emission Line Galaxy sample (ELG) in $0.8<z<1.1$} (denoted as {\tt ELG1} in \cite{KP4}): The ELGs are star-forming galaxies that have strong O${\rm II}$ emission lines. They are generally less clustered than passive galaxies, such as the LRGs. 
The completeness of the final DESI ELG sample will be over 60\%. The DR1 ELG number density is much lower than Y5 because of fiber incompleteness.
We use the Y5 footprint of ELG sample in this study as a representative of high number densities. However, we note that the mocks overestimate the actual DESI Y5 ELG number densities.

\item \textbf{The Quasar Sample (QSO) in $1.6<z<2.1$} (second half of {\tt QSO} in \cite{KP4}): The quasar sample has the highest bias, highest redshift, largest volume, and lowest number density of all DESI tracers. We choose this sample as one of our representatives to 
stress test the algorithms under low number densities and high redshift. 
DESI DR1 analysis had reconstruction on QSO as part of the pipeline, since DESI mock tests found mild improvements on average in BAO constraints when reconstruction was applied. However, both DESI DR1 and eBOSS found that there was virtually no improvement with reconstruction for QSO \cite{KP4,Hou21}.

\item \textbf{The Bright Galaxy Sample (BGS) in $0.1<z<0.4$} (same as {\tt BGS} in \cite{KP4}): BGS occupies the lowest redshift range of DESI. Because of this, the line of sight varies more considerably than other tracers (we refer to this as wide angle). We select this sample to test whether wide-angle effects expose reconstruction algorithms to challenges. We note that what matters is the change of the line of sight, not the size of the solid angle of the survey. BGS is originally a flux-limited sample, although for BAO analyses, we apply a selection in the sample such that the number density is constant throughout its redshift range (so it no longer has the highest number density among all DESI tracers). This selection function contains apparent and absolute magnitude cuts in the $r$ band, with the absolute magnitude cut having a $k$+E correction \cite{KP3}.

\end{itemize}

\begin{figure}
\includegraphics[width=\textwidth]{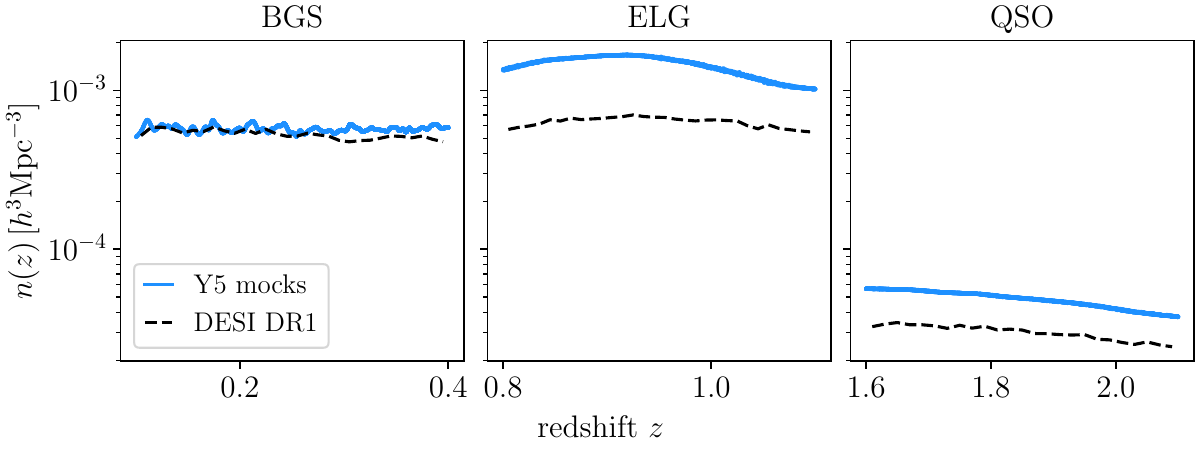}
    \caption{ Galaxy number density in the DESI \textsc{AbacusSummit} mocks with Y5 footprint as well as in the DESI DR1 data as a function of redshift for the selected DESI samples analyzed in this study. 
    We use the entire BGS sample in $0.1<z<0.4$, ELG over $0.8<z<1.1$, and QSO over $1.6<z<2.1$.
    The BGS number density is lower than the actual DESI DR1 sample (not the DR1 sample used for BAO analysis plotted in this figure) because of the magnitude cut.
    The ELG distribution presented here overestimates the real DESI Y5 number densities.
    } 
    \label{fig:nz}
\end{figure}

Figure~\ref{fig:nz} shows the number density as a function of redshift for our selected DESI samples. These samples survey the low (BGS) and high (QSO) redshift, low (QSO) and high (ELG) number density, and wide-angle effects (BGS), in order to thoroughly test the reconstruction algorithms.

This current paper primarily focuses on the power spectrum, i.e.\ the two-point function in Fourier space. We calculate the power spectrum pre- and post-reconstruction with a FKP minimum variance weight \cite{Feldman94}:
\begin{equation}
    w_i=\frac{1}{1+N(z_i)P_0},
\end{equation}
where $N(z)$ is the redshift distribution of the sample. We use $P_0=10000$ (Mpc/$h)^3$ for all tracers.

\section{Optimal setup for reconstruction with DESI data: grid resolution and number of randoms}\label{sec:optimal_setup}
 
Before discussing the performance of each reconstruction algorithm, we find the optimal setup for conducting reconstruction both for this study and for DESI BAO analysis in general. Reconstruction algorithms have a number of input parameters that must be appropriately chosen. These parameters include grid resolution when assigning galaxies onto the grid, number of randoms to be used, reconstruction convention (\textbf{RecSym} and \textbf{RecIso}), and smoothing scales. We address the first two in this section. Reconstruction convention is a choice. We present comparisons of the algorithms for both conventions, while focusing on \textbf{RecSym}, when discussing the reconstruction performance. DESI DR1 analysis \cite{KP4} also opts to use \textbf{RecSym} as the baseline. 
The smoothing scales are determined with BAO fitting 
and are presented in our companion paper \cite{recon1}.

Grid resolution is a parameter we would like to fix in this study. It affects the derivative calculation with finite differences in the MG algorithm and affects the Nyquist frequency for iFFT and iFFTP, although its effect is subdominant to the smoothing scale. For grid resolution tests, we set a 2 Mpc/$h$ cell size as our baseline for comparison with larger cell sizes. Since the lowest smoothing scale for reconstruction in our tests is 15 Mpc/$h$, a 2 Mpc/$h$ cell size is well below necessity, and we would like to reduce computational time without overly losing accuracy.
We test 4 Mpc/$h$ and 6 Mpc/$h$ cell sizes and compare them to the 2 Mpc/$h$ baseline with the reconstructed power spectrum. We show this convergence test with an ELG mock with the application of the MG reconstruction algorithm smoothed at 15 Mpc/$h$ using the \textbf{RecSym} convention. Figure~\ref{fig:convergence} shows that a 4 Mpc/$h$ grid resolution still agrees with the baseline resolution within 0.2\% for monopole and $1\%$ for quadrupole up to $k=0.4\ h$/Mpc, which are within our error budget. A 6 Mpc/$h$ resolution shows more difference, but is still within 0.6\% agreement to the baseline in monopole and 2\% agreement in quadrupole. We note that below $k=0.2\ h$/Mpc, where most of the BAO signals are, the agreement of 4 Mpc/$h$ and 6 Mpc/$h$ with 2 Mpc/$h$ is better than 0.1\% in monopole and 0.5\% in quadrupole. The deviation in the power spectrum in larger $k$ is likely from smoothing out the small-scale features that are present in 2 Mpc/$h$ smoothing. We therefore adopt a 4 Mpc/$h$ grid resolution for the analysis in this paper for both conducting reconstruction and calculating the power spectrum, except for calculating the post-reconstruction power spectrum for QSO, due to memory constraints. We use 6 Mpc/$h$ grid resolution for computing the QSO power spectrum, which is sufficient for the 30 Mpc/$h$ smoothing used for this tracer. Because grid resolution is subdominant to smoothing, we do not expect significant differences from different algorithms or different tracers.  In real DR1 BAO analysis, we also use 4 Mpc/$h$ grid resolution for reconstruction for all tracers, although we use 6 Mpc/$h$ for computing the power spectrum for all.

\begin{figure}
    \centering
    \includegraphics[width=0.49\columnwidth]{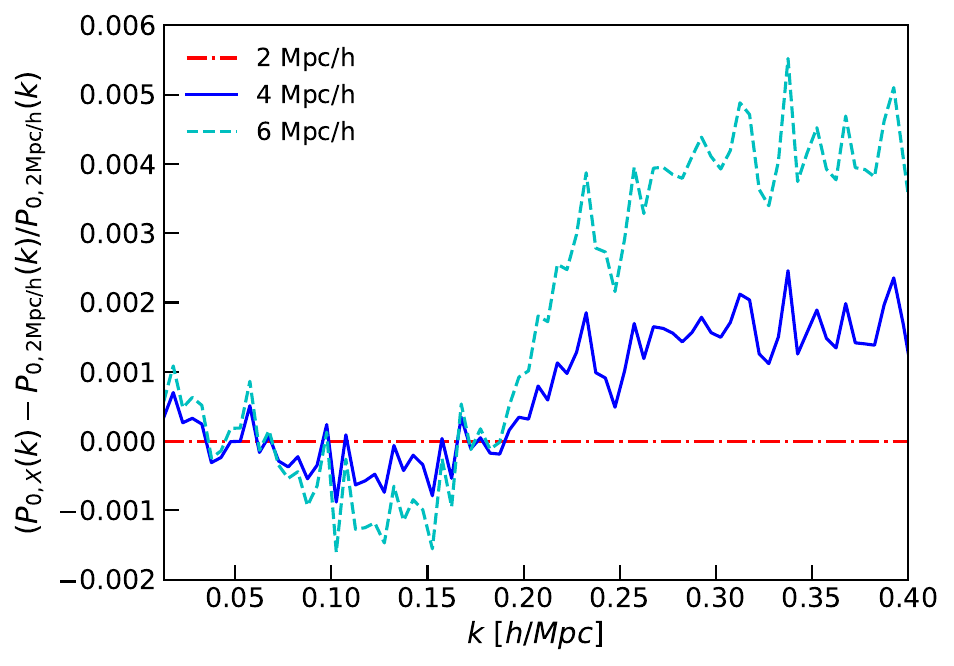}
    \includegraphics[width=0.49\columnwidth]{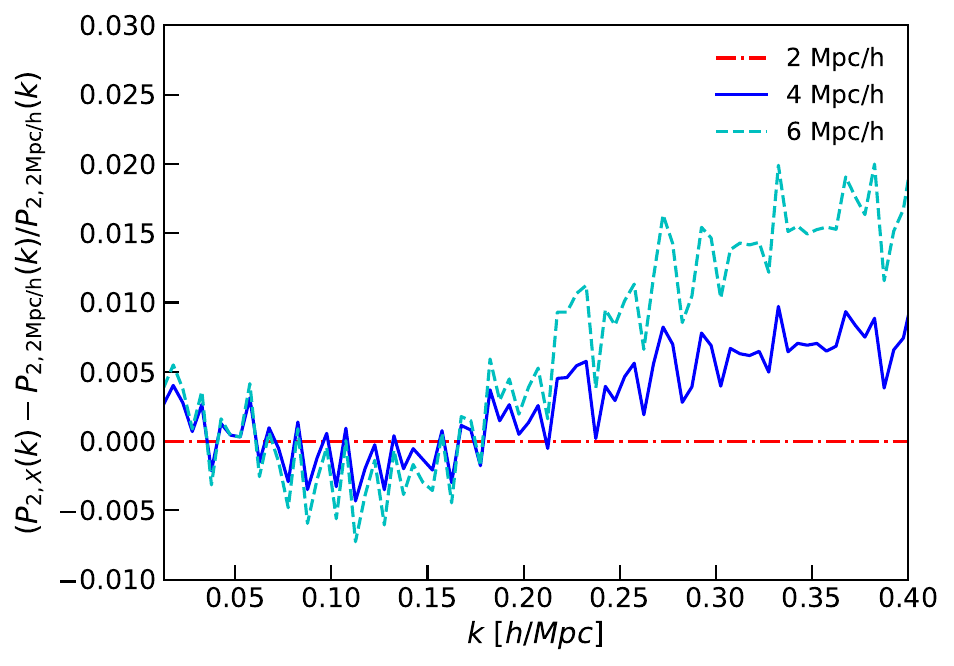}
    \caption{Convergence test for our choice of cell size with monopole (left) and quadrupole (right) power spectrum of the ELG sample for a grid resolution of 4 Mpc/$h$ (blue solid) and 6 Mpc/$h$ (cyan dashed), compared to a reference resolution at 2 Mpc/$h$ (red dash dotted). In the $y$-axis label, $P_{0,2{\rm Mpc}/h}$ and $P_{2,2{\rm Mpc}/h}$ are the monopole and quadrupole power spectra of the reference resolution, 2 Mpc/$h$. Each line here is one simulation. The deviation in the 4 Mpc/$h$ cell size is less than 0.2\% for monopole and less than 1\% for quadrupole, which are within our error budget.  }
    \label{fig:convergence}
\end{figure}

The number of randoms is also an input parameter in reconstruction and we conduct a convergence test for this parameter. 
When the number density of the random particles is not large enough to smoothly trace the survey footprint, it will introduce noise to reconstruction and impact the reconstruction fidelity. We set our baseline for the number of randoms to be 20$\times$ the number of galaxies for comparison with lower numbers of randoms. 
We test to what extent the number of randoms impact the post-reconstruction power spectrum monopole and quadrupole when the number of randoms is reduced to 10$\times$ and 5$\times$ the number of galaxies in Figure~\ref{fig:convergence}. Using the ELG samples with the MG reconstruction algorithm, we find that the difference between 10$\times$ and 20$\times$ are negligible (less than 0.01\% for monopole power spectrum and less than 0.1\% for quadrupole power spectrum). While 5$\times$ appears to introduce relatively more differences, the magnitude of the differences is low, below $\sim 0.13\%$ in monopole and below $\sim 0.6\%$ in quadrupole. Therefore, a safer minimum density of randoms is at least 10$\times$. We use much more than 10$\times$ of randoms in 
real analysis of DESI\footnote{The actual number density of randoms used in DESI is $18\times 2500/{\rm deg}^2$, so the comparison to tracer density varies by tracer.}. In this paper, we use number of randoms at 20$\times$ the number of galaxies for our reconstruction analyses.

\begin{figure}
    \centering
    \includegraphics[width=0.5\columnwidth]{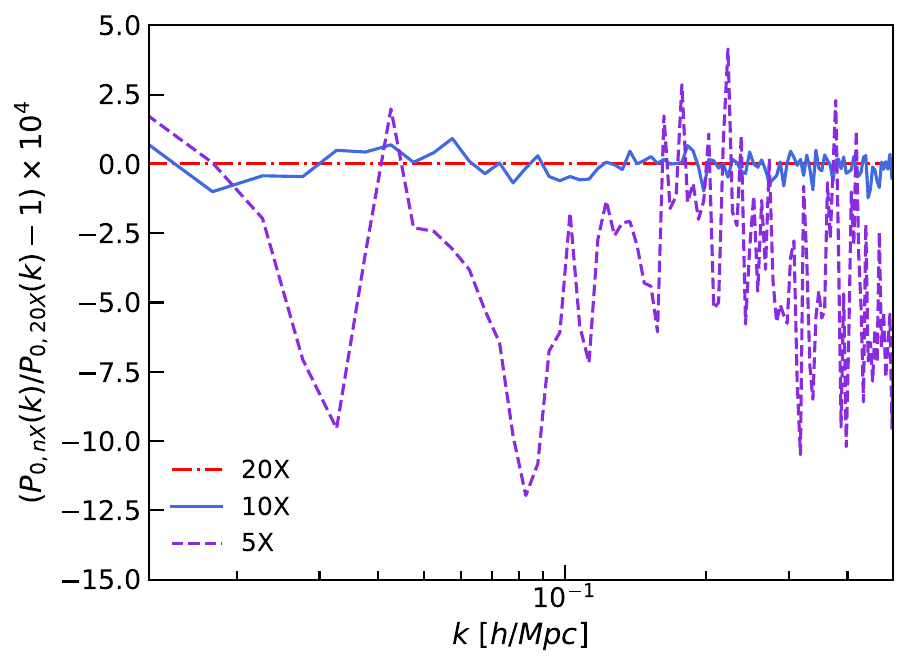}
    \includegraphics[width=0.48\columnwidth]{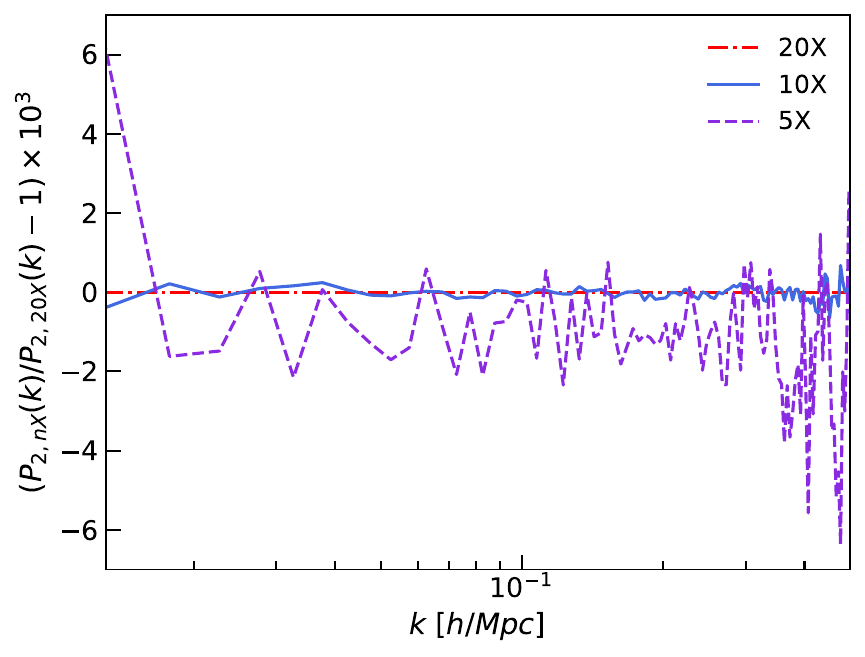}
    \caption{Reconstructed power spectrum monopole (left) and quadrupole (right) using the MG algorithm with number of randoms used being 10$\times$ (blue solid) and 5$\times$ (purple dashed) the number of galaxies in the field, compared to the 20$\times$ randoms baseline (red dash dotted). Each line here is an average of 25 simulations. The power spectra for the number of randoms at 10$\times$ and 20$\times$ are in agreement to better than 0.01\% in monopole and 0.1\% in quadrupole. Although the difference between 5$\times$ and 20$\times$ appears to be larger compared to the difference between 10$\times$ and 20$\times$, the magnitude of the difference is less than 0.13\% in monopole and less than 0.6\% in quadrupole. }
    \label{fig:convergence}
\end{figure}

\section{Results of algorithm comparisons}\label{sec:results_algorithm_comparison}

In this section, we present our tests comparing the three reconstruction algorithms. Since MG is a well-established numerical method to solve differential equations, we use MG as a reference and compare iFFT and iFFTP against MG. This choice is arbitrary, and what we are interested in is the difference among the different algorithms. We show in \ref{appx:vary_MG_parameters} that our default MG setting (damping factor equals to 0.4, number of iterations of Jacobi equals to 5, and number of iterations of V-cycle equals to 6) has converged. So we use this setting throughout this study and compare iFFT and iFFTP against MG under this setting. We note that in our test in BGS cubic boxes, i.e. in plane-parallel approximation, iFFT appears to converge to the single-step FFT reconstruction, while MG deviates from it at the level of 0.01\%, which we consider as negligible. iFFTP in cubic boxes deviates even more, which we will discuss in \cref{sec:discussion}.

We present a complete set of tests for ELG, covering all the metrics we use in this study and for all algorithms. Based on what ELG tests show, we omit certain tests for QSO and BGS and relegate to Appendix~\ref{appx:additional_figures} the test results that are similar to those of ELG. We present ELG results in \cref{sec:ELG_results}. We first closely examine the displacement in \cref{sec:ELG_displacement}, since the three algorithms essentially differ by how the displacement is estimated.
We then study the post-reconstruction two-point clustering statistics, because the two-point correlation function/power spectrum is what is used to constrain BAO parameters in DESI, and we focus on Fourier space and \textbf{RecSym} reconstruction convention. We present power spectrum results in \cref{sec:recsym_pk} and cross-correlation with the initial condition in \cref{sec:recsym_gk}.
We briefly mention \textbf{RecIso} analysis and compare the results with \textbf{RecSym} results in \cref{sec:ELG_reciso}. After presenting the results with ELG in \cref{sec:ELG_results}, we mention the similarities and differences present in the QSO and BGS samples in \cref{sec:QSO_results,sec:BGS_results}, respectively. Finally, we present BAO fitting results comparing the different algorithms. We focus on the sample that exhibits the most differences from analyses in the earlier sections and also include the fitting result from ELG for completeness. We comment on the computing time in \cref{sec:computing_time}.
Our default smoothing scales are 15 Mpc/$h$ for ELG, 30 Mpc/$h$ for QSO, and 15 Mpc/$h$ for BGS.

\subsection{ELG: high number density tracer case study
} \label{sec:ELG_results}
We choose the ELG sample as a representative of high number density in DESI to test the performance of the three algorithms. Smoothing is applied at 15 Mpc/$h$ for all three reconstruction algorithms. We first analyze the displacement in \cref{sec:ELG_displacement}. In \cref{sec:recsym_pk,sec:recsym_gk}, we focus on reconstruction with the \textbf{RecSym} convention. We discuss the results of \textbf{RecIso} in \cref{sec:ELG_reciso}.

\subsubsection{Displacement
}\label{sec:ELG_displacement}

Since MG, iFFT, and iFFTP handle the redshift space component in \cref{eq:poisson} differently,
we investigate whether the displacements of the three algorithms present differences along and perpendicular to the line of sight. 
We decompose the displacements into the usual spherical coordinate unit vectors $\boldsymbol{\hat{r}}$, $\boldsymbol{\hat{\theta}}$, and $\boldsymbol{\hat{\phi}}$. In this decomposition, $\boldsymbol{\hat{r}}$ is parallel to the line of sight, while $\boldsymbol{\hat{\theta}}$ and $\boldsymbol{\hat{\phi}}$ are perpendicular. The choice of the perpendicular to the line of sight directions can be arbitrary, and we choose $\boldsymbol{\hat{\theta}}$ and $\boldsymbol{\hat{\phi}}$ for simplicity. Note that the reconstruction convention is not relevant for the displacement calculation; the convention defines how the density field is calculated after the displacement is obtained. In the following analysis, we divide the line-of-sight displacement by $1+f$ to compare with the perpendicular directions.

\begin{figure}
    \centering
    \includegraphics[width=\linewidth]{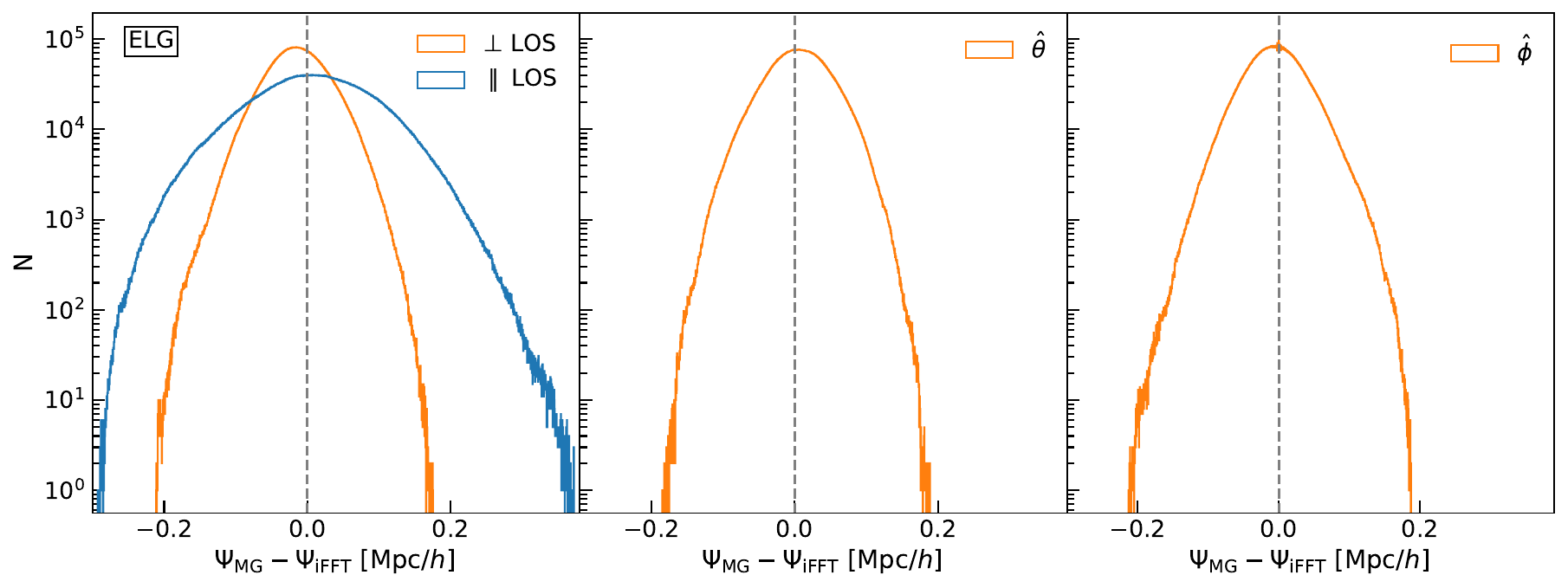}
    \caption{The distribution of displacement vector differences between \textbf{MG} and \textbf{iFFT} (3 iterations) displacements of the \textbf{ELG} sample using one mock. The displacement vectors are decomposed into along the line-of-sight $\boldsymbol{\hat{r}}$ and perpendicular to the line-of-sight directions (left), and the perpendicular to the line-of-sight direction is further decomposed into $\boldsymbol{\hat{\theta}}$ (middle) and $\boldsymbol{\hat{\phi}}$ (right) directions in spherical coordinates. 
    Along the line of sight, the spread of the distribution is larger compared to perpendicular to the line of sight, indicating that the different ways of treating the redshift direction by the two algorithms result in more differences in the displacement in the redshift direction, but the magnitude of the differences is small even in the parallel direction compared to the average line-of-sight displacement of ELG ($\sim$1.9 Mpc/$h$). The two algorithms have less differences perpendicular to the line-of-sight, as shown in the middle and right panels with the component perpendicular to the line of sight further decomposed. }
    \label{fig:ELG_displacement_decompose}
\end{figure}

Figure~\ref{fig:ELG_displacement_decompose} shows the distribution of the magnitude differences between MG and iFFT displacements of all the galaxies in one ELG mock for both along and perpendicular to the line of sight on the left, in $\boldsymbol{\hat{\theta}}$ in the middle, and in $\boldsymbol{\hat{\phi}}$ on the right. As expected, the two have better agreement in the perpendicular to the line of sight directions. Even though the difference has a larger spread in the parallel to the line-of-sight direction than perpendicular to the line-of-sight, the magnitude is not significant compared to the average line-of-sight displacement magnitude of ELG ($\sim$1.9 Mpc/$h$). The distributions in the $\boldsymbol{\hat{\theta}}$ and $\boldsymbol{\hat{\phi}}$ directions are roughly symmetric, i.e. not skew to one direction. We note that we do not compare fractional difference in the figure because it can blow up for small displacements, so it is less informative.
We leave the question of how much these displacement differences impact the summary statistics to the next section.

\begin{figure}
    \centering
    \includegraphics[width=\linewidth]{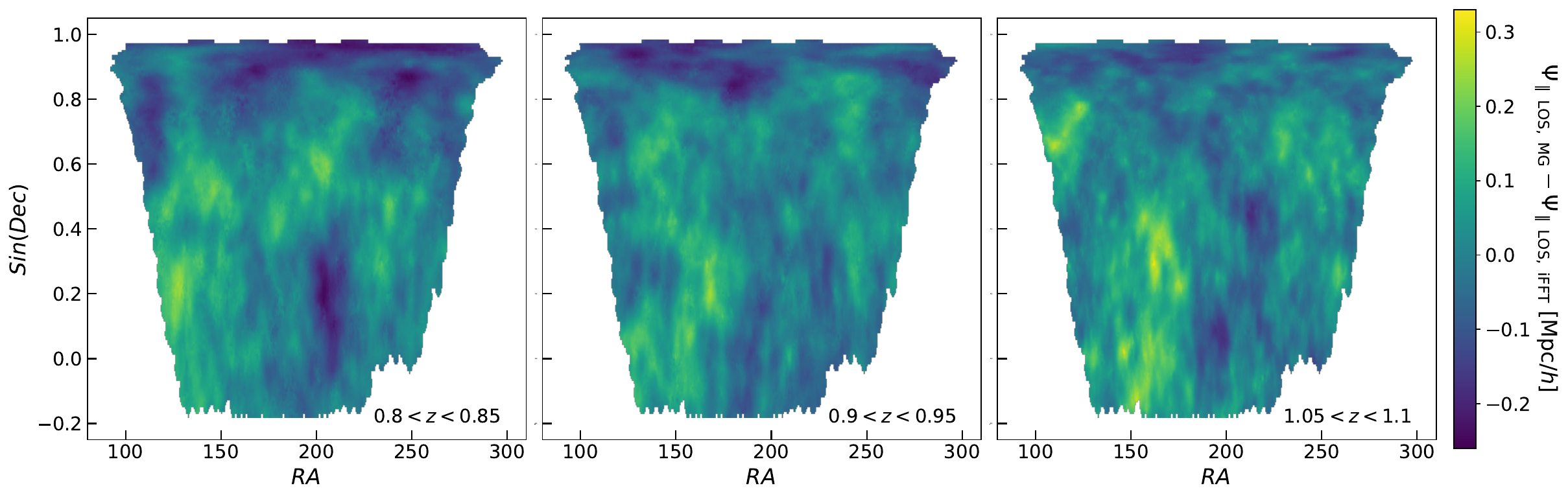}
    \caption{Line-of-sight component of the displacement magnitude difference between \textbf{MG} and \textbf{iFFT} (3 iterations) (both applied with a 15 Mpc/$h$ smoothing) projected to DESI Y5 NGC footprint, shown for \textbf{ELG} $0.8<z<0.85$ (left), $0.9<z<0.95$ (middle) and $1.05<z<1.1$ (right) redshift slices. We show the sine of the declination such that the bins are uniform. The high redshift end shows slightly larger differences. Overall, the magnitude of differences whether on the redshift boundary or not is small compared to the average line-of-sight displacement magnitude of ELG ($\sim$ 1.9 Mpc/$h$). There also does not appear to be larger differences clustered along survey boundaries.}
    \label{fig:los_scalar_footprint_colormap}
\end{figure}

We next examine whether there is a geometry dependence for the displacement differences, especially focusing on the survey boundaries. Along survey boundaries, the density field has regions where there is no data, so we expect larger errors of the displacement along boundaries. 
We therefore plot the differences in the line-of-sight displacement as a function of sky position. The differences here are the differences between MG and iFFT, and not the differences between what one would have obtained had there been data outside the boundary. 
Figure~\ref{fig:los_scalar_footprint_colormap} shows the line-of-sight displacement magnitude differences between MG and iFFT on the sky for three slices of redshift, $0.8<z<0.85$, $0.9<z<0.95$ and $1.05<z<1.1$. These three redshift bins include the redshift boundaries of this ELG sample and one in the middle. There is a slight increase in the difference along the high redshift end, although the magnitude is again small, compared to the average line-of-sight displacement magnitude (maximum $\sim$0.3 Mpc/$h$ compared to $\sim$1.9 Mpc/$h$). We do not observe outlier displacement differences clustered along survey boundaries. We have also examined the differences of angles in the perpendicular to the line-of-sight direction between MG and iFFT on the footprint, but we do not observe any dependence on the survey geometry. These suggest that the large displacement differences are not associated with survey boundaries.

\begin{figure}
    \centering
    \includegraphics[width=\linewidth]{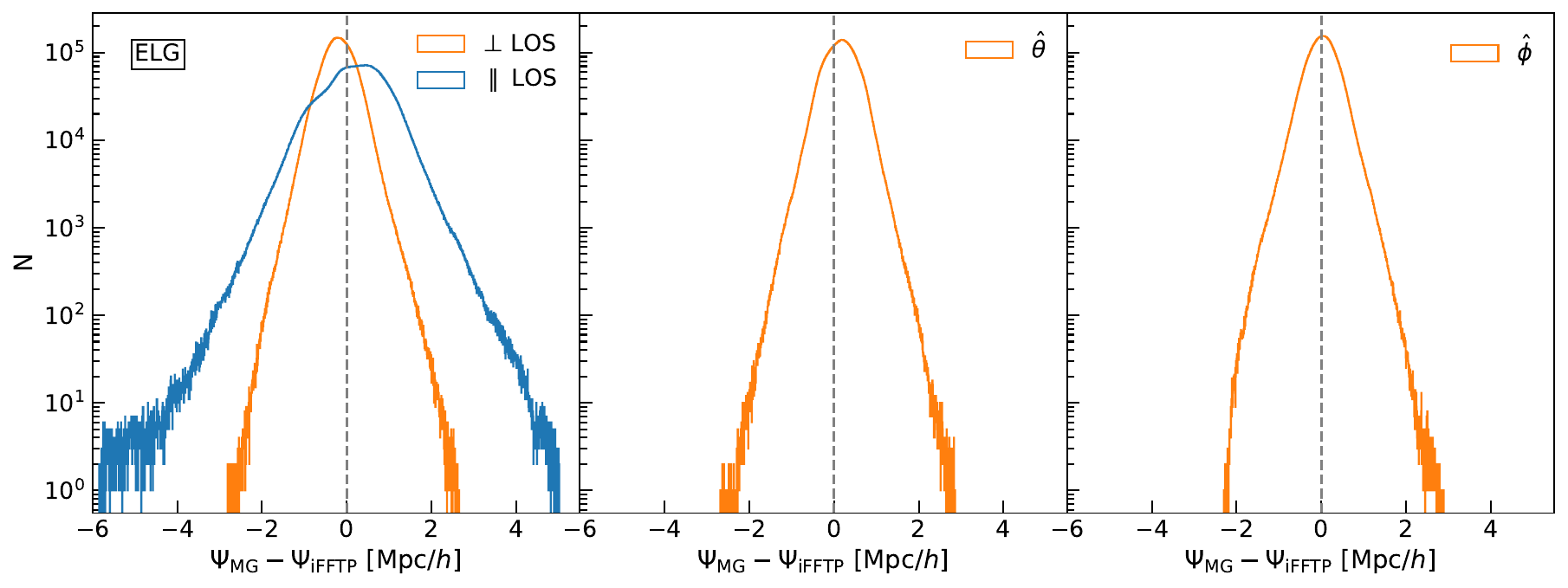}
    \caption{The distribution of displacement vector differences between \textbf{MG} and \textbf{iFFTP} (3 iterations) displacements of the \textbf{ELG} sample using one mock. The displacement vectors are decomposed into along the line-of-sight $\boldsymbol{\hat{r}}$ and perpendicular to the line-of-sight directions (left), and the perpendicular to the line-of-sight direction is further decomposed into $\boldsymbol{\hat{\theta}}$ (middle) and $\boldsymbol{\hat{\phi}}$ (right) directions in spherical coordinates. 
    Along the line of sight, the spread of the distribution is larger compared to perpendicular to the line of sight, indicating that the different ways of treating the redshift direction by the two algorithms result in more differences in the displacement in the redshift direction. The two algorithms have less differences perpendicular to the line-of-sight direction, as shown in the middle and right panels with the component perpendicular to the line of sight further decomposed. Especially along the line of sight, the magnitude of the differences can be larger than the average line-of-sight displacement of ELG ($\sim$1.9 Mpc/$h$). The displacement difference between iFFTP and MG shown here is obviously larger than the displacement difference between iFFT and MG as shown in Figure~\ref{fig:ELG_displacement_decompose}, clearly presenting the discrepancy between the two algorithms.  }
    \label{fig:ELG_displacement_decompose_IFFTP}
\end{figure}

\begin{figure}
    \centering
    \includegraphics[width=\linewidth]{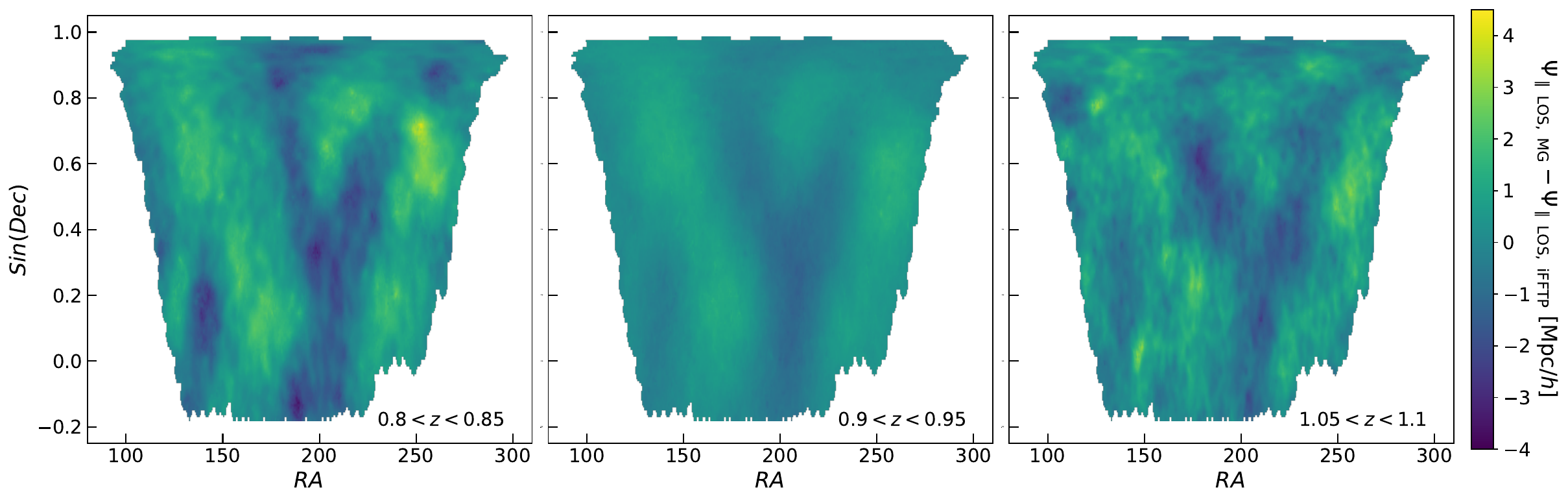}
    \caption{Line-of-sight component of the displacement magnitude difference between \textbf{MG} and \textbf{iFFTP} (3 iterations) (both applied with a 15 Mpc/$h$ smoothing) projected to DESI Y5 NGC footprint, shown for \textbf{ELG} $0.8<z<0.85$ (left), $0.9<z<0.95$ (middle) and $1.05<z<1.1$ (right) redshift slices. We show the sine of the declination such that the bins are uniform. The low and high redshift end both show $ \sim 2-3$ times larger differences than a redshift slice in the middle, indicating that iFFTP has problems at the boundary. The magnitude of differences, especially on the redshift boundary, can be larger than the average line-of-sight displacement magnitude of ELG ($\sim$ 1.9 Mpc/$h$).}
    \label{fig:los_scalar_footprint_colormap_IFFPT}
\end{figure}

Next, we show the displacement comparison between MG and iFFTP in the distribution of the magnitude differences along and perpendicular to the line of sight as well as at survey boundaries. The much larger scales shown in Figures~\ref{fig:ELG_displacement_decompose_IFFTP} and~\ref{fig:los_scalar_footprint_colormap_IFFPT} evidently show that there is significantly more difference between iFFTP and MG. The magnitude difference can be larger than the average line-of-sight displacement of ELG. Figure~\ref{fig:los_scalar_footprint_colormap_IFFPT} also shows that there is $\sim2-3$ times larger differences between MG and iFFTP at redshift boundaries than in redshift slices in the middle, confirming the boundary problems of iFFTP.


\subsubsection{Power spectrum (RecSym)}\label{sec:recsym_pk}

\begin{figure}
    \centering
    \includegraphics[width=0.49\columnwidth]{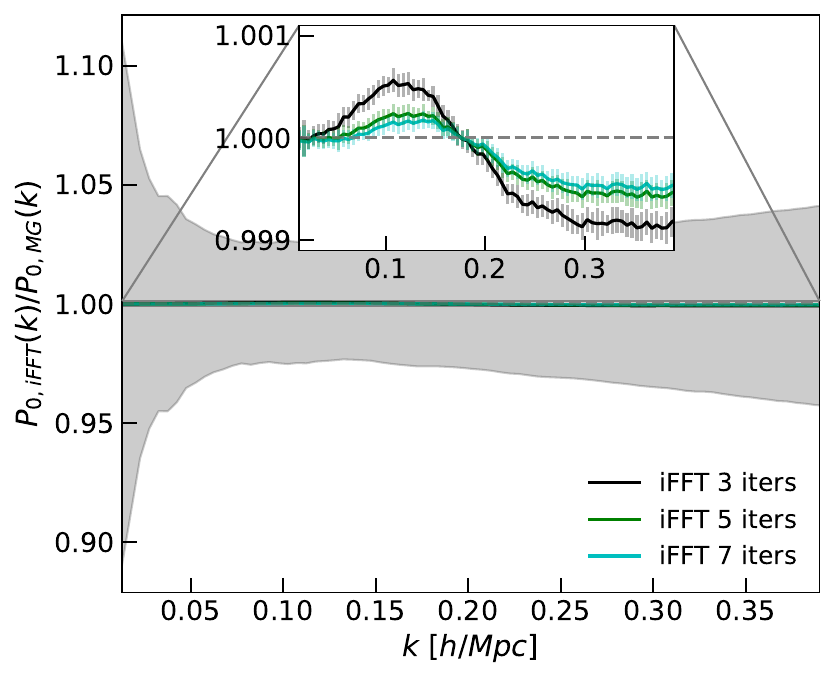}
    \includegraphics[width=0.48\columnwidth]{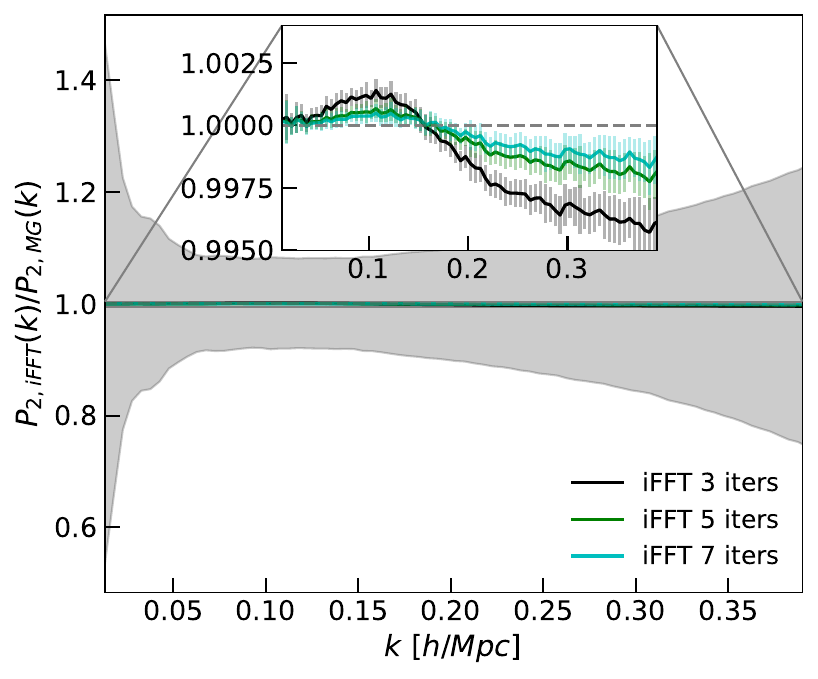}
    \caption{Ratio in the post-reconstruction power spectrum monopole (left) and quadrupole (right) between iFFT (black for 3 iterations, green for 5 iterations, and cyan for 7 iterations) and MG (both applied with a 15 Mpc/$h$ smoothing), using the \textbf{RecSym} convention for the \textbf{ELG} sample, averaged over 25 mocks. We compare the standard deviation of the differences between the two power spectra normalized by the mean of the MG power spectrum (shown as the error bars) with the DESI DR1 ELG power spectrum errors as measured by the square root of the diagonal of its covariance (shown as the grey band). The insets zoom in on the mean of 25 mocks with the error bars. iFFT appears to converge. The differences between the two methods are minimal compared to the DR1 power spectrum errors. Even with 3 iterations, the difference between the mean of iFFT and MG is within 0.1\% for monopole and within 0.4\% for quadrupole. }
    \label{fig:Pk_ELG2_MG_iFFT}
\end{figure}

We now compare iFFT and MG with their \textbf{RecSym} post-reconstruction power spectra.
Figure~\ref{fig:Pk_ELG2_MG_iFFT} shows the post-reconstruction power spectrum monopole and quadrupole ratios, between these two algorithms for the ELG sample, for the average of 25 mocks. We show the results in linear space (rather than log space) to focus on the BAO scale.
We show as error bars the standard deviation of the difference of the two algorithms in the 25 DESI Y5 mocks divided by the mean of the MG power spectra, and we compare these errors to the DESI DR1 power spectrum errors, shown as the grey shaded regions in the figure. The DR1 power spectrum errors are taken from the covariance matrix of the post-reconstruction power spectrum multipoles.
The insets zoom in on the mean of the 25 mocks with the aforementioned error bars. 
We observe that iFFT appears to be converging, since the more iterations the iFFT goes through, the smaller the differences there are between iFFT and MG, although it is not converging to the MG result. The difference, however, is small. The typical differences between the two algorithms are significantly smaller than the DR1 power spectrum errors. The maximum difference of about 0.1\% in monopole and 0.4\% in quadrupole (with 3 iterations of iFFT) is only about 2.5\% of the DR1 measurement error at the same scale. The differences between the two algorithms are also smaller than the approximate Y5 power spectrum errors, which can be obtained by downscaling the DR1 errors by a factor of 1.7 (the volume difference between Y5 and DR1 is about a factor of 2-3). The minimal differences shown here suggest that the impact on the BAO constraints due to using one algorithm versus the other would be insignificant for DESI DR1 as well as for the final DESI data. 
We find that for all choices of the number of iterations used (3, 5 or 7) for iFFT, 
the differences between the two algorithms are all very small. Even with only 3 iterations, the difference between iFFT and MG is within 0.1\% for monopole and 0.4\% for quadrupole, within our error budget. So, the the power spectrum appears to be robust against the small displacement differences between MG and iFFT shown in \cref{sec:ELG_displacement}.

\begin{figure} 
    \centering
    \includegraphics[width=0.7\columnwidth]{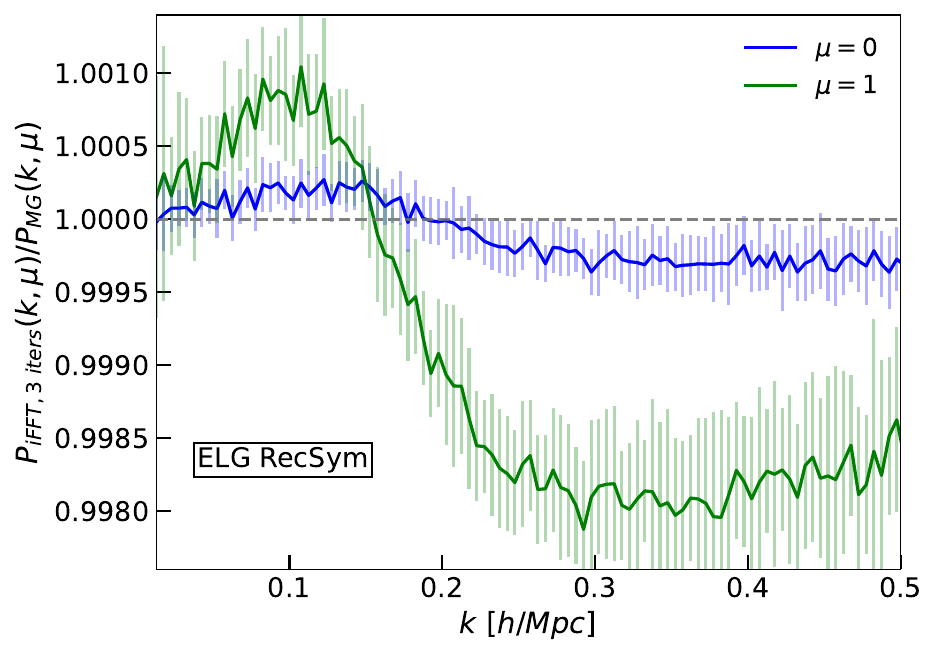}
    \caption{
    The ratio of post-reconstruction power spectra between iFFT (3 iterations) and MG for the \textbf{ELG} sample, for $\mu=0$ (blue) and 1 (green), where $\mu$ is the cosine of the angle between the line of sight and $\boldsymbol{k}$. Both algorithms are applied with a 15 Mpc/$h$ smoothing and using the \textbf{RecSym} convention. Here $\mu=0$ is perpendicular to the line of sight and $\mu=1$ is along the line of sight. Each line is an average of 25 mocks. We observe that both cases show differences between iFFT and MG within 0.3\%. However, along the line of sight, the differences between the two algorithms are larger, especially on smaller scales ($k\gtrsim 0.2 h$/Mpc).  }
    \label{fig:Pk_angle}
\end{figure}

We further examine the post-reconstruction power spectrum along and perpendicular to the line of sight, plotting it as a function of $k$ and $\mu$, where $\mu$ is the cosine of the angle between the line of sight and the wavenumber $\boldsymbol{k}$. This is calculated by using the multipoles we measure
\begin{equation}
    P_s(k,\mu)=\sum_{l=0}^{4}P_{s,l}(k)\mathcal{P}_{l}(\mu),
\end{equation}
where $\mathcal{P}_{l}(\mu)$ is the Legendre polynomial of order $l$. The subscript $s$ denotes redshift space.
Figure~\ref{fig:Pk_angle} shows the ratio of power spectra between iFFT (3 iterations) and MG when $\mu=0$ (perpendicular to the line of sight) and when  $\mu=1$ (along the line of sight). We use the multipoles $l=0$, 2, and 4 to compute $P(k,\mu)$. The differences overall are very small, i.e. at the subpercent level, but closer to the line of sight as well as towards smaller scales, we observe slightly increased differences in the power spectrum. These differences are at the maximum $\sim$0.3\%.

\begin{figure}
    \centering
    \includegraphics[width=0.49\columnwidth]{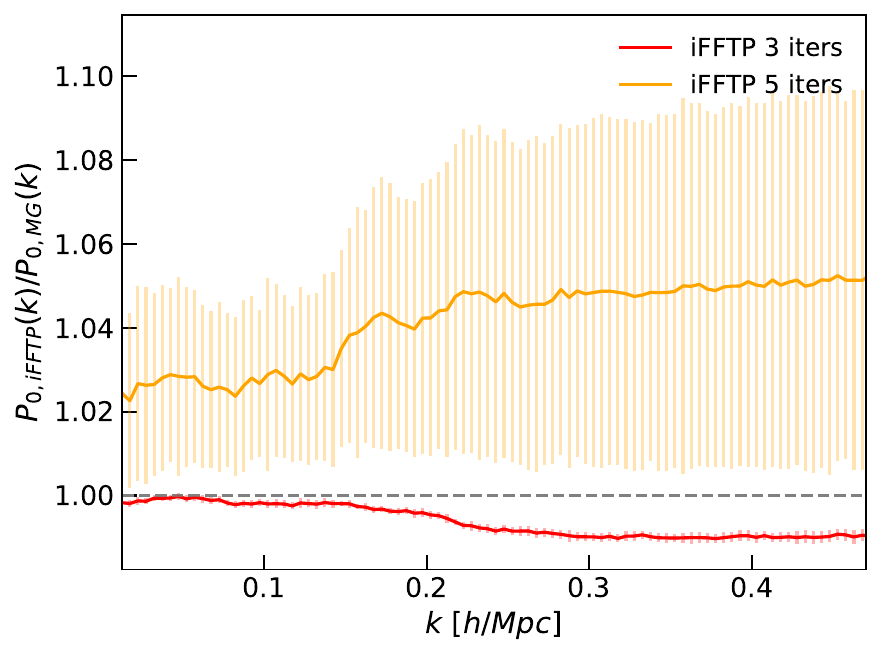}
    \includegraphics[width=0.49\columnwidth]{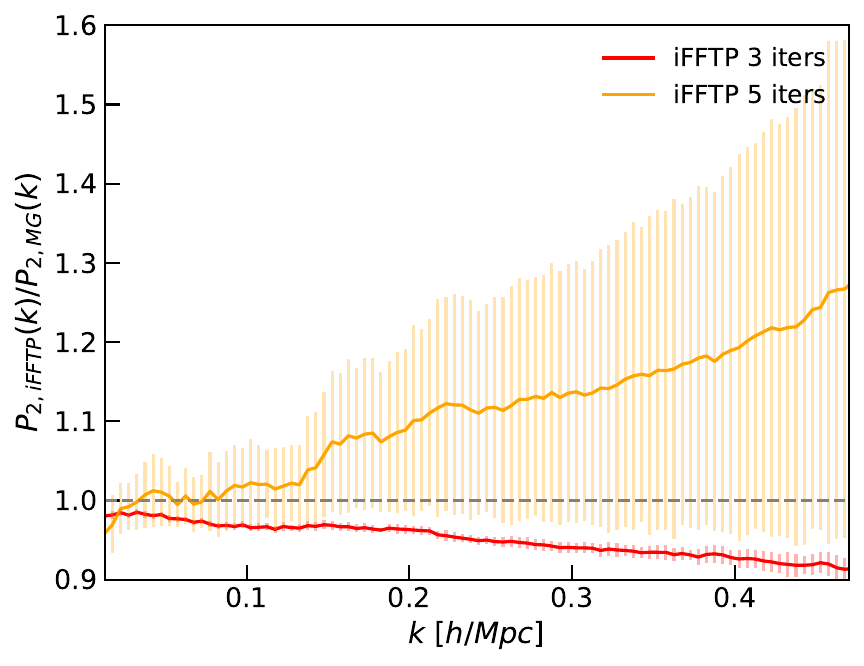}
    \caption{Ratio of post-reconstruction monopole (left) and quadrupole (right) power spectra between iFFTP (red for 3 iterations, orange for 5 iterations) and MG (both applied with a 15 Mpc/$h$ smoothing), using the \textbf{RecSym} convention for the \textbf{ELG} sample, averaged over 25 mocks. The iFFTP algorithm blows up with more iterations, so we do not include results with more iterations here. For 3 iterations, the agreement between iFFP and MG is within 1\% for monopole and 10\% for quadrupole. For 5 iterations, the agreement decreases to 5\% for monopole and 25\% for quadrupole. }
    \label{fig:Pk_ELG2_MG_iFFTP}
\end{figure}

Now we compare the performance of iFFTP and MG in the power spectrum. In Figure~\ref{fig:Pk_ELG2_MG_iFFTP}, we show the ratio of post-reconstruction power spectra by the two algorithms. eBOSS analyses used 3 iterations of iFFTP, so our consideration of number of iterations starts at 3.
Unlike MG and iFFT, the iFFTP algorithm appears to diverge with increasing numbers of iterations.
With 3 iterations, the agreement between iFFTP and MG is within 1\% for monopole and 10\% for quadrupole for the scales typically used for BAO fitting. This is at a much larger disagreement than what is between iFFT and MG. With 5 iterations, the agreement between iFFTP and MG go down to within 5\% and 25\% for monopole and quadrupole, respectively. The power spectrum from 7 iterations of iFFTP agrees with MG even worse. The reason why iFFTP diverges is that moving galaxies iteratively to remove RSDs without matched randoms can move some galaxies outside the survey boundaries, which results in errors in the density estimates at the boundaries, thus affecting the displacement calculation, as shown in \cref{sec:ELG_displacement}. 
The boundary effect is worse in the redshift direction, as indicated by the quadrupole (right panel). This problem could be improved by moving the randoms during the iterations to match the galaxies at each step, but we do not consider such an extension, as the iFFT algorithm avoids this problem by removing the RSDs without moving the galaxies. We also note that this is a problem with applying the iFFTP algorithm in realistic simulations having survey boundaries and not in simulations with periodic boundary conditions, which we will discuss in \cref{sec:discussion}.

The eBOSS analyses \cite{Bautista18,Bautista21} did use iFFTP (with 3 iterations). 
As indicated in the re-analysis of eBOSS in comparison to DESI DR1 pipeline \cite{KP4}, the difference caused by using the iFFTP algorithm versus iFFT together with other changes of BAO fitting procedure is at most a $\sim$0.2$\sigma$ shift in the constraints of the BAO parameters. While not a substantial deviation, in the current era of precision cosmology, we need to caution such systematics. Considering that iFFTP does not converge and needs care at the survey boundaries, together with Figure~\ref{fig:Pk_ELG2_MG_iFFTP} that shows considerable deviation from other algorithms, we recommend against the use of iFFTP in BAO survey analysis without further development.

\subsubsection{Cross-correlation with the initial density field (RecSym)}\label{sec:recsym_gk}
To assess how well reconstruction recovers the linear density field, we can cross-correlate the reconstructed density field with the initial condition in the simulation. 
Here, we cross-correlate with the initial condition field in the presence of the survey geometry.
The initial condition fields are created in two steps. The \textsc{AbacusSummit} simulations generate Gaussian random fields as initial condition fields at $z=99$ by using the linear theory power spectrum at $z=1$ and scaling it back to $z=99$ in a way such that the simulations arrive at $z=1$ with the correct linear power spectrum for the non-neutrino component \cite{Maksimova21}. To generate initial condition DESI mocks, similar to snapshots at other redshifts, these initial fields are tiled with multiple copies to cover the DESI volume (because the simulation boxes are smaller than the observed volume) \cite{KP3}. In this analysis of cross-correlation with the initial condition, we focus on the ``propagator'', which is a cross-correlation normalized only by the initial power spectrum. This metric is commonly used in BAO analysis, because it characterizes the nonlinear damping of the BAO feature. The angle-averaged propagator is defined as
\begin{equation}\label{eq:gk}
G(k)=\frac{\left<\delta_{\rm recon}(k)^{*}\delta_{\rm ini}(k)\right>}{\left<\delta_{\rm ini}(k)^2\right>}.
\end{equation}
For \textbf{RecSym}, where galaxies and randoms are displaced by the same amount, hence RSD is not removed, an effective reconstruction will result in the propagator asymptotically converging to the Kaiser factor on large scales
(assuming the large-scale bias is removed). 
The angle-averaged propagator is then expected to converge to 
$(1+1/3\beta)$ on large scales in the plane-parallel approximation. For \textbf{RecIso}, where RSD is removed, an effective reconstruction will lead to the propagator at unity on large scales. Examining the propagator at different angles from the line of sight, as measured by $\mu$, 
we can also assess the recovery of the linear density field along different directions. We compute the angle-dependent propagator by
\begin{equation}
    G(k,\mu)=\sum_{l=0}^{\infty}G_l(k)\mathcal{P}_l(\mu).
\end{equation}
Here, ${P}_l(\mu)$ is again the Legendre polynomial of order $l$. Qualitatively, the slower the propagator drops from the perfect correlation, 
the smaller the BAO damping scale is. This damping scale can be obtained by fitting the propagator with a Gaussian or modified Gaussian model \cite{Seo16}.

\begin{figure}
    \centering
    \includegraphics[width=\columnwidth]{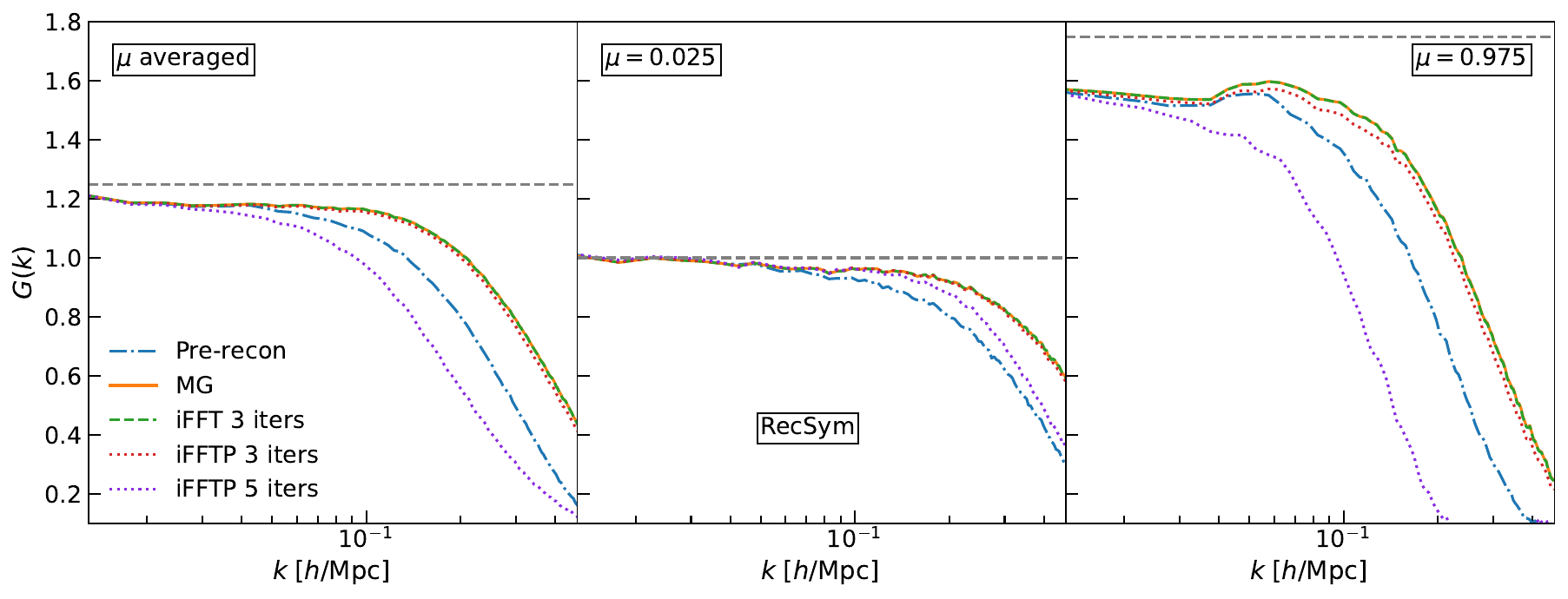}
    \caption{Propagator for the three reconstruction algorithms for one \textbf{ELG} mock with the \textbf{RecSym} convention when $\mu$ is integrated over (left), and in $\mu=0.025$ (middle) and 0.975 (right) bins, in comparison of pre-reconstruction propagator (blue dash dotted). The $\mu$-bin width is 0.05. The horizontal dashed lines show expected Kaiser approximation in the plane-parallel approximation, $(1+1/3\beta)=1.25$ for the first panel and $(1+\beta\mu^2)=1.0$ and 1.75 for the middle and right panels, respectively. We shift the propagators by 8\% to correct for misestimate of the linear bias. MG, iFFT and iFFTP (3 iterations) return improved propagator, with MG and iFFT performing slightly better than iFFTP with 3 iterations. MG and iFFT overlap for all three cases. iFFTP with 3 iterations shows noticeable differences from MG and iFFT, while still having improvement over pre-reconstruction; however, iFFTP with 5 iterations shows degraded propagator along the line of sight and when $\mu$ is integrated over. }
    \label{fig:ELG_Gk_recsym}
\end{figure}

In Figure~\ref{fig:ELG_Gk_recsym}, we show the propagator $G(k)$ when integrating over the angle to the line of sight (i.e. when angle-averaged),
and in two $\mu$ bins in bin width of 0.05, when $0<|\mu|<0.05$ (perpendicular to the line of sight) and when $0.95<|\mu|<1$ (along the line of sight). We shift the amplitude of all propagators by about $8\%$ to correct for inaccurate measurement of linear bias such that the $\mu=0.025$ panel (perpendicular to the line of sight, where RSDs do not contribute) is at 1 on large scales\footnote{DESI DR2 BAO analysis shows that BAO is unbiased with a 10\% offset in the linear bias estimate \cite{Andrade25}.}. We observe that MG, iFFT (3 iterations), as well as iFFTP (3 iterations) do an excellent job of recovering the linear density. MG and iFFT again agree well. iFFTP with 3 iterations shows noticeable differences and is slightly worse than MG or iFFT. However, iFFTP with 5 iterations is considerably worse than the other two. Along the line of sight, iFFTP with 5 iterations is even worse than pre-reconstruction. So we see similar trends in this propagator analysis that the iFFTP results get progressively worse with more iterations, due to the boundary effect discussed above. 
When angle is integrated over, 
the expected amplitude of the propagator in the large-scale limit in the plane-parallel approximation is around $1+1/3\beta=1.25$, for ELG with $\beta=0.75$. However, the measured propagator amplitude at low $k$ in the left panel of Figure~\ref{fig:ELG_Gk_recsym} is close to but below this factor. The disagreement is stronger along the line of sight, as shown in the right panel (the expected amplitude in the plane-parallel approximation is $1+\beta=1.75$ for the line-of-sight direction). This is likely due to the fact that in real surveys, lines of sight vary, so the Kaiser factor differs from that for a fixed line of sight.
It could also be affected by an inaccurate estimate of $f$. 
Despite the mismatch with the expected asymptotic behavior, the result of our comparison still holds.

\subsubsection{RecIso}\label{sec:ELG_reciso}
In this section, we discuss some differences that present in the post-reconstruction power spectrum and propagator of the \textbf{RecIso} convention. Because the estimate of the displacement is exactly the same between \textbf{RecSym} and \textbf{RecIso}, the differences in the summary statistics are purely differences in the convention, i.e. how the randoms are moved along the line of sight. In particular, in \textbf{RecIso}, randoms are displaced by a different amount in comparison to galaxies, by omitting the additional displacement for RSDs, thereby removing the RSD effect.
Nonetheless, these differences in the summary statistics show the different sensitivities of the conventions to the algorithm differences, which help us evaluate the conventions.

\begin{figure}
    \centering
    \includegraphics[width=0.488\linewidth]{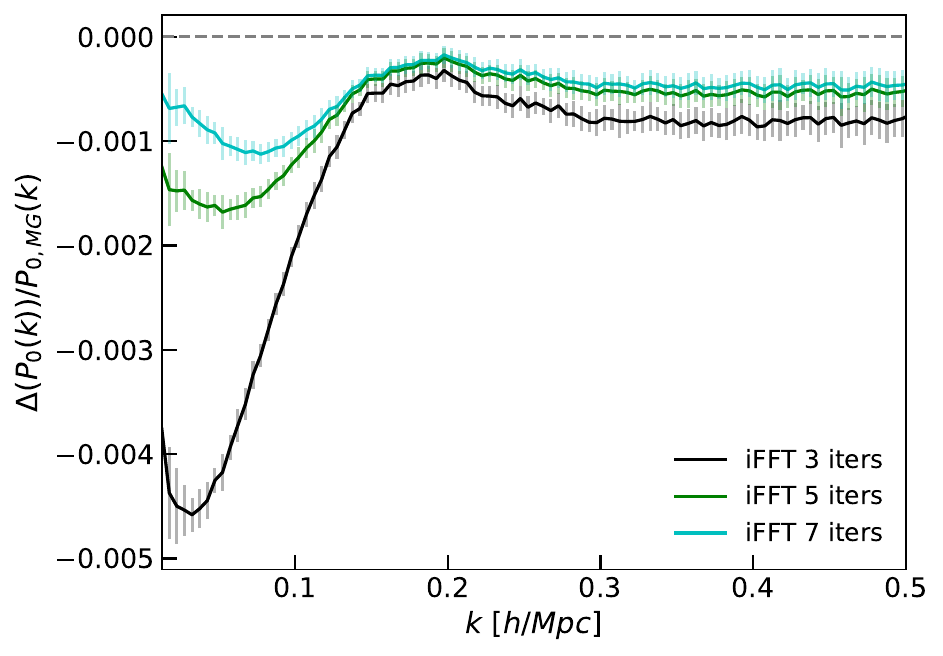}
    \includegraphics[width=0.492\linewidth]{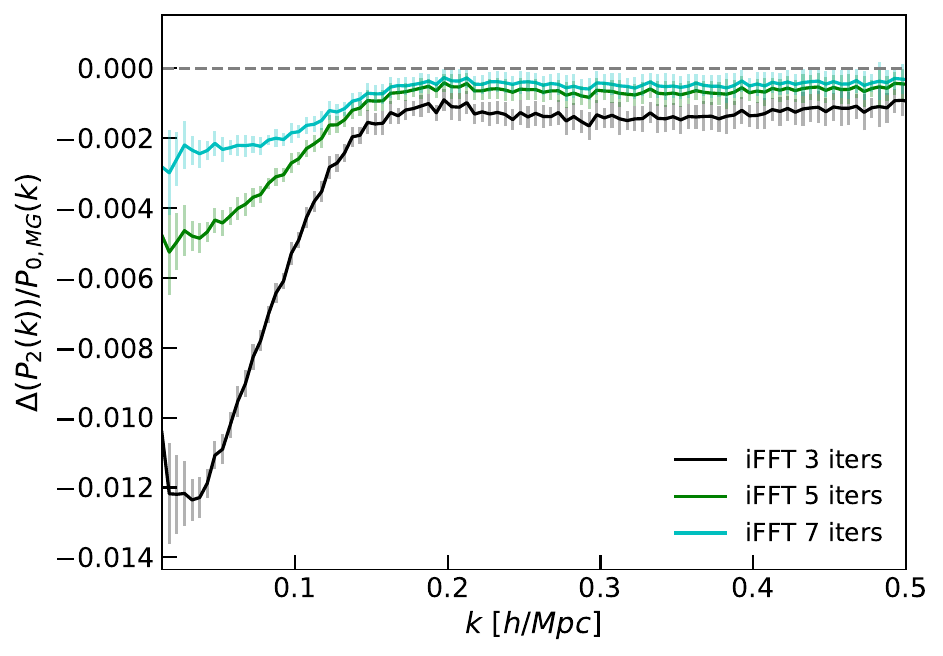}
    \caption{The monopole (left) and quadrupole (right) power spectrum differences between iFFT (black for 3 iterations, green for 5 iterations, and cyan for 7 iterations) and MG (both applied with a 15 Mpc/$h$ smoothing) divided by the MG monopole, using the \textbf{RecIso} convention for the \textbf{ELG} sample. Each line is an average of 25 mocks. The differences appear to be larger on larger scales. The maximum discrepancy is about 0.5\% in monopole and 1.2\% in quadrupole (w.r.t. monopole) with 3 iterations of iFFT. }
    \label{fig:ELG_Iso}
\end{figure}

We first show the ELG \textbf{RecIso} post-reconstruction power spectrum comparison for the MG and iFFT algorithms in Figure~\ref{fig:ELG_Iso}. We show the difference with respect to the (MG) monopole because of the zero-crossing in the quadrupole. We observe that iFFT with 3 iterations can differ from MG at 0.5\% for monopole and the difference is the largest on large scales. The quadrupole displays the same trend, and iFFT with 3 iterations can differ from MG at 1.2\% on large scales (w.r.t. monopole).  Since \textbf{RecIso} attempts to remove large-scale RSDs\footnote{Because the galaxy field, i.e. the displaced field, removes the large-scale RSDs in both conventions, so technically \textbf{RecIso} achieves not removing RSDs by not adding the RSDs back to the density field when moving randoms.}
, the imperfectness of the estimate of the displacement more easily presents on large scales in \textbf{RecIso}. \textbf{RecSym}, on the other hand, has to converge to linear theory on large scales. This suggests that the differences in the estimated line-of-sight displacements by the different algorithms surface more clearly with \textbf{RecIso}; thus, \textbf{RecIso} is more sensitive to the accuracy of reconstruction along the line of sight.

\begin{figure}
    \centering
    \includegraphics[width=0.7\columnwidth]{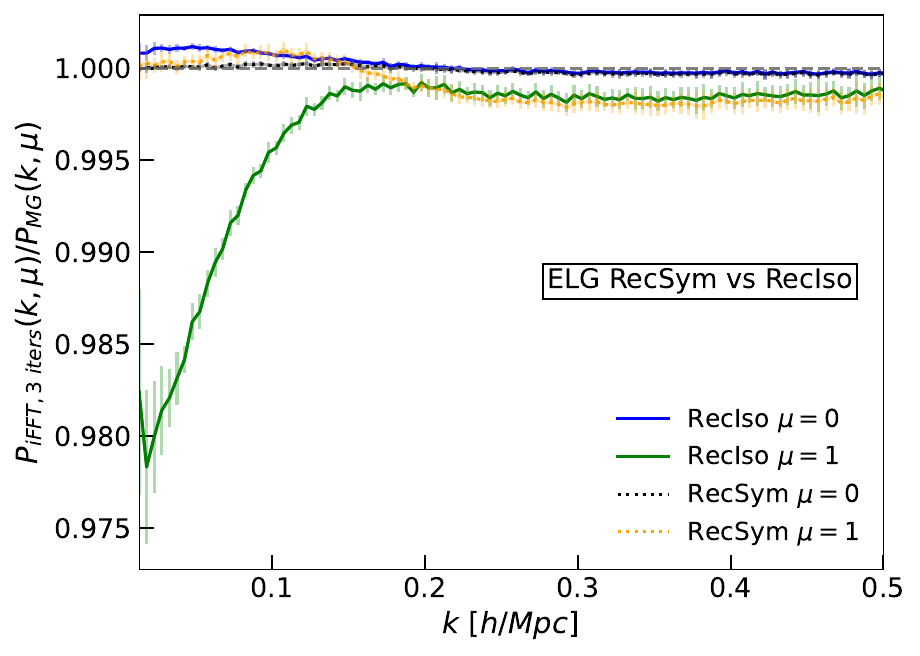}
    \caption{
    Ratio of power spectrum between iFFT and MG for the \textbf{ELG} sample with both algorithms applied with a 15 Mpc/$h$ smoothing, comparing the \textbf{RecIso} convention (solid) with the \textbf{RecSym} convention (dotted), for $\mu=0$ (perpendicular to the line of sight, blue for \textbf{RecIso} and black for \textbf{RecSym}) and 1 (parallel to the line of sight, green for \textbf{RecIso} and yellow for \textbf{RecSym}).
    The \textbf{RecSym} lines are the same as those shown in Figure~\ref{fig:Pk_angle}. 
    Each line is an average of 25 mocks. We observe discrepancies along the line of sight between the two algorithms more in the \textbf{RecIso} convention than the \textbf{RecSym} convention, and on large scales, where the differences can be as large as 2\%. However, at $k\gtrsim0.15\ h$/Mpc, the two conventions return close results. }
    \label{fig:Pkmu_ELG_Iso}
\end{figure}

Next, we show \textbf{RecIso} $P(k,\mu)$ in Figure~\ref{fig:Pkmu_ELG_Iso} to further examine the performance along and perpendicular to the line of sight. We observe larger differences in the power spectrum towards larger scales and along the line of sight with \textbf{RecIso}, indicated by the clear deviation at $\mu=1$, about 2\% at its maximum. In \textbf{RecSym}, the maximum difference in $P(k,\mu=1)$ is 0.3\% (shown as yellow dotted line in this figure for comparison). This again suggests that the differences in the line-of-sight displacement between three iterations of iFFT and MG more easily present in \textbf{RecIso} (in $k\lesssim 0.1 h$/Mpc), 
although again the difference is of small magnitude compared to the power spectrum value.

\begin{figure}
    \centering
    \includegraphics[width=0.495\linewidth]{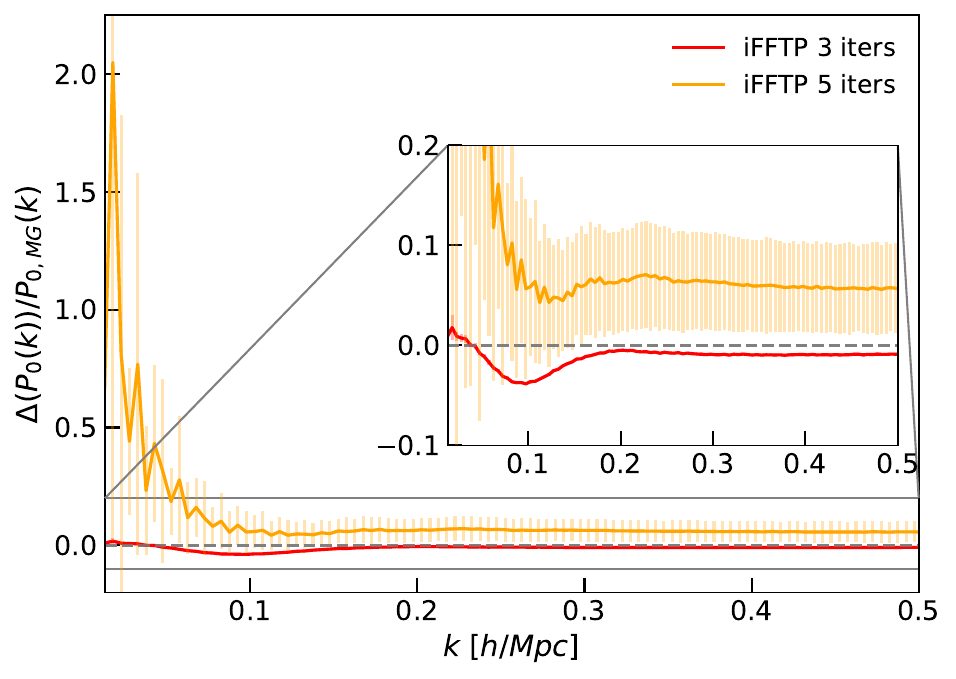}
    \includegraphics[width=0.49\linewidth]{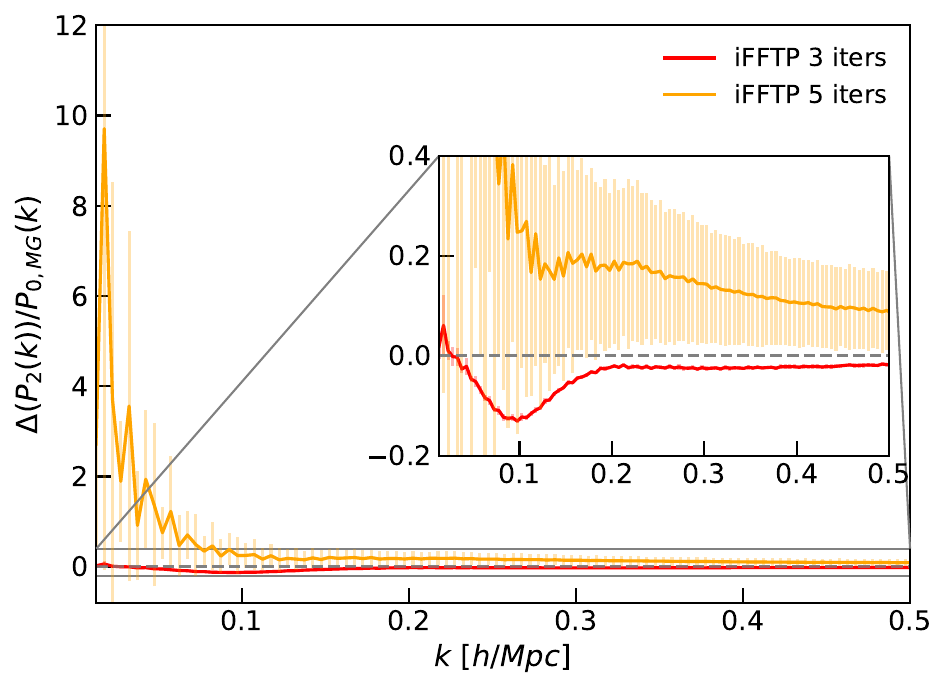}
    \caption{Ratio in power spectrum monopole between iFFTP (red for 3 iterations, orange for 5 iterations) and MG (both applied with a 15 Mpc/$h$ smoothing),  using the \textbf{RecIso} convention for the \textbf{ELG} sample. The iFFTP exhibits more differences in \textbf{RecIso} than in \textbf{RecSym}. For 3 iterations, the disagreement between iFFTP and MG is at most 5\% in monopole and 10\% in quadrupole (both w.r.t.\ MG monopole) on large scales. For 5 iterations, the disagreement increases to a factor of 2 in monopole and a factor of 10 in quadrupole on large scales. On scales more relevant to BAO, however, five iterations of iFFTP gives around 7\% difference in monopole and around 15\% difference in quadrupole. 
}
    \label{fig:ELG_iFFTP_P0_Iso}
\end{figure}

The differences are also larger between iFFTP and MG in \textbf{RecIso} power spectrum compared to \textbf{RecSym}, as shown in Figure~\ref{fig:ELG_iFFTP_P0_Iso}. In monopole, three iterations of iFFTP differs from MG at most 5\% on large scales; in quadrupole, the differences can be at most 10\%. 
For 5 iterations of iFFTP, \textbf{RecIso} shows dramatic difference on large scales; in monopole, the differences between iFFTP and MG are a factor of 2 and in quadrupole, the differences can be almost a factor of 10. On scales more relevant for BAO, however, five iterations of iFFT gives around 7\% difference in monopole and 15\% difference in quadrupole. For comparison, the maximum differences with \textbf{RecSym} are around 1\% in monopole and 10\% in quadrupole with 3 iterations and 5\% and 25\% with 5 iterations, with no extreme differences present on the largest scales.

\begin{figure}
    \centering
    \includegraphics[width=\linewidth]{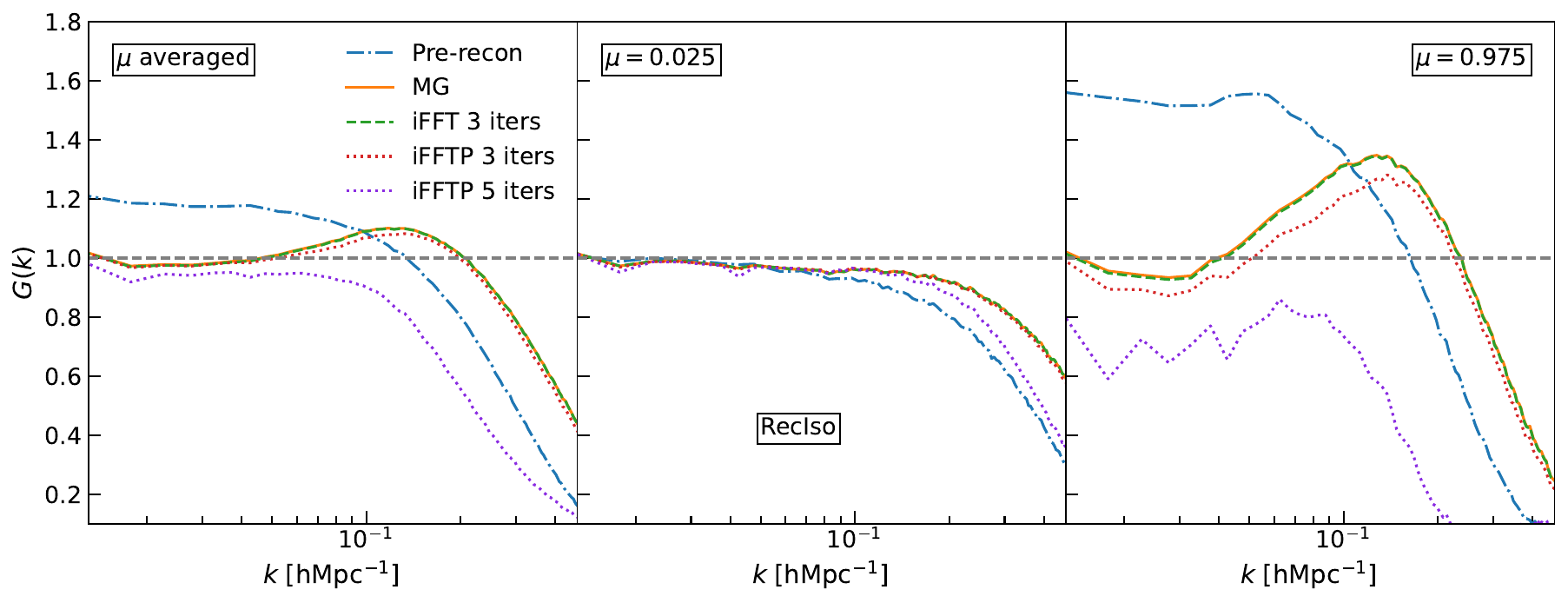}
    \caption{Propagator for the three reconstruction algorithms for one \textbf{ELG} mock with the \textbf{RecIso} convention when $\mu$ is integrated over (left), and in $\mu=0.025$ (middle) and 0.975 (right) bins, in comparison of pre-reconstruction propagator (blue dash dotted). The horizontal dashed lines show the expected amplitude (middle) or the expected amplitude after removing RSDs (left and right). We shift the propagators by 8\% to correct for misestimate of the linear bias. All three reconstruction algorithms return improved propagator, with MG and iFFT perform slightly better than iFFTP (3 iterations). MG and iFFT are on top of each other for all three cases, while iFFTP (3 iterations) shows noticeable differences, which are larger compared to \textbf{RecSym}. iFFTP with 5 iterations again shows degraded propagator along the line of sight and when $\mu$ is integrated over, which also appears to be worse than \textbf{RecSym}. The shapes of the propagator by all three algorithms in the $\mu=0.975$ bin show distortions on large scales, suggesting that the estimate of the line-of-sight displacement is challenging for all three algorithms.  }
    \label{fig:ELG_Gk_reciso}
\end{figure}

We also explore the propagator for \textbf{RecIso} reconstruction. The expected shape for the propagator in \textbf{RecIso} is different from that in \textbf{RecSym} due to the effect of RSDs. In particular, the propagator is expected to be at unity on large scales, if the removal of RSDs is perfect. In addition, there is a slight angle-dependent bump at the turn of the propagator due to RSDs \cite[e.g.][]{Seo16b,recon}, which can be characterized by $(1+\beta[1-\Sigma(k)]\mu^2)$ where $\Sigma(k)$ is the same smoothing kernel used in reconstruction \cite[][]{Seo16b}. The smaller scales, however, should be identical to \textbf{RecSym}. As with \textbf{RecSym} propagators shown in Figure~\ref{fig:ELG_Gk_recsym}, we also shift the propagators here to correct for misestimate of the linear bias. As shown in Figure~\ref{fig:ELG_Gk_reciso}, MG, iFFT (3 iterations), and iFFTP (3 iterations) show that the reconstructed densities are closer to the initial condition than the pre-reconstructed density, indicating that reconstruction is effective in these cases. In this figure, MG and iFFT (3 iterations) are on top of each other, while iFFTP (3 iterations) exhibits clear differences, more prominent than those in the \textbf{RecSym} propagator (Figure~\ref{fig:ELG_Gk_recsym}), especially along the line of sight. iFFTP with 5 iterations is noticeably worse. While still reducing the BAO damping scale perpendicular to the line of sight, iFFTP with 5 iterations increases the damping scale along the line-of-sight direction. 
The differences between MG and iFFT remain small in the propagator, same as in the \textbf{RecSym} case. The shape of the propagator is not constant on large scales in the $\mu=0.975$ bin, suggesting that the estimate of the line of sight displacement is challenging in all three algorithms, although iFFTP is behaving the worst.

\begin{figure}
    \centering
    \includegraphics[width=1\linewidth]{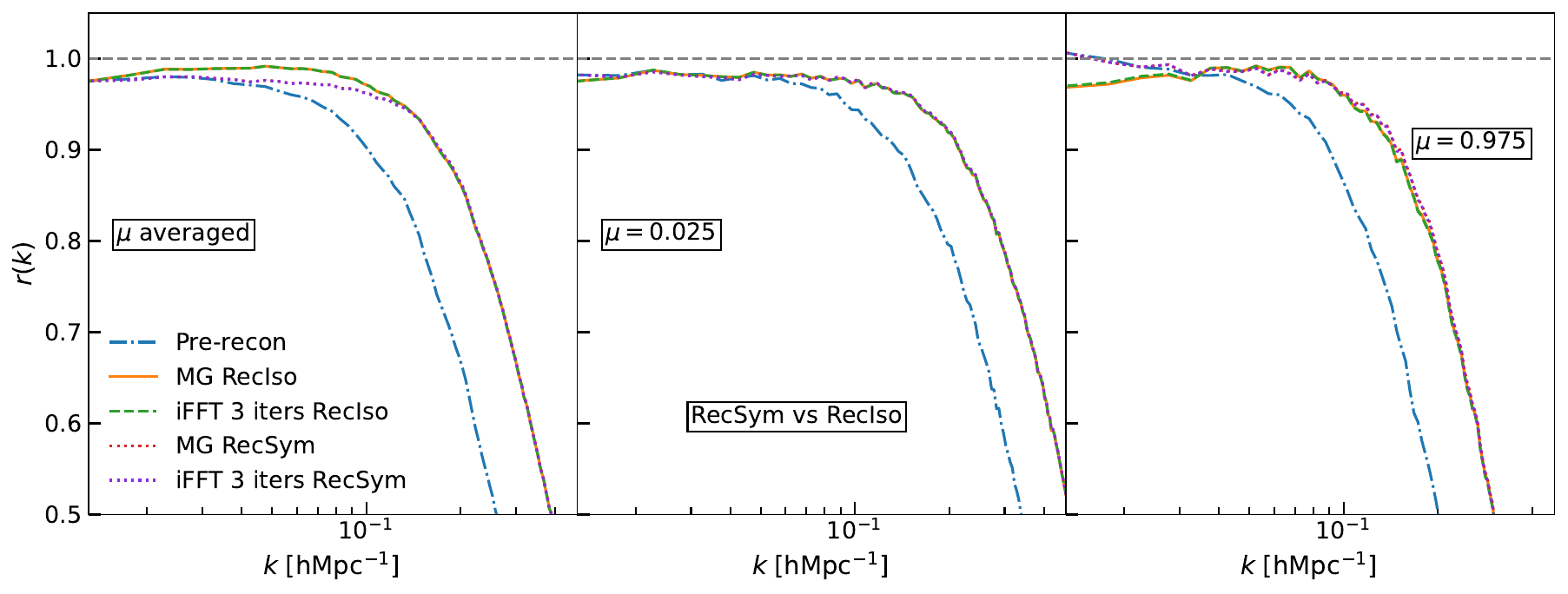}
    \includegraphics[width=1\linewidth]{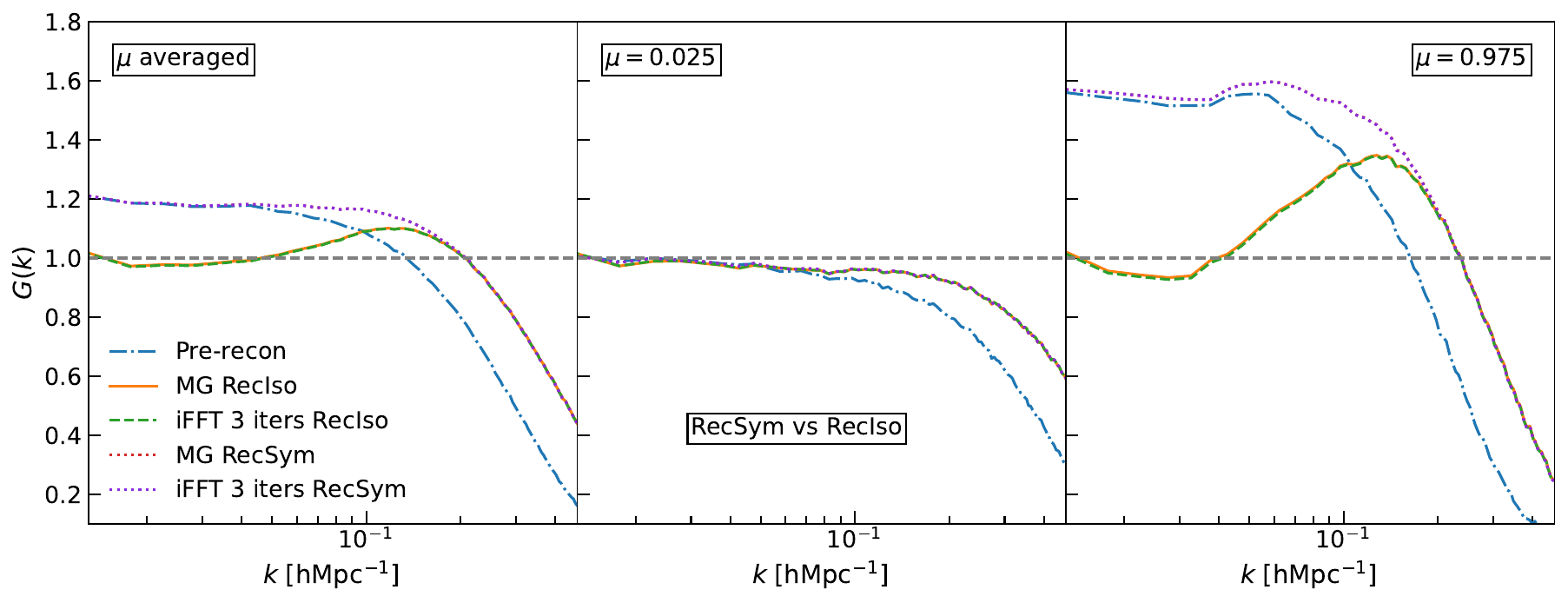}
    
    \caption{
    Cross-correlation coefficient $r(k)$ (top) and propagator $G(k)$ (bottom) for \textbf{ELG} comparing the \textbf{RecSym} and \textbf{RecIso} conventions, focusing on the MG and the iFFT (3 iterations) algorithms. Here the blue dash-dotted line is the pre-reconstruction case, the orange solid line is the MG \textbf{RecIso}, the green dashed line is the iFFT \textbf{RecIso}, the red dotted line is the MG \textbf{RecSym}, and the purple dotted line is the iFFT \textbf{RecSym}. In $r(k)$, \textbf{RecSym} is closer to unity on large scales along the line of sight. In $G(k)$, \textbf{RecSym} does not produce a bump that presents in \textbf{RecIso}.
    }
    \label{fig:recsym_reciso}
\end{figure}

We also more closely compare the \textbf{RecSym} and \textbf{RecIso} conventions by placing the two conventions together for the propagator as well as the usual cross-correlation coefficient, focusing on the MG and iFFT. The usual cross-correlation coefficient is defined as 
\begin{equation}
r(k)=\frac{\left<\delta^{*}_{\rm recon}(k)\delta_{\rm ini}(k)\right>}{\sqrt{\left<\delta_{\rm ini}(k)^2\right>\left<\delta_{\rm recon}(k)^2\right>}}.
\end{equation}
Compared to the propagator defined in \cref{eq:gk}, $r(k)$ does not capture the amplitude information, but it can present the phase differences more clearly. A perfect reconstruction would return a $r(k)$ at unity for both reconstruction conventions. Because whether or not the RSDs are removed happens on large scales, we expect that the differences will show up only on large scales for the comparison. Figure~\ref{fig:recsym_reciso} upper right shows that \textbf{RecSym} behaves better on large scales,
indicated by the fact that the cross-correlation coefficient $r(k)$ is closer to 1 in \textbf{RecSym} than in \textbf{RecIso} on large scales along the line-of-sight direction. \textbf{RecIso} is slightly worse than the pre-reconstruction case. In the left and middle panels, that $r(k)$ is below 1 on large scales is likely due to shot noise.
In $G(k)$, \textbf{RecSym} by construction does not produce a bump that presents in \textbf{RecIso}. The two conventions are nearly identical on smaller scales as well as perpendicular to the line of sight in both $r(k)$ and $G(k)$. 

The $r(k)$ and $G(k)$ analysis here does not clearly show the advantage of \textbf{RecSym} over \textbf{RecIso}, besides $r(k)$ is slightly closer to 1 on large scales along the line of sight with \textbf{RecSym} and that iFFTP shows more differences from the other two algorithms in \textbf{RecIso} than in \textbf{RecSym} in $G(k)$ (Figure~\ref{fig:ELG_Gk_reciso} compared to Figure~\ref{fig:ELG_Gk_recsym}). However, from our power spectrum test results, \textbf{RecSym} is preferred because the differences between the two algorithms are clearly smaller in \textbf{RecSym} than in \textbf{RecIso} on large scales, i.e. \textbf{RecSym} is less sensitive to reconstruction errors. Therefore, the \textbf{RecSym} convention is more robust against potential systematics due to differences in reconstruction algorithms.
The advantages of \textbf{RecSym} are discussed from a more theoretical point of view in \cite{White15} (where \textbf{RecSym} was introduced originally), \cite{Chen19} (more analytical developments), and \cite{KP4s2-Chen} (theoretical systematics studies in support of DESI DR1 BAO \cite{KP4}). For the most part, it is easier to model the reconstructed two-point statistics, if the RSDs are not removed. Additionally, because \textbf{RecSym} displaces galaxies and randoms by the same amount, any larger boundary effect is leveraged by displacing randoms in the same manner. Our DR1 BAO analysis adopts the \textbf{RecSym} convention.

Through tests with the ELG sample, we have demonstrated that the iFFTP algorithm in its current form has worse performance than MG or iFFT. This worse performance is largely due to the flaw in the current form of the algorithm that it is unstable at the boundary. Therefore, we do not test iFFTP in the following two tracers. We will discuss iFFTP in \cref{sec:discussion}, where we explain the flaw and how it might be corrected in the future.

\subsection{QSO: large volume tracer case study}\label{sec:QSO_results}
We apply the same metrics detailed previously for reconstruction on the QSO sample. In general, QSO behaves similarly to or better than ELG in the metrics we consider. The differences between MG and iFFT are generally smaller in QSO (note that this does not mean reconstruction fidelity is higher in QSO than in ELG; in fact, it is not, due to the lower number density of QSO). We do not detect anything abnormal with the application of iFFT and MG to QSO. We show similar figures in \cref{appx:additional_figures} for conciseness and better readability. We reconstruct using a default smoothing scale of 30 Mpc/$h$ (larger than the ELGs to account for the lower number density of the QSO sample).

For the displacement test on the dependence of survey geometry, we use the randoms rather than the galaxies to reduce noise. We also observe slightly larger differences in displacement between MG and iFFT near the low redshift end survey boundary, as shown in Figure~\ref{fig:los_scalar_footprint_colormap_QSO}, but these differences are insignificant compared to the average line-of-sight displacement of QSO (maximum $\sim$0.1 Mpc/$h$ vs $\sim$1.4 Mpc/$h$). In power spectrum, we find that iFFT converges slightly faster in QSO than in ELG. The difference between three iterations of iFFT and MG is less than 0.02\% on average for both monopole and quadrupole (both w.r.t. MG monopole) for the \textbf{RecSym} reconstruction convention, shown in Figure~\ref{fig:Pk_QSO_MG_iFFT}. The two algorithms almost entirely agree within error bars. We show the difference w.r.t. monopole due to the zero-crossing in the quadrupole. The QSO \textbf{RecSym} propagator also presents a similar behavior to ELG, with MG and iFFT (3 iterations) overlapping. The improvement over pre-reconstruction is less compared to ELG, suggesting lower reconstruction fidelity in QSO. Similar to the ELG sample, the \textbf{RecIso} convention for the QSO sample shows more differences on large scales in the power spectrum, with three iterations of iFFT being different by about 0.04\% than MG on large scales. 

The better agreement between MG and iFFT for the QSO sample is likely due to the fact that QSO is at a higher redshift, which reduces the impact of line-of-sight variations.

\subsection{BGS: wide angle tracer case study}\label{sec:BGS_results}
Because the goal of all three algorithms are trying to estimate the displacement when the plane-parallel approximation does not hold, testing them with a sample where the line of sight varies considerably can expose the algorithms to more challenges. Errors in the code might be more easily to show up. Issues associated with not using a fine enough grid might also be more easily to show up. Moreover, using a sample where the plane-parallel approximation does not hold, we can empirically test whether iFFT converges; \cref{eq:iFFT_general_form_org} and \cref{eq:iFFT_general_form_adp} show convergence for the plane-parallel case, but there is no easy form for the non-plane-parallel case. We note that ``wide angle'' here means that the line of sight changes widely at a fixed scale, and the size of the solid angle of the survey does not come into play.

We observe slightly more discrepancy between MG and iFFT displacements in the BGS sample. So we discuss differences in the displacement together its impact on two-point statistics in this current section.
We show the \textbf{RecSym} results here and show \textbf{RecIso} results in \cref{appx:additional_figures}. 
Our default setting for the BGS sample is smoothing at 15 Mpc/$h$.

\begin{figure}
    \centering
    \includegraphics[width=\linewidth]{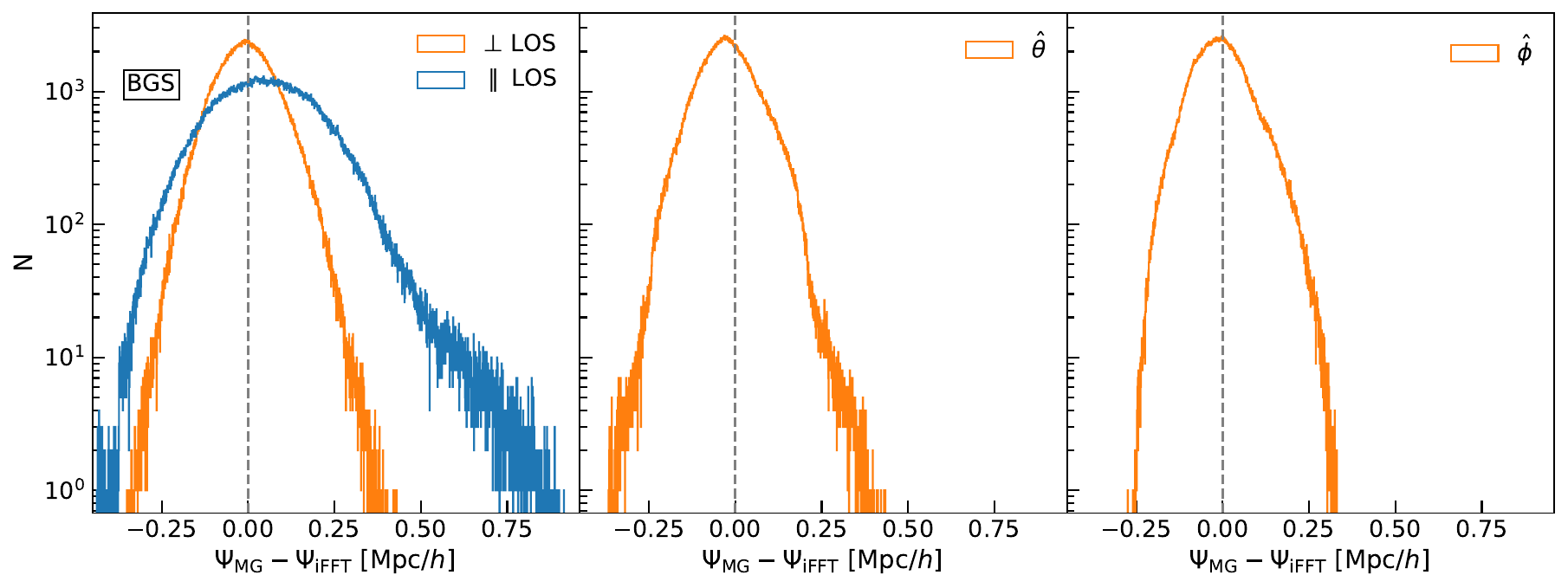}
    \caption{The distribution of displacement vector differences between \textbf{MG} and \textbf{iFFT} displacements with the \textbf{BGS} sample using one mock.
    The displacement vectors are decomposed into three spherical coordinate directions as in Figure~\ref{fig:ELG_displacement_decompose}.
    The line of sight direction spread is skew towards the right side and is wider than the spread of the perpendicular to the line of sight. Compared to the average line-of-sight displacement magnitude ($\sim$2.9 Mpc/$h$), the fraction can be larger than the ELG and QSO cases.
    }
    \label{fig:BGS_MG_iFFT_diff_distribution}
\end{figure}

In the analysis of the displacement, we find that MG and iFFT (3 iterations) have more differences in the displacement magnitude, at the largest ($\sim$0.92 Mpc/$h$) about 2.5 times more than ELG ($\sim$0.37 Mpc/$h$) and 9 times more than QSO ($\sim$0.10 Mpc/$h$), shown in the magnitude distribution in Figure~\ref{fig:BGS_MG_iFFT_diff_distribution}. Compared to the average displacement along the line of sight of BGS ($\sim$2.9 Mpc/$h$), the maximum fractional displacement difference in BGS is $\sim$32\%, which is larger than the maximum in ELG ($\sim$19\%) and in QSO ($\sim$7\%). 
On the left panel of Figure~\ref{fig:BGS_MG_iFFT_diff_distribution}, the parallel to the line of sight distribution is skew to the right. When the perpendicular to the line of sight component is decomposed into $\boldsymbol{\hat{\theta}}$ and $\boldsymbol{\hat{\phi}}$, we observe that these two distributions are also more irregularly shaped compared to ELG and QSO.

\begin{figure}
    \centering
    \includegraphics[width=\linewidth]{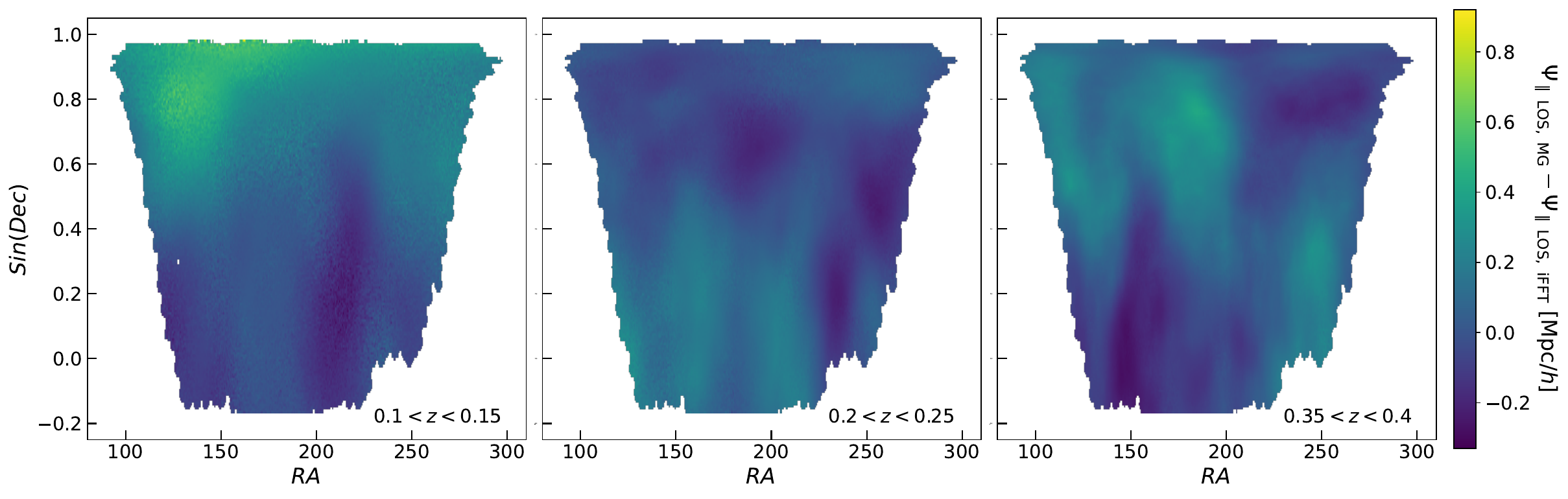}
    \caption{
    Line-of-sight component magnitude difference between \textbf{MG} and \textbf{iFFT} (both applied with a 15 Mpc/$h$ smoothing) projected to DESI Y5 NGC footprint, shown for \textbf{BGS} randoms in $0.1<z<0.15$ (left), $0.2<z<0.25$ (middle) and $0.35<z<0.4$ (right) redshift slices. The low redshift end boundary shows more differences between the two algorithms than a middle redshift range or the high redshift end.
    These differences are not negligible compared to the average line-of-sight displacement magnitude of BGS ($\sim$2.9 Mpc/$h$).
    }
    \label{fig:los_scalar_footprint_colormap_BGS}
\end{figure}

We also observe slightly more dependence on the survey geometry for the line-of-sight displacement differences as shown in Figure~\ref{fig:los_scalar_footprint_colormap_BGS}. Here we use the distribution of BGS randoms, rather than galaxies, because there are fewer objects in the BGS sample compared to the other two samples.
We show $0.1<z<0.15$, $0.2<z<0.25$ and $0.35<z<0.4$ for the effects of survey sky area edges as well as for the redshift boundary effects by comparing to a redshift slice in the middle of the BGS redshift range. Both sky area boundaries and redshift boundaries have some effects on the displacement estimates with the BGS sample. For the sky area, the extreme differences appear to be close to the edges in a couple of locations. The magnitude of the displacement difference can also be larger than ELG and QSO, especially at the low redshift end. Compared to the average line of sight displacement magnitude of BGS ($\sim$2.9 Mpc/$h$), the differences in the low redshift end can be non-negligible. Indeed the maximum fractional difference of 31\% occurs at the low redshift end. This suggests that the line-of-sight differences between MG and iFFT show up more in BGS. 

Two factors can contribute to the patterns shown in these two figures.
The wide-angle effects, i.e. more significant changes of lines of sight, associated with the BGS sample can mildly affect the performance of both algorithms. Because the BGS sample is at very low redshifts, the change of line-of-sight direction at a fixed scale is more significant in BGS than in other samples. 
It could also be that the BGS sample has a greater boundary/volume ratio and so these boundary effects have a larger impact. These can amplify the differences by the two algorithms.

\begin{figure}
    \centering
    \includegraphics[width=0.492\columnwidth]{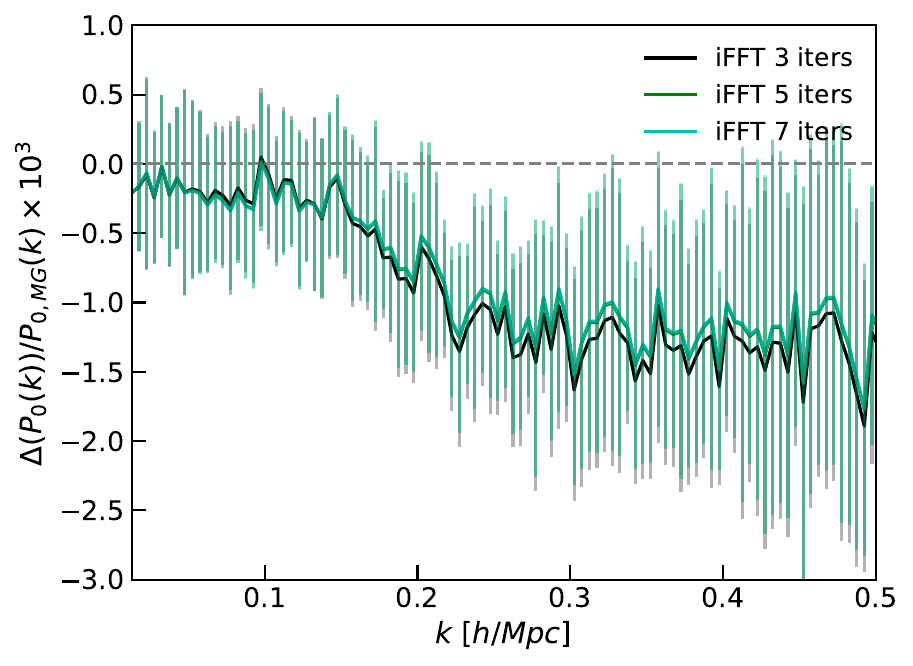}
    \includegraphics[width=0.48\columnwidth]{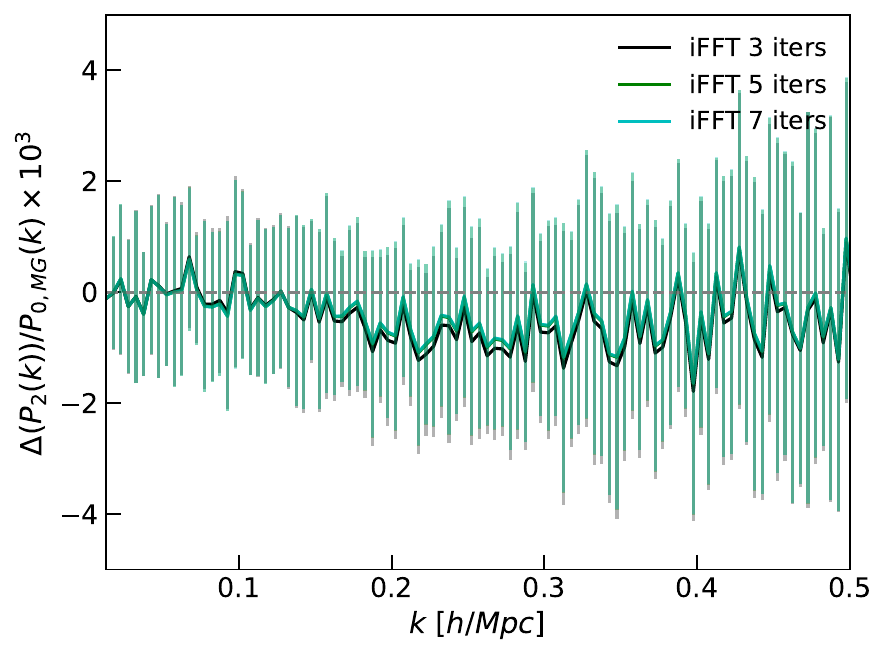}
    \caption{Ratio in the power spectrum monopole (left) and quadrupole (right) between iFFT (black for 3 iterations, green for 5 iterations, and cyan for 7 iterations) and MG (both applied with a 15 Mpc/$h$ smoothing), using the \textbf{RecSym} convention for the \textbf{BGS} sample. Each line is an average of 25 mocks. Even with 3 iterations, the differences between iFFT and MG are within 0.2\% on average for monopole and quadrupole (both w.r.t.\ MG monopole). The two agree within error bars in quadruole and some scales in monopole. 
    }
    \label{fig:Pk_BGS_MG_iFFT}
\end{figure}

\begin{figure} 
    \centering
    \includegraphics[width=0.7\columnwidth]{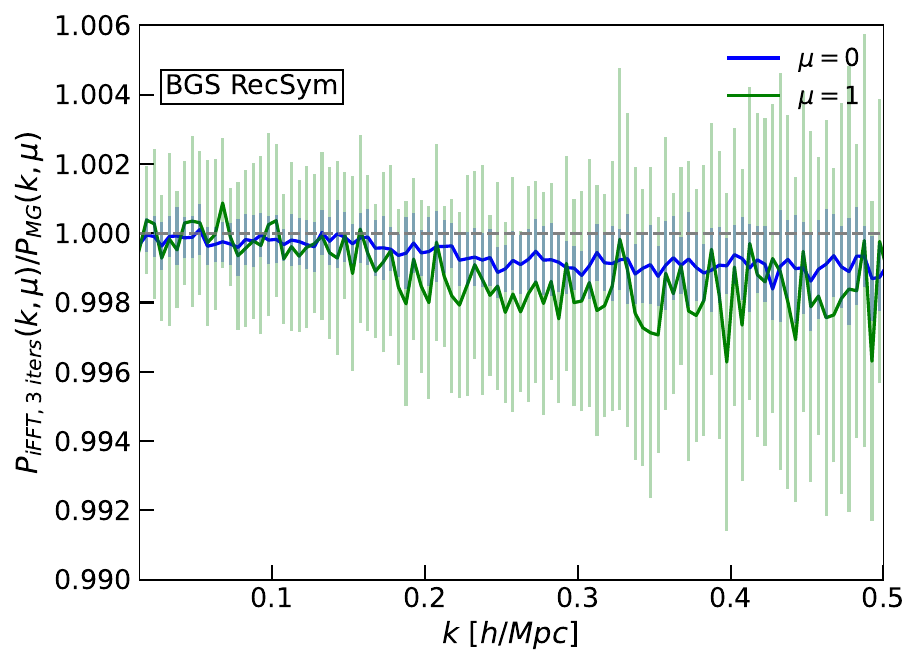}
    \caption{
    Ratio of power spectrum between iFFT and MG for the \textbf{BGS} sample with both algorithms applied with a 15 Mpc/$h$ smoothing and using the \textbf{RecSym} convention, for $\mu=0$ (blue) and 1 (green), where $\mu$ is cosine of the angle between the line of sight and $\boldsymbol{k}$. Here $\mu=0$ is perpendicular to the line of sight and $\mu=1$ is along the line of sight. Each line is an average of 25 mocks. We observe that both cases show differences between iFFT and MG within 0.4\% on average. However, along the line of sight the differences between the two algorithms are larger, especially on smaller scales ($k\gtrsim 0.2\ h$/Mpc), but they agree within error bars. 
    }
    \label{fig:Pk_angle_BGS}
\end{figure}

\begin{figure}
    \centering
    \includegraphics[width=\columnwidth]{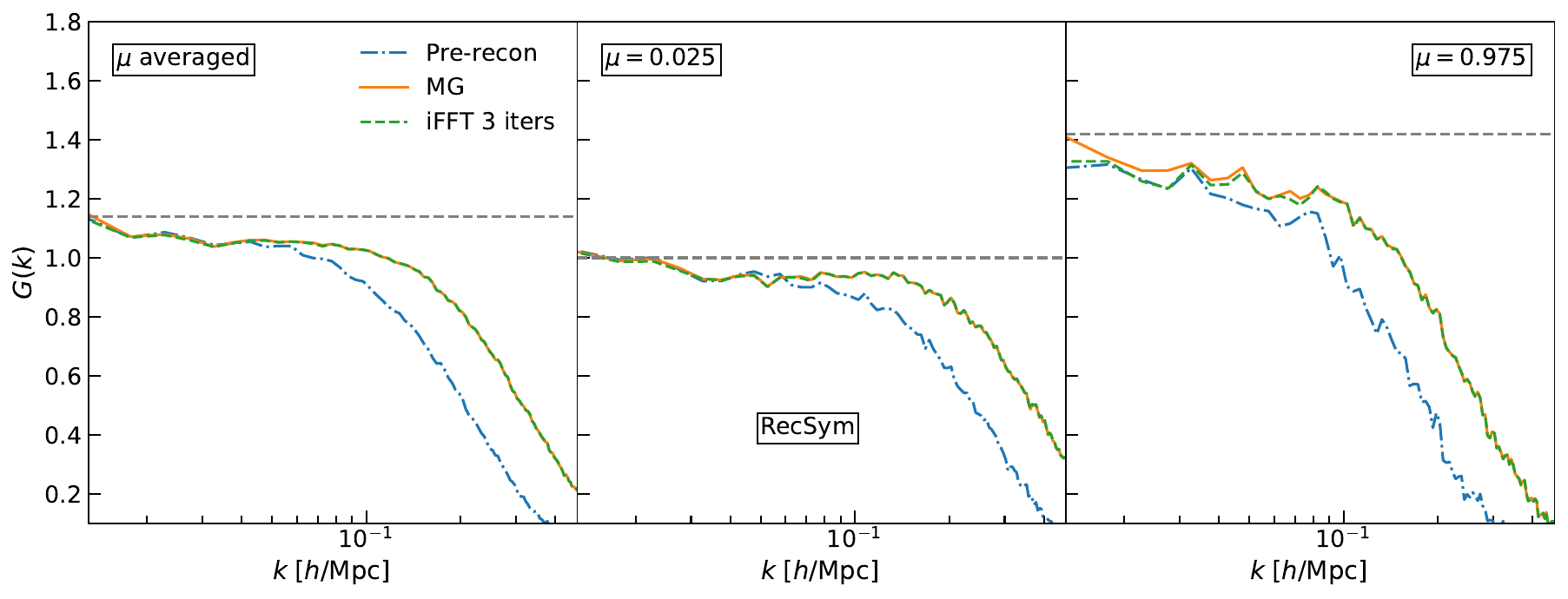}
    \caption{Propagator for the three reconstruction algorithms for one \textbf{BGS} mock with the \textbf{RecSym} convention when $\mu$ is integrated over (left), and in $\mu=0.025$ (middle) and 0.975 (right) bins, in comparison of pre-reconstruction propagator (blue dash dotted). The $\mu$-bin width is 0.05. The horizontal dashed lines show expected Kaiser approximation in the plane-parallel approximation, $(1+1/3\beta)=1.14$ for the first panel and $(1+\beta\mu^2)=1.0$ and 1.42 for the middle and right panels, respectively. We shift the propagators by 14\% to correct for misestimate of the linear bias. MG and iFFT (3 iterations) return improved propagator. However, they do not overlap on large scales along the line of sight.  }
    \label{fig:BGS_Gk_recsym}
\end{figure}

The effect as shown in the power spectrum, however, is negligible. Figure~\ref{fig:Pk_BGS_MG_iFFT} shows the difference between the power spectra by iFFT and MG divided by the MG monopole in the \textbf{RecSym} convention. We show the difference due to the zero-crossing in the quadrupole of BGS. The trend is qualitatively similar to that in ELG and the two methods differ at the maximum 0.2\% on average in both the monopole and the quadrupole (w.r.t. MG monopole). We note that although qualitatively similar, a smaller large-scale discrepancy was observed in BGS than in ELG. However, the two algorithms agree within error bars in quadrupole and on some scales in monopole as well. There is also less difference among 3, 5, and 7 iterations, suggesting that iFFT converges slightly faster in BGS than in ELG. However, iFFT still does not converge to MG.
In the comparison of the line of sight power spectrum with \textbf{RecSym} in Figure~\ref{fig:Pk_angle_BGS}, the line of sight direction still exhibits more differences than perpendicular to the line of sight, but more towards smaller scales ($k\gtrsim 0.2\ h$/Mpc). However, the differences are within 0.4\% on average for both perpendicular and parallel to the line of sight, slightly larger than that in ELG, but they almost agree within error bars here. In general, the more differences we observe in the displacement analysis do not have a significant impact on the power spectrum. On the propagator, MG and iFFT (3 iterations) agree well, except that, interestingly, there is a slight difference between them on large scales along the line of sight, as shown in \cref{fig:BGS_Gk_recsym}. When averaged over angles, however, the differences on large scales are much smaller.

We show similar results of BGS \textbf{RecIso} power spectrum and propagator comparisons in \cref{appx:additional_figures}.


\subsection{BAO constraints
}
Motivated by the more noticeable discrepancy shown in the displacement test for BGS and in the caution of unnoticeable correlated errors in the power spectrum building up, we perform BAO fitting to test whether this larger discrepancy leads to bias in the BAO distance estimates. We include the fitting results for iFFTP as a reference. For completeness, we also include BAO constraints for the ELG sample, which is the sample we have used to test all algorithms and all metrics in this study. 

Our fitting procedure largely follows that used in the DESI 2024 BAO analysis \cite{KP4}, which represents the most robust fitting procedure. We briefly summarize the steps below.

We fit the following model for the post-reconstruction power spectrum
\begin{equation} \label{eq:generic_model}
    P(k, \mu)= \mathcal{B}(k, \mu) P_{\rm nw}(k) + \mathcal{C}(k, \mu)P_{\rm w}(k) + \mathcal{D}(k)\,,
\end{equation}
where $P_{\rm w}(k)$ and $P_{\rm nw}(k)$ are the BAO wiggle and no BAO wiggle components of the linear power spectrum, which is obtained from \textsc{CLASS}\footnote{\url{https://github.com/lesgourg/class\_public}} using the \textsc{AbacusSummit} fiducial cosmology (detailed in \cref{sec:mocks}). The $\mathcal{B}(k, \mu)$, $\mathcal{C}(k, \mu)$, and $\mathcal{D}(k)$ encompass various physical and non-physical features of the power spectrum. 
The term $\mathcal{B}(k, \mu)$ models the smooth component of the galaxy clustering using quasi-linear theory, 
\begin{equation}
    \mathcal{B}(k,\mu) = \left(b_{1}+f\mu^{2}[1 - \Sigma(k)]\right)^{2} F_{\rm fog}\,
    \label{eqn:Bkmu}
\end{equation}
where the first term is a modified form of the Kaiser factor \cite{Kaiser87} that accounts for the effect of reconstruction in the \textbf{RecIso} convention. In \textbf{RecSym}, $\Sigma(k)=0$, since $\textbf{RecSym}$ does not remove large-scale RSDs. The second term models the Finger of God (FoG) effect by $F_{\rm fog}=(1+k^2\mu^2\Sigma_s^2/2)^{-2}$, where $\Sigma_s$ is the FoG parameter.
The term $\mathcal{C}(k, \mu)$ describes the BAO damping caused by nonlinear evolution, 
\begin{equation}
    \mathcal{C}(k,\mu) = \left(b_{1}+f\mu^{2}[1 - \Sigma(k)]\right)^{2}\exp\left[-\frac{1}{2}k^2\biggl(\mu^{2}\Sigma_{\parallel}^2 + (1-\mu^{2})\Sigma^{2}_{\perp}\biggl)\right],
    \label{eq:baodamping}
\end{equation}
where $\Sigma_{\parallel}^2$ and $\Sigma^{2}_{\perp}$ are the BAO damping scales parallel and perpendicular to the line of sight, respectively. Reconstruction reduces the values of these two parameters by partially removing gravitational nonlinearity. The first term is similarly the Kaiser term. The FoG is dropped here due to high degeneracy with the $\Sigma_{\parallel}$ parameter in fitting.
Lastly, $\mathcal{D}(k)$ captures additional nonlinearities and observational effects, parametrized by a spline function
\begin{equation} \label{eq:spline_basis}
    \mathcal{D}_\ell(k) = \sum_{n=-1}^{n_{\mathrm{max}} }a_{\ell,n} W_3\left(\frac{k}{\Delta} - n\right)\,
\end{equation}
where $W_3$ is the piecewise cubic spline kernel. The term $\Delta$ is a $k$ width chosen such that the spline function matches the broadband shape of the power spectrum and does not reproduce the BAO signal. The free coefficient $a_{l,n}$ is then guaranteed to be not degenerate with the BAO parameters.

This power spectrum template is fitted to the observed wavenumbers, $k_{\rm obs}$, which are related to the true wavenumbers, $k^\prime$, by 
\begin{align}
    k^\prime  =  \frac{\alpha_\text{AP}^{1/3}}{\alpha_\text{iso}}\left[ 1+\mu^2_{\rm obs} \left( \frac{1}{\alpha_\text{AP}^2}-1 \right) \right]^{1/2} k_{\rm obs}
    \label{eq:dilationk}
\end{align}
and
\begin{align}
    \mu^\prime  = \frac{\mu_{\rm obs}}{\alpha_\text{AP}}\left[ 1+\mu_{\rm obs}^2 \left( \frac{1}{\alpha_\text{AP}^2}-1 \right) \right]^{-1/2},
\label{eq:dilationmu}
\end{align}
where $\alpha_{\rm iso}$ and $\alpha_{\rm AP}$ are the isotropic and Alcock-Paczynski dilation BAO parameters that capture the differences between the true and the observed wavenumbers. These two BAO parameters in terms of the Hubble parameter and the angular diameter distance are
\begin{equation}
    \qquad \qquad \alpha_{\rm iso} =  (\alpha_\parallel\alpha_\perp^2)^{1/3},
 \qquad \qquad \alpha_{\rm AP} = \alpha_\parallel/\alpha_\perp. 
    \label{eqn:aisoap_defs}
\end{equation}
where
\begin{equation}
    \qquad \qquad \alpha_{||} = \frac{H^{\mathrm{fid}}(z)r^{\mathrm{fid}}_{d}}{H(z)r_{d}}, \qquad \qquad \alpha_{\perp} = \frac{D_{A}(z)r^{\mathrm{fid}}_{d}}{D^{\mathrm{fid}}_{A}(z)r_{d}}.
    \label{eqn:alpha_defs}
\end{equation}
So in terms of power spectrum multipole, we have the model
\begin{align}
    P_{\ell, \rm obs}(k_{\rm obs}) = \frac{2\ell+1}{2} \int_{-1}^1 &d\mu_{\rm obs}\ \mathcal{L}_\ell(\mu_{\rm obs}) \big[\mathcal{B}(k_{\rm obs}, \mu_{\rm obs}) P_{\rm nw, obs}(k_{\rm obs}) \nonumber \\
    &+ \mathcal{C}(k^\prime(k_{\rm obs},\mu_{\rm obs}), \mu^\prime(k_{\rm obs},\mu_{\rm obs}))P_{\rm w}(k^\prime(k_{\rm obs},\mu_{\rm obs}))\big] + \mathcal{D}_\ell(k_{\rm obs}).
    \label{eq:pow_spec_multipoles}
\end{align}

In this test, we focus on fitting $\alpha_{\rm iso}$ and $\alpha_{\rm AP}$, using the monopole and quadrupole post-reconstruction power spectra, while marginalizing all other parameters. We set Gaussian priors for the FoG parameter at $\mathcal{N}(\Sigma_{s}^{\rm fid}= 2\  {\rm Mpc}/h, 2\  {\rm Mpc}/h)$ and damping parameters at $\mathcal{N}(\Sigma_{\parallel}^{\rm fid}= 8\ {\rm Mpc}/h, 2\ {\rm Mpc}/h)$ and $\mathcal{N}(\Sigma_{\perp}^{\rm fid}=3\ {\rm Mpc}/h, 1\ {\rm Mpc}/h)$ for BGS. For ELG, these priors are $\mathcal{N}(\Sigma_{s}^{\rm fid}= 0\  {\rm Mpc}/h, 2\  {\rm Mpc}/h)$, $\mathcal{N}(\Sigma_{\parallel}^{\rm fid}= 5.35\ {\rm Mpc}/h, 1.4\ {\rm Mpc}/h)$, and $\mathcal{N}(\Sigma_{\perp}^{\rm fid}=3\ {\rm Mpc}/h, 1\ {\rm Mpc}/h)$.
We use the covariance matrix specifically computed for the fitting power spectra. We also include window convolution for DESI Y5 footprint in the fitting procedure.
We fit in the $k$ range $[0.02,0.3]\ h$/Mpc with a $k$-bin width of 0.005 $h$/Mpc. We use the official DESI fitting pipeline \textsc{desilike}\footnote{\url{https://github.com/cosmodesi/desilike}} to conduct the fits by posterior profiling using \texttt{minuit} \cite{James75}.

Figure~\ref{fig:BGS_fits} shows the fits for $\alpha_{\rm iso}$ and $\alpha_{\rm AP}$ for each of the 25 BGS mocks and the mean of the 25 fits. In all cases, the fits are nearly identical between MG and iFFT post-reconstruction power spectra. The mean of the 25 fits is at $\alpha_{\rm iso}=1.0004\pm 0.0025$ and $\alpha_{\rm AP}=1.0021\pm 0.0058$ with MG, $\alpha_{\rm iso}=1.0005\pm 0.0025$ and $\alpha_{\rm AP}=1.0021\pm 0.0059$ with iFFT, and $\alpha_{\rm iso}=1.0005\pm 0.0025$ and $\alpha_{\rm AP}=1.0009\pm 0.0056$ with iFFTP (the errors are the errors of the mean). All three cases are consistent with 1. So the fits are unbiased. We note that the mean of the $\alpha_{\rm AP}$ fit with iFFTP is noticebly different from the other two, although in this case, it is closer to 1. This test shows that even when the displacements exhibit noticeable differences, these differences are not affecting the BAO fits. All algorithms return unbiased BAO estimates within the expected precision of the DESI final data set. 

BAO constraints in the ELG mocks show similar results. In ELG, the mean of the 25 fits is at $\alpha_{\rm iso}=1.0014\pm 0.0009$ and $\alpha_{\rm AP}=1.0021\pm 0.0025$ with MG, $\alpha_{\rm iso}=1.0014\pm 0.0009$ and $\alpha_{\rm AP}=1.0027\pm 0.0027$ with iFFT, and $\alpha_{\rm iso}=1.0014\pm 0.0009$ and $\alpha_{\rm AP}=1.0035\pm 0.0025$ with iFFTP (the errors are the errors of the mean). A couple of fits here are away from 1 by slightly more than 1$\sigma$, but we do not consider them as a systematic offset, as DESI uses a threshold of 3$\sigma$ to declare a systematic effect \cite{KP4,Andrade25}. We note that the mean of the $\alpha_{\rm AP}$ fit with iFFTP noticeably deviates more from 1 than the other two algorithms in this case. However, it is still within 2$\sigma$. The obvious deviation in iFFTP as shown in displacement and two-point statistics result in minimal differences in BAO constraints. However, we still caution the use of iFFTP, for the aforementioned flaws in the algorithm and deviations shown earlier.

Consistent BAO results in spite of differences present in other metrics we have shown demonstrates that BAO constraints are robust against algorithm differences. Nevertheless, the more obvious differences we show in the displacement, power spectrum, and propagator remain valid as we enter the late Stage IV era and Stage V where ever higher precision will be needed. They will also be useful for application of reconstruction beyond BAO analysis.

\begin{figure}
    \centering
    \includegraphics[width=1\linewidth]{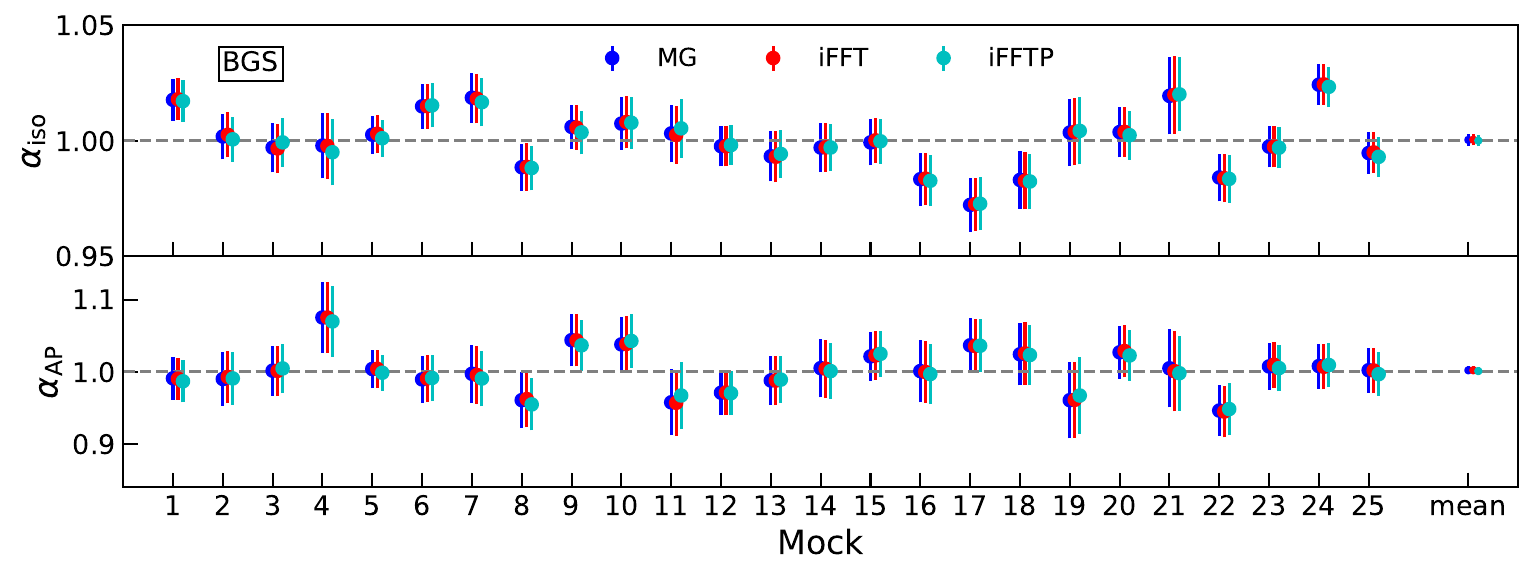}
    \caption{Anisotropic BAO fits for 25 \textbf{BGS} mocks as well as the mean of 25 fits. The covariance used here is specifically calculated for the mean of the power spectra for each algorithm. The error bar on the data point for the mean is the error on the mean. For the mean, $\alpha_{\rm iso}=1.0004\pm 0.0025$ and $\alpha_{\rm AP}=1.0021\pm 0.0058$ with MG, $\alpha_{\rm iso}=1.0005\pm 0.0025$ and $\alpha_{\rm AP}=1.0021\pm 0.0059$ with iFFT, and $\alpha_{\rm iso}=1.0005\pm 0.0025$ and $\alpha_{\rm AP}=1.0009\pm 0.0056$ with iFFTP. Here the fits for MG, iFFT, and iFFTP post-reconstruction power spectra give almost identical BAO parameter constraints. The mean is also consistent with 1. }
    \label{fig:BGS_fits}
\end{figure}

\begin{figure}
    \centering
    \includegraphics[width=1\linewidth]{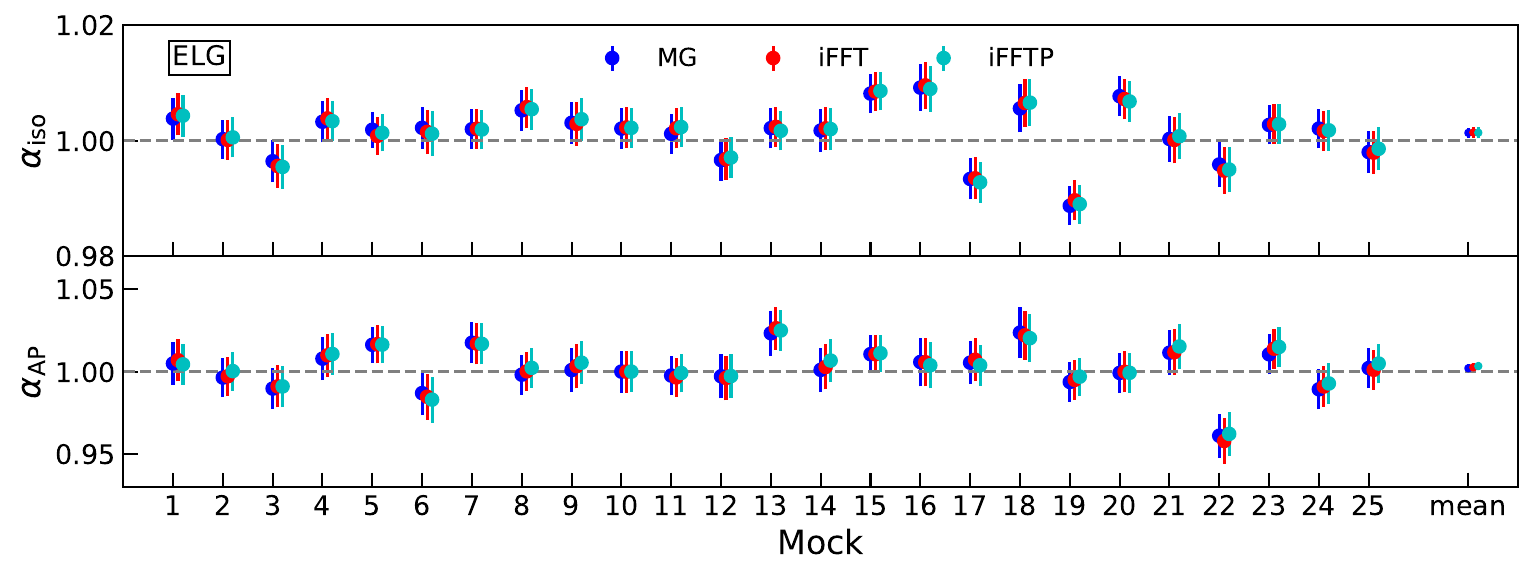}
    \caption{Anisotropic BAO fits for 25 \textbf{ELG} mocks as well as the mean of 25 fits. The covariance used here is specifically calculated for the mean of the power spectra for each algorithm. The error bar on the data point for the mean is the error on the mean. For the mean, $\alpha_{\rm iso}=1.0014\pm 0.0009$ and $\alpha_{\rm AP}=1.0021\pm 0.0025$ with MG, $\alpha_{\rm iso}=1.0014\pm 0.0009$ and $\alpha_{\rm AP}=1.0027\pm 0.0027$ with iFFT, and $\alpha_{\rm iso}=1.0014\pm 0.0009$ and $\alpha_{\rm AP}=1.0035\pm 0.0025$ with iFFTP. Here the fits for MG, iFFT, and iFFTP post-reconstruction power spectra give almost identical BAO parameter constraints. The mean is also consistent with 1 within 2$\sigma$. }
    \label{fig:BGS_fits}
\end{figure}

\subsection{Computing time}\label{sec:computing_time}
Having established that we recommend against the iFFTP algorithm, in this section we compare the computing time between MG and iFFT only. Both algorithms are parallelized using MPI. Using 128 \textsc{AMD-EPYC-7763} CPU cores with MPI parallelization on the Perlmutter cluster on NERSC\footnote{\url{https://docs.nersc.gov/systems/perlmutter/architecture/}} and using the ELG sample,\footnote{In this test, we did not set the grid size to be exactly the same for MG and iFFT, and MG was run on a slightly larger grid, due to different prime factorizations required for the algorithms. For the main tests presented in this paper, we did keep the grid size the same, i.e. using the MG grid size for iFFT and iFFTP.} it takes $\sim$1300 seconds for the MG code to compute the displacement potential with our default MG parameters and $\sim$150 seconds for the iFFT code to conduct 3 iterations. 
Hence, it is computationally much more efficient to perform the iFFT algorithm. Due to its computational efficiency and nearly identical results to MG, the DESI DR1 BAO analysis adopted the iFFT as the default algorithm. 
However, we note that we have not attempted to optimize the MG code, whereas we are using a highly optimized FFT library for iFFT. It is likely that the MG running speed could be dramatically improved.

\section{Discussion}\label{sec:discussion}

As shown in the last section, the iFFTP algorithm is not as effective as the other algorithms examined and can be unstable, especially with more iterations, but iFFT and MG agree to a very high degree in the tests we have conducted. In general, 
the difference between iFFT and MG is the smallest for
the QSO sample, followed by ELG and then BGS. This suggests that iFFT and MG work well in high redshift and both low and high number densities, although they face slightly more challenges when the lines of sight change considerably. While the algorithms appear to be susceptible to wide-angle effects in the BGS sample in the displacement tests, the magnitude of the differences is relatively low such that it does not affect the power spectrum and consequently the BAO parameters.

In this section, we discuss the iFFTP algorithm in a little more depth and we also discuss the outlook of going beyond the standard reconstruction.

\subsection{iFFTP}
While we observe a failure of the current iFFTP algorithm when applied to realistic mocks, there are still interesting questions to explore about this algorithm. Addressing these questions is beyond the scope of the current paper, but we briefly discuss them here. First, while iFFTP does not converge in realistic simulations that have survey boundaries, we note that it does fine in cubic box simulations, where periodic boundary conditions are applied. In cubic boxes, galaxies will not be moved off boundaries but they will in realistic simulations, which causes divergence. There are ways to mitigate this problem in real surveys, for example by shifting or smoothing the randoms. 

We also find that in cubic boxes and in terms of the propagator, iFFTP performs slightly better than iFFT at the same number of iterations. In other words, even though we observe no improvement in the BAO damping scale after reconstruction with iFFTP at 5 iterations (and more), the same algorithm in the cubic boxes reduces BAO damping more than the other two algorithms (though the difference is not large). There are two possible subtle differences in iFFTP in comparison to iFFT. The different ways that they use to remove RSDs lead to these differences. The first is that iFFT and iFFTP likely achieve the same level of convergence to real-space density at different rates (\cref{eq:iFFT_general_form_org} and~\cref{eq:iFFT_general_form_adp} may not apply to iFFTP). The second is that if iFFTP and iFFT obtain the identical displacement $\boldsymbol{\mathcal{S}}_{\rm iFFTP,disp}=\boldsymbol{\mathcal{S}}_{\rm iFFT,disp}$, they are applied to different fields. Because galaxies have been moved to real space locations after removing RSDs in iFFTP, $\boldsymbol{\mathcal{S}}_{\rm iFFTP,disp}$ is evaluated at and applied to real space galaxy positions, whereas $\boldsymbol{\mathcal{S}}_{\rm iFFT,disp}$ i.e. \cref{eq:iFFT_disp} (together with the RSD part, i.e. $\boldsymbol{\mathcal{S}}_{\rm iFFT,disp}+f(\boldsymbol{\mathcal{S}}_{\rm iFFT,disp}\cdot \boldsymbol{\hat{r}})\boldsymbol{\hat{r}}$) is applied to redshift space galaxies, with $\boldsymbol{\mathcal{S}}_{\rm iFFT,disp}$ evaluated at these redshift space positions as well. Potentially the former leads to better performance, because the displacement $\boldsymbol{\Psi}^{(1)}$ is by definition in real space. However, more study is needed to understand the effects of these differences.

\subsection{Beyond standard reconstruction}
The reconstruction algorithms we analyze here are all within the framework of standard reconstruction, which is approximately the Zel'dovich displacement and is thus often modeled with the Zel'dovich displacement. In this section, we switch from the continuity equation language and discuss in the context of Zel'dovich displacement. One question one might wonder is whether it is useful for constraining BAO to estimate higher-order displacement than Zel'dovich and even the full displacement. Firstly, we need to emphasize that while the standard reconstruction algorithm estimates the Zel'dovich displacement, this displacement is not the actual Zel'dovich displacement. There are two factors. One is indicated in \cref{sec:review_recon} -- we use the nonlinear, Eulerian density in place of linear, Lagrangian density to calculate the displacement using the Zel'dovich formula. Therefore, it is an estimate, although with sufficient smoothing, the two densities are rather close. The other factor is that we are calculating a displacement that starts from the Eulerian positions and goes back to Lagrangian positions. 
The displacement that goes from Lagrangian positions to Eulerian positions (which is the Zel'dovich displacement) may not be exactly invertible when two Lagrangian positions go to the same Eulerian position. However, estimating the displacement starting from the Eulerian position gives the damping scale for the shift field the same as the damping scale of the displaced field in the modeling. 
So it is necessary to start from the Eulerian position. This point is detailed in the our companion paper on theoretical systematics \cite{KP4s2-Chen}.
There is another complication -- the Zel'dovich displacement gives a Zel'dovich field, not a fully nonlinear field. If we started from a Zel'dovich field, this displacement would be exactly invertible. What we observe, however, is a fully nonlinear field. So even if we get a perfect Zel'dovich displacement, we cannot use this displacement to go back to the initial position beyond large scales. Not only because this is only first order displacement, but also the coordinates are different ($\boldsymbol{q}=\boldsymbol{x}$ only at the first order).

Let us now discuss the outlook of going beyond the approximate Zel'dovich displacement. Within the standard reconstruction framework, that is, estimating the displacements, and moving the galaxies and randoms back (following one of the \textbf{RecSym} or \textbf{RecIso} conventions), it is the step of estimating the displacement that can be adapted, and the rest stays the same. If we improve order by order, we will still solve for the first-order displacement, which will be used to estimate the second order displacement. This will quickly become cumbersome as we go beyond the second order. Alternatively, we can solve the full displacement. This means that we will need to solve for a more complicated equation than \cref{eq:poisson} or~\ref{eq:poisson_phi}. A natural question is how much can be gained by estimating higher-order or the full displacements. For the second order, Refs.~\cite[e.g.][]{Seo10,Schmittfull17,recon} have shown that adding second-order displacement has a negligible effect. This is explicitly shown with BAO parameter fits in Ref.~\cite{recon}. Even at the displacement level, the changes are small. The reason for mild effects can be explained as follows. The Zel'dovich displacement to a high accuracy moves from Lagrangian to Eulerian space, which is to a large extent invertible (which is why standard reconstruction is effective), but the higher-order displacements are more responsible for the deformation of the Lagrangian region into the final Eulerian region, which is a smaller scale effect \cite{recon}. The bulk flows most responsible for damping the BAO peak are mostly due to larger scale nonlinearities, as opposed to smaller scale nonlinearities, such as effects of galaxy formation; therefore, the most contribution to BAO damping comes from the Zel'dovich displacement. In addition, recall that the standard reconstruction uses the final density field to estimate the Zel'dovich displacement, which is not perfect; hence, it is even harder to obtain the correct second-order displacement, because it depends on the first-order estimation.

It is also unclear whether having a full displacement estimate will lead to a significant improvement over standard reconstruction. The full displacement is harder to solve from a full nonlinear continuity equation, although the optimization algorithm, optimal transport, in the fully discrete case provides an approximate solution to that equation \cite[e.g.][]{Brenier91,Frisch02,Brenier03}. A semi-discrete optimal transport algorithm also gives a close estimate \cite{Nikakhtar24}. However, although not directly shown with BAO parameter fits, the sharpened BAO peak from an approximate full displacement is almost identical to the one resulted from Zel'dovich displacement (the actual Zel'dovich, not inverse Zel'dovich), as indicated in \cite{Nikakhtar24}. Therefore, there may not be a significant gain for BAO fits even if we have the full displacement. This again supports the robustness of BAO. The Fourier space comparison, e.g. power spectrum and cross-correlations, however, can show more disagreement among different reconstruction methods. The solution to this paradox is that there is broadband power that the standard reconstruction does not recover well but other reconstruction methods recover better. However, the broadband is not fitted with physics; only the BAO wiggles contain BAO information and standard reconstruction already recovers the wiggles very well. Nevertheless, even if the improvement may not be significant, there may still be opportunities to further tighten the BAO parameter errors by estimating more than the Zel'dovich displacement. This will still be helpful in the era of precision cosmology. Of course, for a new method of reconstruction to be useful, we will need an accurate and well-founded theory model as well.

\section{Conclusion}\label{sec:conclusion}
In this study, we closely examine three algorithms that estimate the displacements with changing lines of sight from the linearized redshift space continuity equation to conduct reconstruction for BAO. The MG algorithm solves the differential equation in configuration space for the potential of the displacement. The iFFT and iFFTP algorithms both iteratively remove RSDs and estimate the real space linear density, but iFFT achieves so by updating the potential iteratively, while iFFTP algorithm removes RSDs by moving galaxies iteratively. We examine two new codes (MG and iFFT) and one applied to survey data before (iFFTP). 
We use realistic mock simulations of DESI samples in the Y5 footprint to conduct the analysis and we focus on the ELG, QSO and BGS samples that cover the low and high redshift, low and high number density, and potential wide-angle effects.

Our findings are as follows:
\begin{itemize}

     \item We first scrutinize the MG, iFFT, and iFFTP algorithms by analyzing the displacements, since the algorithms essentially differ in how they estimate the displacement. 
    The displacement tests show that
    \begin{itemize}
        \item There is a much larger difference in the displacement estimate between iFFTP and MG than between iFFT and MG, both along and perpendicular to line of sight.
        
        \item There is a wider spread of the magnitude difference in the line-of-sight displacements for all three tracers for both iFFTP vs MG and iFFT vs MG comparisons.
        
        \item The displacement difference between iFFTP and MG is much larger at redshift boundaries than in a middle redshift slice, suggesting that iFFTP has problems at boundaries and especially along the line of sight. 
        
        \item All three samples display slight dependence on survey geometry for the line-of-sight component magnitude differences between iFFT and MG. Near redshift boundaries, the differences are slightly larger. However, these differences are much smaller than the difference shown in the iFFTP and MG comparison.
    \end{itemize}
    
    \item MG and iFFT power spectra are comparable to a high extent, in all samples, with both \textbf{RecSym} and \textbf{RecIso} conventions. The differences between the two (with iFFT having 3 iterations) are less than 0.4\% in \textbf{RecSym} in both monopole and quadrupole post-reconstruction power spectrum as well as power spectrum along the line of sight, for all three samples. The LRG sample, which we did not test in this study, is not extreme in number densities, redshift, or wide-angle effects, so we expect LRG to behave well with these two algorithms as well.
    The differences are larger in \textbf{RecIso}, but are still within 1.2\% in both monopole and quadrupole (maximum 2\% along the line of sight). The extreme differences are on the largest scales for \textbf{RecIso}. 
    For ELG with \textbf{RecSym}, comparing to DR1 power spectrum errors as measured by DESI DR1 covariance matrix, these differences are within the power spectrum errors and will not affect BAO constraints. The algorithm differeces are also within estimated Y5 errors. 
    
    \item For BGS and ELG with \textbf{RecSym}, we also explicitly tested BAO fits and found negligible differences between MG and iFFT for a Y5 precision. DESI DR1 BAO analysis opts to use iFFT with 3 iterations.
    
    \item The analysis of the power spectrum along and perpendicular to the line of sight shows the dependence of algorithm performance on the line of sight. Along the line of sight and in both \textbf{RecSym} and \textbf{RecIso}, the power spectra of iFFT and MG agree less compared to that in the perpendicular to the line of sight case on all scales. For \textbf{RecSym}, the differences are especially on smaller scales for all three tracers. For \textbf{RecIso}, the differences are especially on larger scales (larger than BAO scales). This is not the case in BGS \textbf{RecIso} monopole and quadrupole, but the BGS power spectrum parallel to the line of sight still has similarly larger discrepancy on large scales. 

    \item The \textbf{RecSym} convention presents smaller differences between different reconstruction algorithms than \textbf{RecIso} does overall. Because \textbf{RecIso} attempts to remove large-scale RSDs, the imperfect removal more easily presents in the \textbf{RecIso} convention. In comparison, \textbf{RecSym} is less sensitive to reconstruction errors. DESI DR1 BAO analysis opts to use the \textbf{RecSym} convention.

    \item iFFTP appears to be not convergent and exhibits more differences from MG and iFFT in the power spectrum. With the ELG sample using the \textbf{RecSym} convention, the difference between iFFTP with 3 iterations and MG is about 1\% in monopole and 10\% in quadrupole power spectrum. With 5 iterations, the differences increase to 5\% and 25\% for monopole and quadrupole, respectively. With the \textbf{RecIso} convention, the monopole difference is around 5\% and the quadrupole difference is around 10\% (both w.r.t.\ MG monopole) with 3 iterations. The differences become larger with 5 iterations, with the monopole having around 7\% differences and the quadrupole having around 15\% differences (without considering the largest scales). These deviations can potentially cause systematics in BAO fitting. More importantly, the iFFTP algorithm is not convergent, which presents instability to BAO analysis. Considering these, we recommend against using this algorithm for BAO constraints without further developments. As mentioned in the Discussion section, the algorithm performs fine in periodic boundary conditions, and the divergence of the algorithm is due to survey boundaries in realistic mocks. This is fixable in principle, but we do not consider practical fixes for this problem in this work.
   
    \item BGS exhibits more differences between iFFT and MG than ELG or QSO. This suggests that number density and volume have relatively mild effects on the reconstruction performance, but wild-angle effects (i.e. widely changing lines of sight), related to this, low redshift, have a stronger impact on the performance.
    Although the summary statistics as well as BAO fits do not show large deviations, the displacement comparison has more differences (at the maximum 32\% along the line of sight compared to 19\% and 7\% for the other two samples).
    The distributions of decomposition of the displacements to parallel and perpendicular to the line of sight show more irregularly shaped distributions than ELG or QSO does. The survey boundary effects are also slightly stronger in BGS. Hence, wide angle effects as well as a larger boundary-to-volume ratio can expose the algorithms to more challenges and cause more discrepancy between MG and iFFT. These larger differences, however, do not appear to affect BAO constraints. 

\end{itemize}

The results of this study suggest that the differences in the reconstruction algorithms do not lead to sizable bias for BAO constraints, thus they do not contribute to the systematic budget in the DESI DR1 BAO analysis \cite{KP4}. In our companion papers, other sources of systematics are also carefully examined - theoretical \cite{KP4s2-Chen} and observational \cite{KP3s3-Krolewski,KP3s4-Yu,KP3s5-Pinon} systematics, and HOD \cite{KP4s11-Garcia-Quintero,KP4s10-Mena-Fernandez}, fiducial cosmology \cite{KP4s9-Perez-Fernandez}, as well as covariance \cite{KP4s7-Rashkovetskyi,KP4s6-Forero-Sanchez} systematics - some of which also do not contribute to the overall systematics. These systematics result in a total of 0.245\% and 0.3\% error in the BAO parameters $\alpha_{\rm iso}$ and $\alpha_{\rm AP}$ in DESI DR1. Thus, the BAO technique remains one of the most robust probes of cosmic expansion history and provides stringent constraints on dark energy.

Reconstruction powerfully decreases the errors in the BAO measurements by reducing the nonlinear damping. It will remain a standard procedure in the upcoming DESI BAO analyses as well as all BAO analyses in future surveys, such as {\it Euclid} \cite{Euclid13} and {\it Roman} \cite{LSST12,LSST19}. Looking to the future, as nonstandard reconstruction attempts to be applied to surveys, similar care needs to be taken to ensure further reduction of the BAO errors without introducing bias.

\section{Data Availability}
The data used in this analysis will be made public along the Data Release 1 (details in \url{https://data.desi.lbl.gov/doc/releases/}).

\acknowledgments

We thank Jiamin Hou and Yu Yu for illuminating discussions. We also thank an anonymous referee for helpful suggestions. 
XC is supported by Future Investigators in NASA Earth and Space Science and Technology (FINESST) grant (award \#80NSSC21K2041).
ZD acknowledges support from the National Key R\&D Program of China (2023YFA1607800, 2023YFA1607802), the National Science Foundation of China (grant numbers 12273020 and 11621303), and the science research grant from the China Manned Space Project with NO. CMS-CSST-2021-A03. H-JS acknowledges support from the U.S. Department of Energy, Office of Science, Office of High Energy Physics under grant No. DE-SC0023241. H-JS also acknowledges support from Lawrence Berkeley National Laboratory and the Director, Office of Science, Office of High Energy Physics of the U.S. Department of Energy under Contract No. DE-AC02-05CH1123 during the sabbatical visit.  NP
is supported in part by DOE DE-SC0017660.
SN acknowledges support from an STFC Ernest Rutherford Fellowship, grant reference ST/T005009/2. CGQ acknowledges support provided by NASA through the NASA Hubble Fellowship grant HST-HF2-51554.001-A awarded by the Space Telescope Science Institute, which is operated by the Association of Universities for Research in Astronomy, Inc., for NASA, under contract NAS5-26555.

This material is based upon work supported by the U.S. Department of Energy (DOE), Office of Science, Office of High-Energy Physics, under Contract No. DE–AC02–05CH11231, and by the National Energy Research Scientific Computing Center, a DOE Office of Science User Facility under the same contract. Additional support for DESI was provided by the U.S. National Science Foundation (NSF), Division of Astronomical Sciences under Contract No. AST-0950945 to the NSF’s National Optical-Infrared Astronomy Research Laboratory; the Science and Technology Facilities Council of the United Kingdom; the Gordon and Betty Moore Foundation; the Heising-Simons Foundation; the French Alternative Energies and Atomic Energy Commission (CEA); the National Council of Humanities, Science and Technology of Mexico (CONAHCYT); the Ministry of Science and Innovation of Spain (MICINN), and by the DESI Member Institutions: \url{https://www.desi.lbl.gov/collaborating-institutions}. Any opinions, findings, and conclusions or recommendations expressed in this material are those of the author(s) and do not necessarily reflect the views of the U. S. National Science Foundation, the U. S. Department of Energy, or any of the listed funding agencies.

The authors are honored to be permitted to conduct scientific research on Iolkam Du’ag (Kitt Peak), a mountain with particular significance to the Tohono O’odham Nation.

\appendix
\section{Linear equation $\mathbf{A}\phi=\delta_s$ for multigrid} \label{appx:algebra}
The multigrid algorithm aims to solve a linear equation of the potential of the displacement, $\phi$, \cref{eq:poisson_phi}. Here we provide details for how to turn this equation into the form of $\boldsymbol{A}\phi=\delta_s$\footnote{More details can be found in the companion notes to the MG code in \url{https://github.com/martinjameswhite/recon_code/blob/master/notes.pdf}.}. With the divergence applied in spherical coordinates in the $\boldsymbol{\hat{\boldsymbol{r}}}$ direction, we can evaluate $\boldsymbol{\hat{r}}\cdot \nabla\phi$ and its $r$ derivative in Cartesian coordinates. We will obtain
\begin{equation}
    \nabla\cdot(\boldsymbol{\hat{r}}\cdot \nabla\phi)\boldsymbol{\hat{r}}=\sum_{ij}r_ir_j\partial_i\partial_j\phi+\frac{2}{r}\sum_{i}r_i\partial_i\phi,
\end{equation}
where $i$ and $j$ run through $x$, $y$ and $z$.

From here, we need to evaluate first and second derivatives on a grid. A first derivative in the $\boldsymbol{\hat{x}}$ direction can be approximated as
\begin{equation}\label{eq:dev_phi}
    \partial_x\phi\approx\frac{\phi_{i+1,j,k}-\phi_{i-1,j,k}}{2h},
\end{equation}
where $h$ is the grid resolution.
A second derivative can be approximated as
\begin{equation}
    \partial_x^2\phi\approx\frac{\phi_{i-1,j,k}+\phi_{i+1,j,k}-2\phi_{i,j,k}}{h^2}
\end{equation}
and
\begin{equation}
    \partial_x\partial_y\phi\approx\frac{\phi_{i+1,j+1,k}+\phi_{i-1,j-1,k}-\phi_{i+1,j-1,k}-\phi_{i-1,j+1,k}}{4h^2}.
\end{equation}
With these derivatives, we can then rearrange to obtain the matrix $\boldsymbol{A}$.

\section{Varying multigrid parameters}\label{appx:vary_MG_parameters}
\begin{figure}
    \centering
    \includegraphics[width=0.49\linewidth]{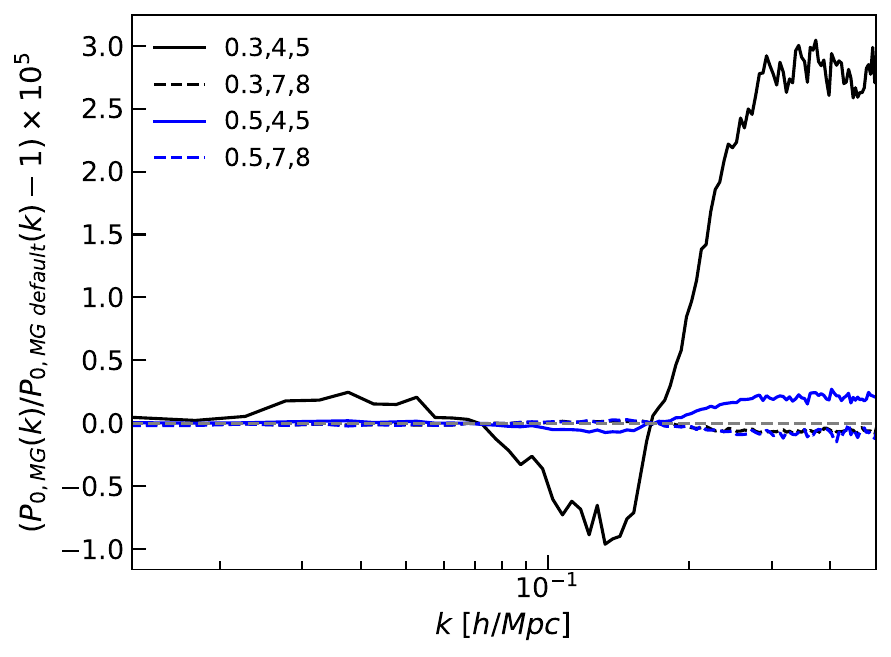}
    \includegraphics[width=0.48\linewidth]{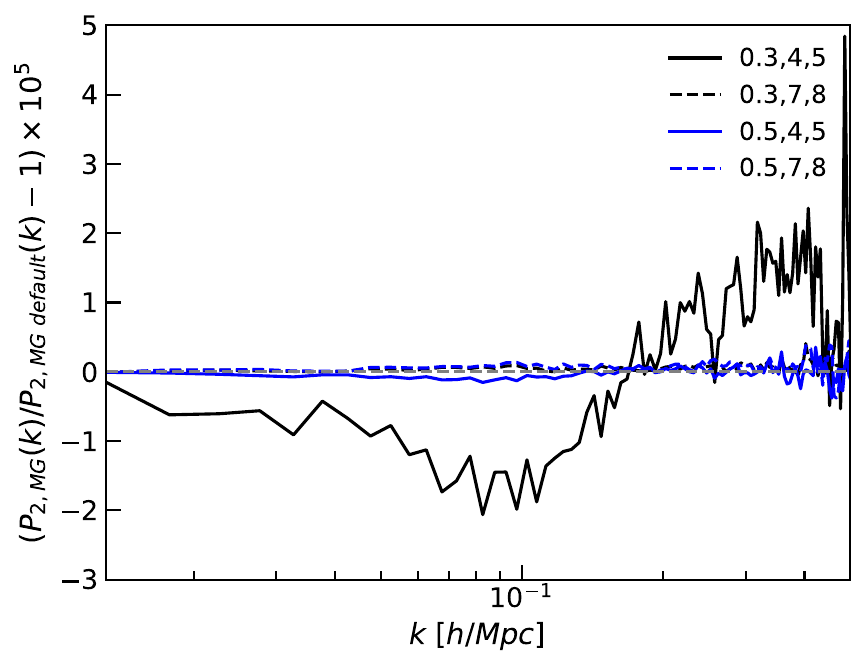}
    \caption{Power spectrum monopole (left) and quadrupole (right) ratio between MG with varying parameters and MG with default parameters using one ELG mock. The default is Jacobi damping factor equals to 0.4, 5 iterations of Jacobi, and 6 iterations of V-cycle. Here the numbers in the legend in order denote damping factor, number of Jacobi iterations, and number of V-cycles. There is a slight deviation from the default when the damping factor and numbers of iterations are lower than default, although changes are small. Higher damping and iterations do not make a substantial difference and are thus not necessary. }
    \label{fig:MG_varyparameters}
\end{figure}

We test the convergence of MG by varying three parameters: the damping factor for Jacobi, the number of iterations for damped Jacobi, and the number of iterations for the V-cycle, using one mock of the ELG sample. Figure~\ref{fig:MG_varyparameters} shows the results of varying these parameters with respect to the default case, which has the damping factor for Jacobi equal to 0.4, 5 iterations of damped Jacobi, and 6 iterations of V-cycle. The algorithm converges quickly. As long as the damping factor or iteration numbers are sufficient, there is very little difference between varying parameters and our default case.

\section{Additional figures}\label{appx:additional_figures}
\subsection{QSO}
\begin{figure}
    \centering
    \includegraphics[width=\linewidth]{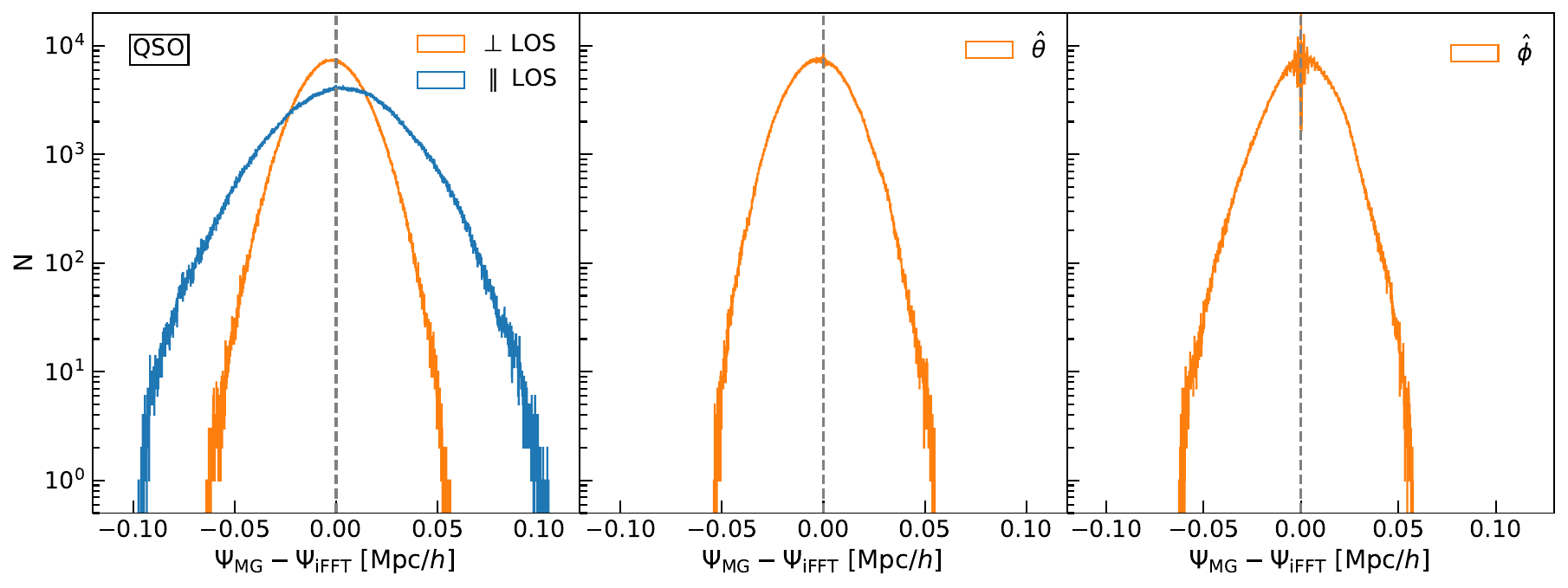}
    \caption{The magnitude difference between \textbf{MG} and \textbf{iFFT} (3 iterations) displacements with the \textbf{QSO} sample using one mock, for along the line of sight and perpendicular to the line of sight (left) and with the perpendicular to the line of sight further decomposed into $\boldsymbol{\hat{\theta}}$ (middle) and $\boldsymbol{\hat{\phi}}$ (right) directions. Even though the spread of the difference along the line of sight is slightly wider, the magnitude is very small, compared to the mean displacement magnitude along the line of sight for QSO, which is around 1.4 Mpc/$h$. }
    \label{fig:QSO_displacement_decomp}
\end{figure}

\begin{figure}
    \centering
    \includegraphics[width=\linewidth]{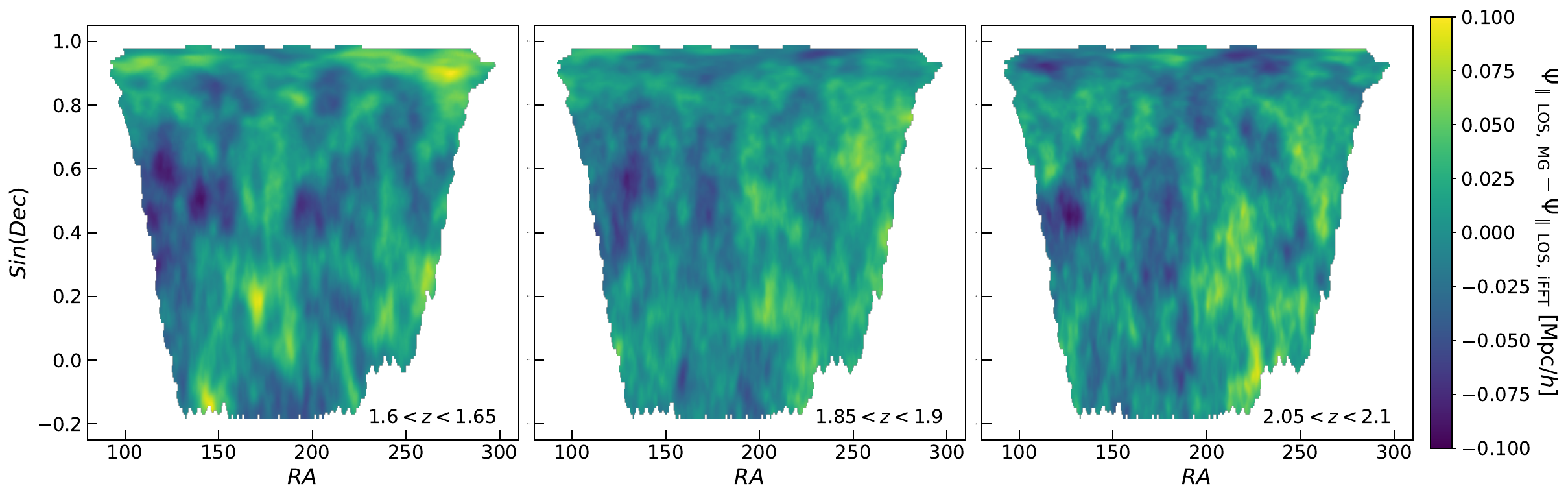}
    \caption{Line of sight component magnitude difference between \textbf{MG} and \textbf{iFFT} (both applied with a 30 Mpc/$h$ smoothing) projected to DESI Y5 footprint, shown for \textbf{QSO} $1.6<z<1.65$ (left), $1.85<z<1.9$ (middle), and $2.05<z<2.1$ (right) redshift slices, using the randoms catalog. 
    The low redshift end shows slightly more differences, but the magnitude is very small compared to the mean line-of-sight displacement of QSO ($\sim$1.4 Mpc/$h$).}
    \label{fig:los_scalar_footprint_colormap_QSO}
\end{figure}

Below we include additional figures for reconstruction performance in QSO. Figures~\ref{fig:QSO_displacement_decomp} and~\ref{fig:los_scalar_footprint_colormap_QSO} show the results of displacement comparison in one QSO mock or in one QSO randoms catalog. These are qualitatively similar to the ELG sample. Namely, there is a slightly larger spread in the difference between the two algorithms along the line of sight, and the magnitude differences along the line of sight can be slightly larger near the redshift boundary (although QSO has slightly larger differences at the low redshift end while ELG has slightly larger differences at the high redshfit end). There are still no obvious extreme displacement differences clustered along the survey boundaries in the QSO sample, as shown in Figure~\ref{fig:los_scalar_footprint_colormap_QSO}. However, the displacement difference spread is smaller in QSO than in ELG, as shown in Figure~\ref{fig:QSO_displacement_decomp}. Here we use a QSO randoms catalog to have lower noise in the figure. The line of sight displacement difference spread is less than 10\% of the average line of sight displacement magnitude of QSO ($\sim$1.4 Mpc/$h$).

\begin{figure}
    \centering
    \includegraphics[width=0.49\columnwidth]{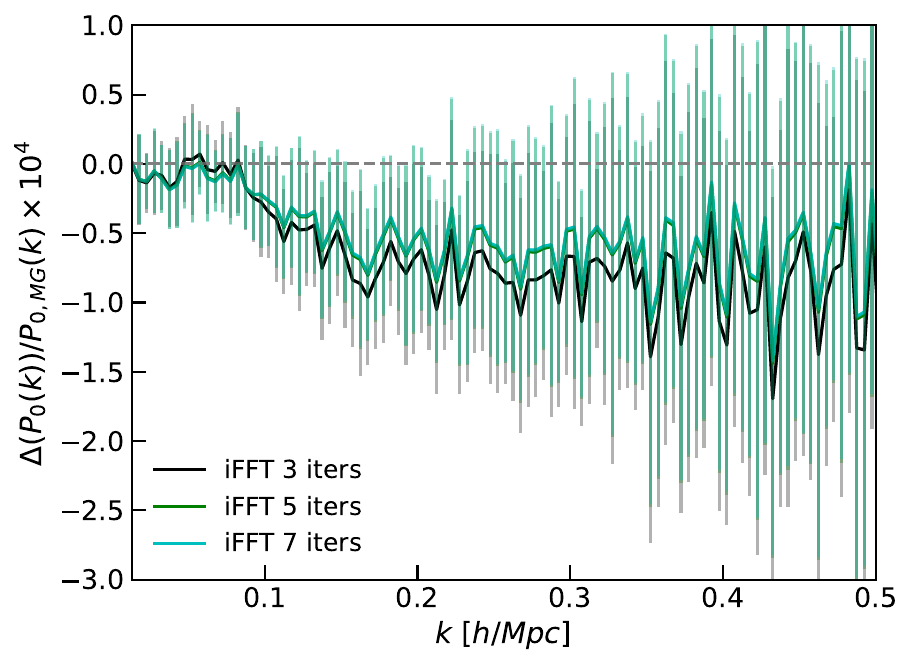}
    \includegraphics[width=0.48\columnwidth]{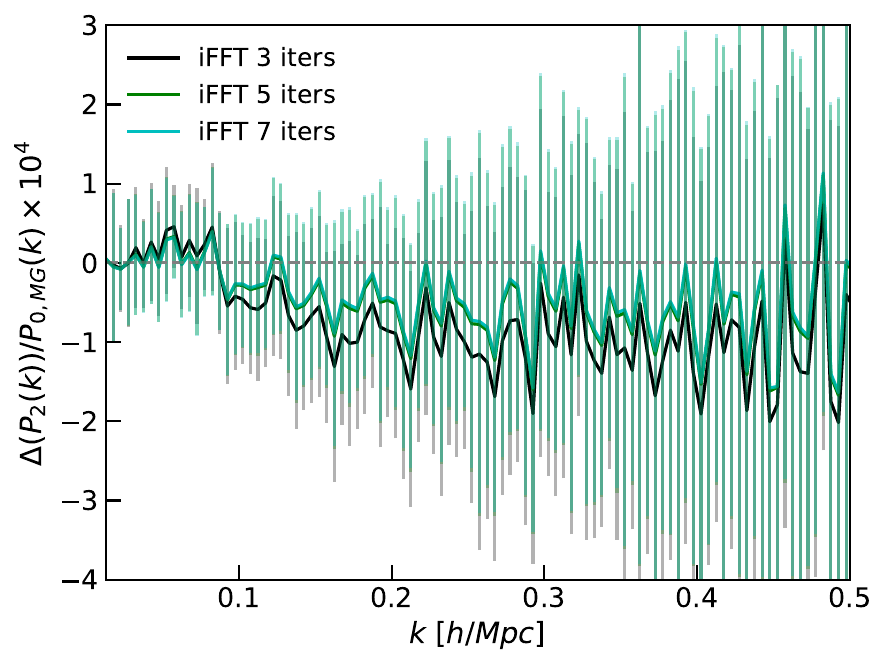}
    \caption{Ratio in power spectrum monopole (left) and quadrupole (right) between iFFT (black for 3 iterations, green for 5 iterations, and cyan for 7 iterations) and MG (both applied with a 30 Mpc/$h$ smoothing), using the \textbf{RecSym} convention for the \textbf{QSO} sample. Even with 3 iterations, the difference between iFFT and MG is within 0.02\% on average for both monopole and quadrupole. The two algorithms almost entirely agree within error bars. The difference between 5 and 7 iterations are negligible, suggesting that the convergence in QSO is faster than ELG. }
    \label{fig:Pk_QSO_MG_iFFT}
\end{figure}

\begin{figure} 
    \centering
    \includegraphics[width=0.7\columnwidth]{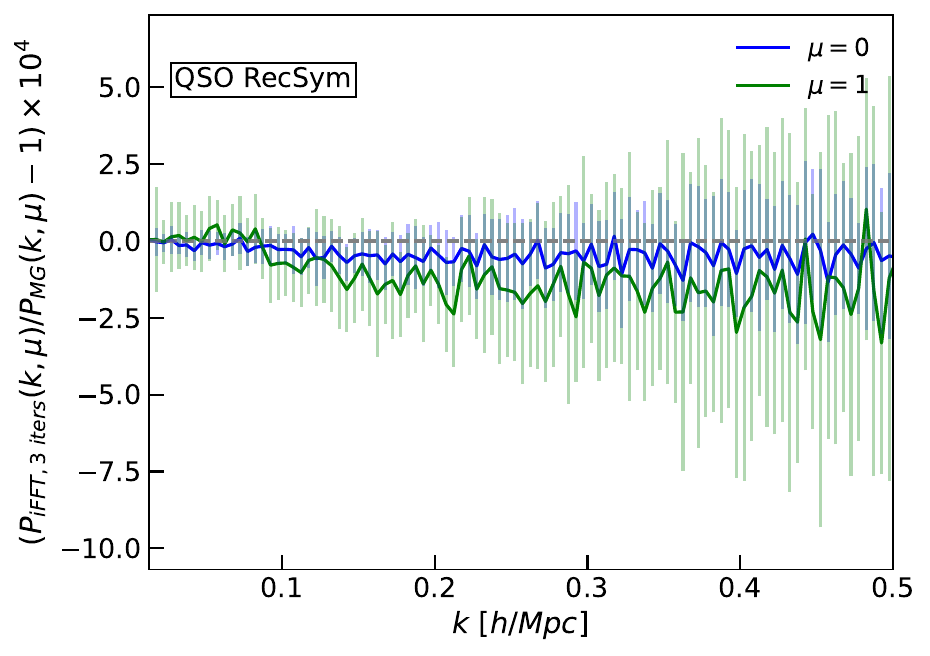}
    \caption{
    Ratio of power spectrum between iFFT and MG for the \textbf{QSO} sample with both algorithms applied with a 30 Mpc/$h$ smoothing and using the \textbf{RecSym} convention, for $\mu=0$ (blue) and 1 (green), where $\mu$ is cosine of the angle between the line of sight and $\boldsymbol{k}$. Here $\mu=0$ is perpendicular to the line of sight and $\mu=1$ is along the line of sight. Each line is an average of 25 mocks. We observe that both cases show differences between iFFT and MG within 0.03\% on average. However, along the line of sight the differences between the two algorithm are larger, especially on smaller scales ($k\gtrsim 0.2 h$/Mpc). However, they almost agree within error bars. }
    \label{fig:Pk_angle_QSO}
\end{figure}

Figure~\ref{fig:Pk_QSO_MG_iFFT} shows the difference between the iFFT and MG power spectra w.r.t.\ the MG monopole in the \textbf{RecSym} convention. The differences between the two algorithms here are smaller than those in the ELG or BGS samples and almost entirely agree within error bars in the scales shown here. 
In Figure~\ref{fig:Pk_angle_QSO}, the power spectra along and perpendicular to the line of sight show qualitatively the same pattern as that in ELG; there are more differences between iFFT and MG along the line of sight, but in QSO the two algorithms agree within error bars.

\begin{figure}
    \centering
    \includegraphics[width=\columnwidth]{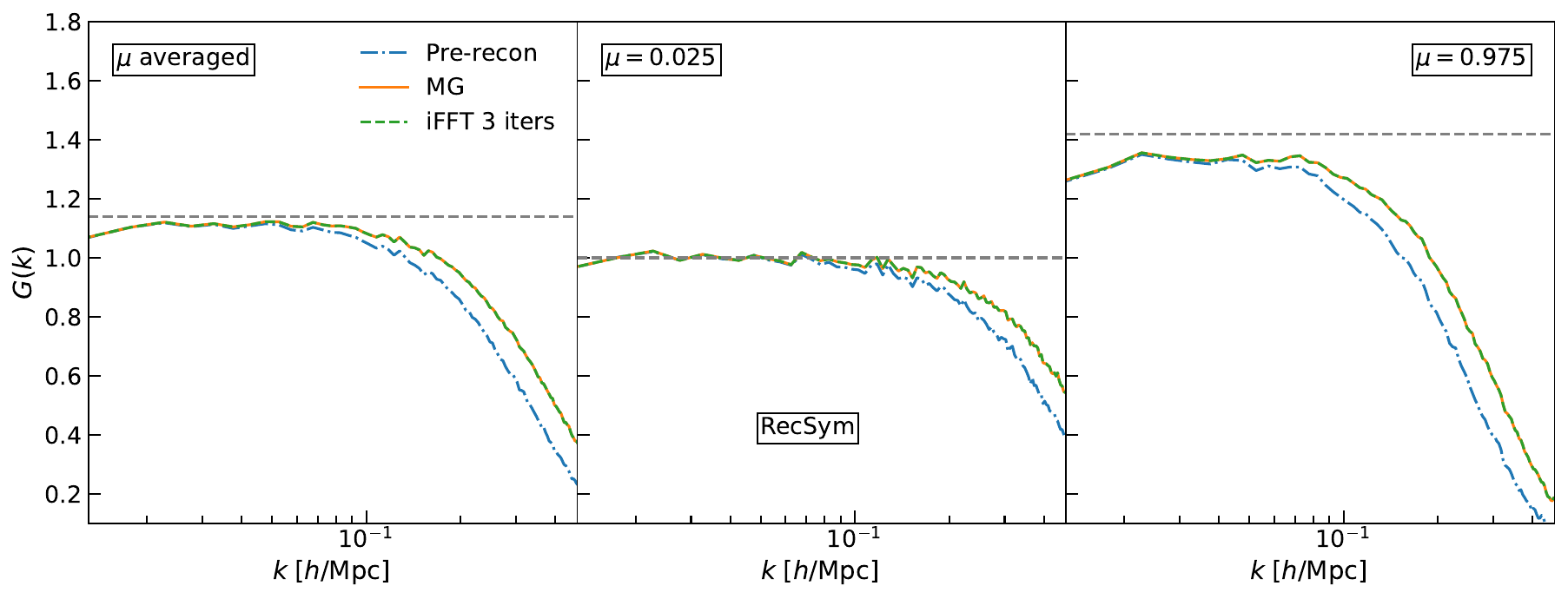}
    \caption{Propagator for the three reconstruction algorithms for one \textbf{QSO} mock with the \textbf{RecSym} convention when $\mu$ is integrated over (left), and in $\mu=0.025$ (middle) and 0.975 (right) bins, in comparison of pre-reconstruction propagator (blue dash dotted). The $\mu$-bin width is 0.05. The horizontal dashed lines show expected Kaiser approximation in the plane-parallel approximation, $(1+1/3\beta)=1.14$ for the first panel and $(1+\beta\mu^2)=1.0$ and 1.42 for the middle and right panels, respectively. We shift the propagators by 5\% to correct for misestimate of the linear bias. MG and iFFT (3 iterations) return improved propagator and they overlap for all three cases.  }
    \label{fig:QSO_Gk_recsym}
\end{figure}

The propagator in the \textbf{RecSym} convention in QSO exhibits qualitatively the same behavior as that in ELG as well, as shown in \cref{fig:QSO_Gk_recsym}. MG and iFFT (3 iterations) overlap. Reconstruction fidelity for QSO is, however, lower than ELG, as shown by the less improvement over pre-reconstruction by both algorithms.

\begin{figure}
    \centering
    \includegraphics[width=0.488\linewidth]{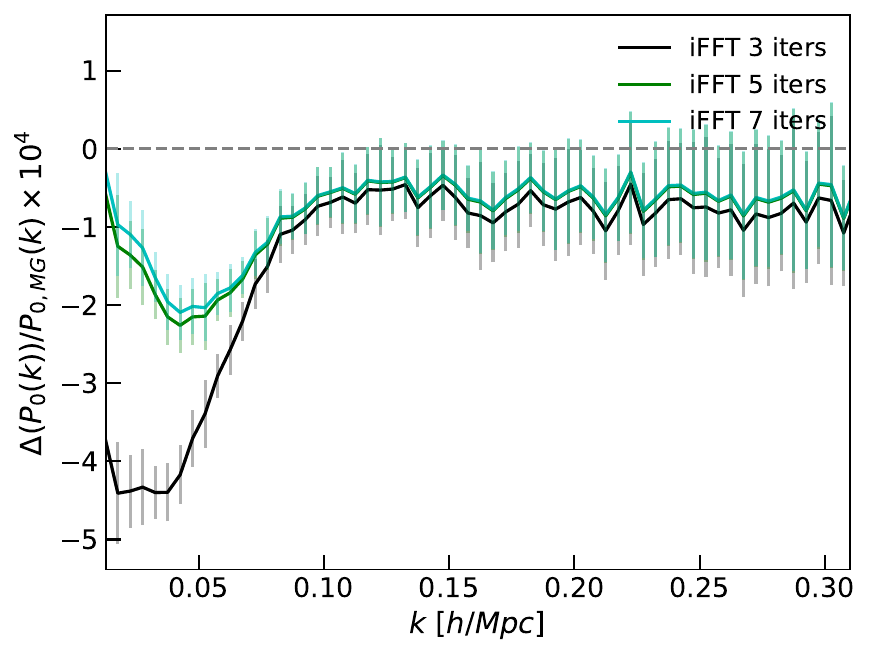}
    \includegraphics[width=0.497\linewidth]{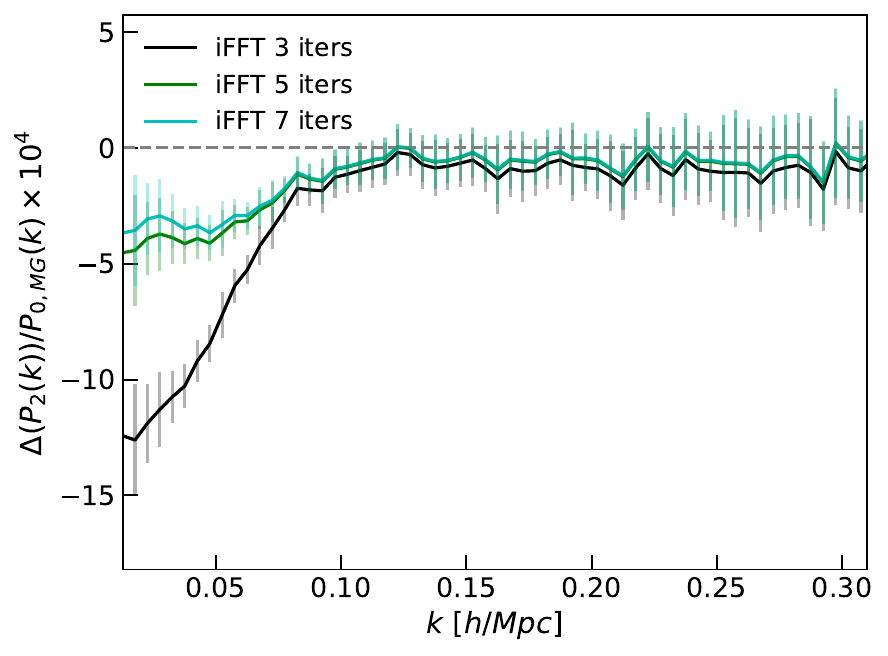}
    \caption{The monopole (left) and quadrupole (right) power spectrum difference between iFFT (black for 3 iterations, green for 5 iterations, and cyan for 7 iterations) and MG (both applied with a $30\hMpc$ smoothing) divided by the MG monopole for the \textbf{QSO} sample, using the \textbf{RecIso} convention. Each line is an average over 25 mocks. Large scales show more discrepancies, but the magnitude is small compared to the power spectrum value.
    We observe the discrepancies at most 0.05\% in monopole and 0.12\% in quadrupole (both w.r.t.\ MG monopole) on average on large scales. }
    \label{fig:QSO_Iso}
\end{figure}

\begin{figure}
    \centering
    \includegraphics[width=0.7\columnwidth]{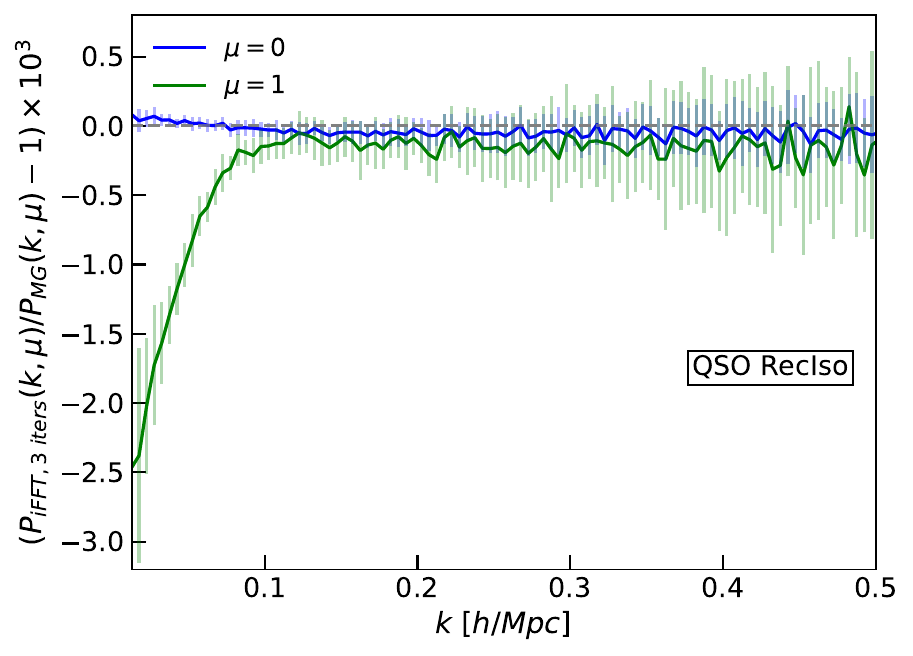}
    \caption{Ratio of power spectrum between iFFT and MG for the \textbf{QSO} sample with both algorithms applied with a 30 Mpc/$h$ smoothing and using the \textbf{RecIso} convention, for $\mu=0$ (blue) and 1 (green), where $\mu$ is cosine of the angle between line of sight and $\boldsymbol{k}$. Here $\mu=0$ is perpendicular to the line of sight and $\mu=1$ is along the line of sight. We again observe discrepancies along the line of sight between the two algorithms on large scales, but the differences are smaller compared to ELG and BGS, within about 0.25\% on average. However, they almost agree within error bars. }
    \label{fig:Pk_angle_ratio_QSO_Iso}
\end{figure}

Figures~\ref{fig:QSO_Iso} and~\ref{fig:Pk_angle_ratio_QSO_Iso} are power spectrum results for the QSO \textbf{RecIso} reconstruction case. The power spectrum differences again show qualitatively similar patterns to those in ELG -- there are more differences between iFFT and MG on larger scales and along the line of sight. However, the differences between the two algorithms are generally smaller here.

\begin{figure}
    \centering
    \includegraphics[width=\linewidth]{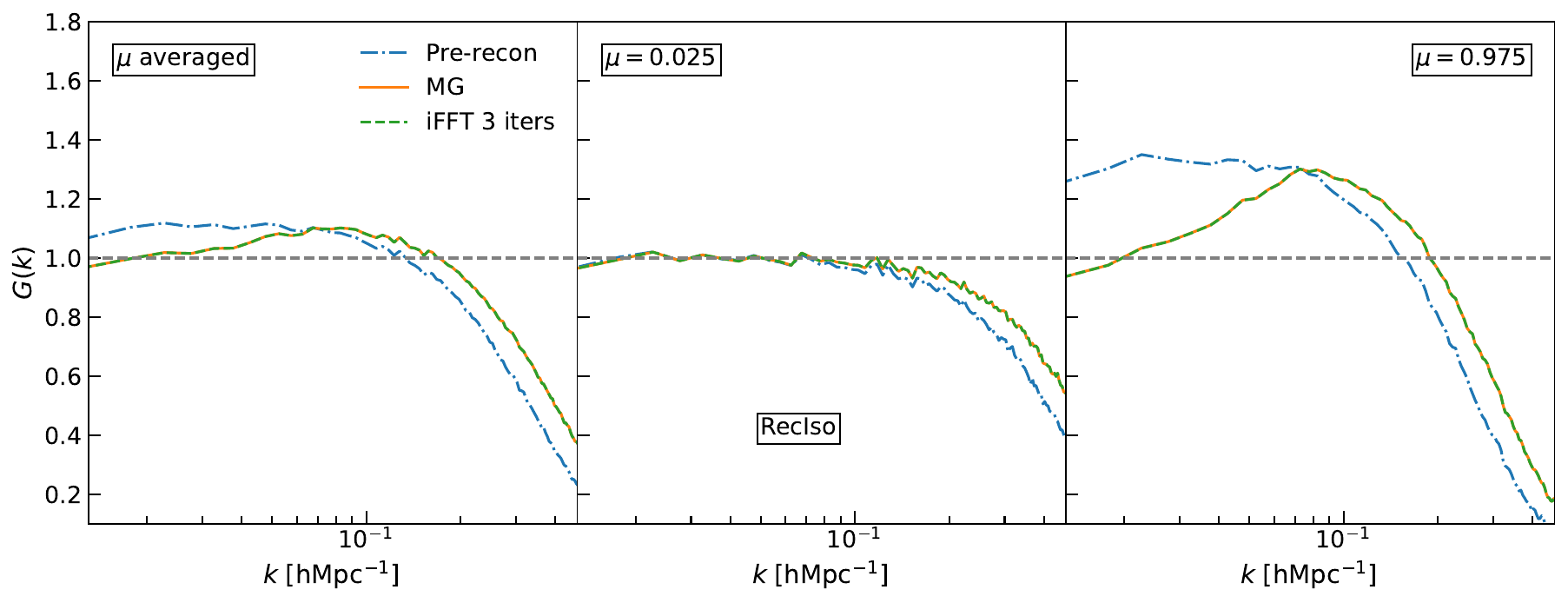}
    \caption{Propagator for the three reconstruction algorithms for one \textbf{QSO} mock with the \textbf{RecIso} convention when $\mu$ is integrated over (left), and in $\mu=0.025$ (middle) and 0.975 (right) bins, in comparison of pre-reconstruction propagator (blue dash dotted). The horizontal dashed lines show the expected amplitude (middle) or the expected amplitude after removing RSDs (left and right). We shift the propagators by 5\% to correct for misestimate of the linear bias. Both reconstruction algorithms return improved propagator, and MG and iFFT (3 iterations) are on top of each other for all three cases. The shapes of the propagator by both algorithms in the $\mu=0.975$ bin show distortions on large scales, suggesting that the estimate of the line-of-sight displacement is challenging for both algorithms.  }
    \label{fig:QSO_Gk_reciso}
\end{figure}

The \textbf{RecIso} propagator behavior in QSO is also similar to that in ELG, with MG and iFFT (3 iterations) overlapping, as shown in \cref{fig:QSO_Gk_reciso}. The improvement over pre-reconstruction is again less than what presents in the case of ELG, suggesting the lower reconstruction fidelity in QSO.

\subsection{BGS}
\begin{figure}
    \centering
    \includegraphics[width=0.50\linewidth]{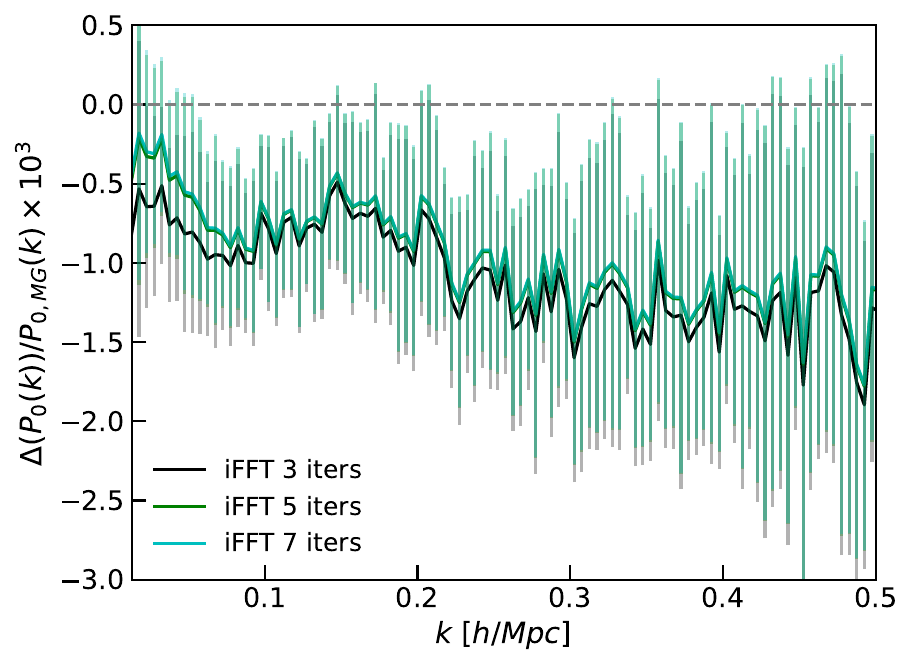}
    \includegraphics[width=0.48\linewidth]{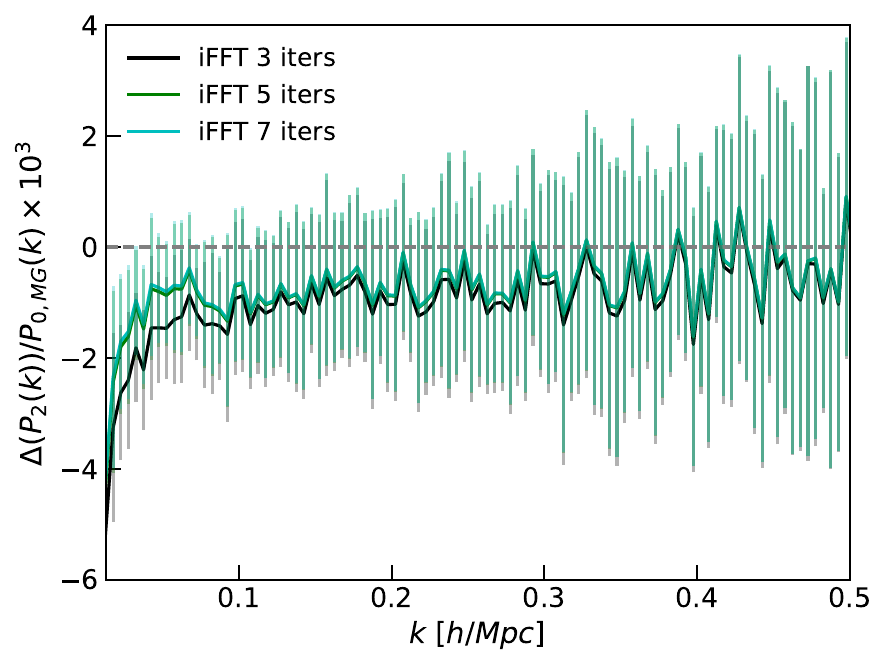}
    \caption{The monopole (left) and quadrupole (right) power spectrum difference between iFFT (black for 3 iterations, green for 5 iterations, and cyan for 7 iterations) and MG (both applied with a 15 Mpc/$h$ smoothing) divided by the MG monopole, using the \textbf{RecIso} convention for the \textbf{BGS} sample. Each line is an average of 25 mocks. Unlike ELG and QSO, where large scales exhibit more differences, that feature does not exist in the BGS monopole and is less prominent in the BGS quadrupole. The differences between the two algorithms are within $\sim$0.3\% w.r.t. MG monopole on average, 
    for both the monopole and the quadrupole on BAO scales. Considering larger scales, the differences in the quadrupole can be 0.5\% on average.}
    \label{fig:BGS_Iso}
\end{figure}

\begin{figure}
    \centering
    \includegraphics[width=0.7\columnwidth]{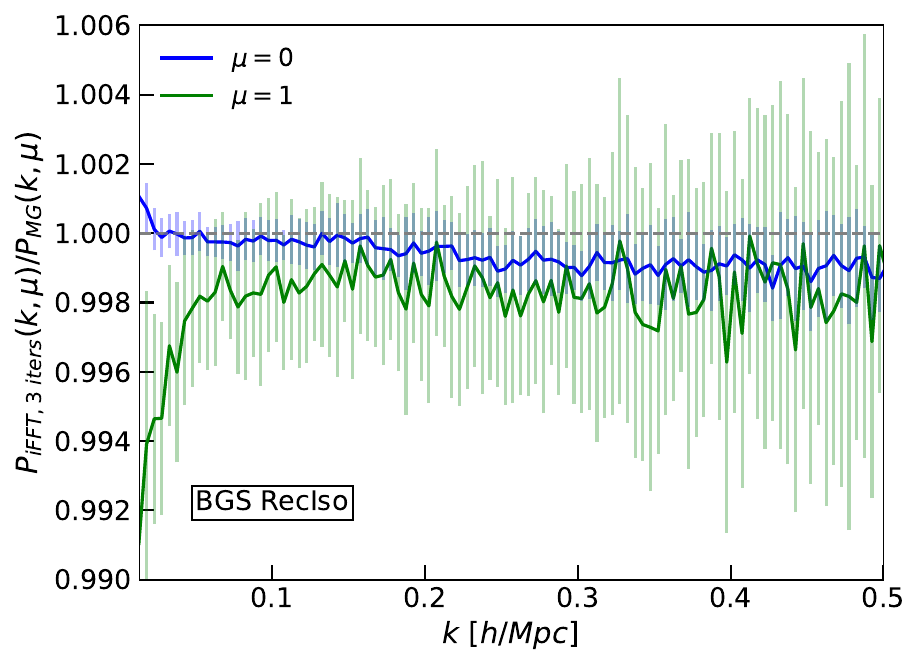}
    \caption{Ratio of power spectrum between iFFT and MG for the \textbf{BGS} sample with both algorithms applied with a 15 Mpc/$h$ smoothing and using the \textbf{RecIso} convention, for $\mu=0$ (blue) and 1 (green), where $\mu$ is cosine of the angle between line of sight and $\boldsymbol{k}$. Here $\mu=0$ is perpendicular to the line of sight and $\mu=1$ is along the line of sight. We observe that both cases show differences between iFFT and MG within 0.8\% on average, where the maximum happens along the line of sight on large scales. As with \textbf{RecIso} for the other two tracers, along the line of sight the differences between the two algorithms are larger, and more so on large scales. 
    Considering only the BAO scales, the differences between the two algorithms are within $\sim$ 0.3\% on average
    for both along and perpendicular to the line of sight, and they almost agree within error bars. 
    }
    \label{fig:BGS_2d_iso}
\end{figure}

\begin{figure}
    \centering
    \includegraphics[width=\columnwidth]{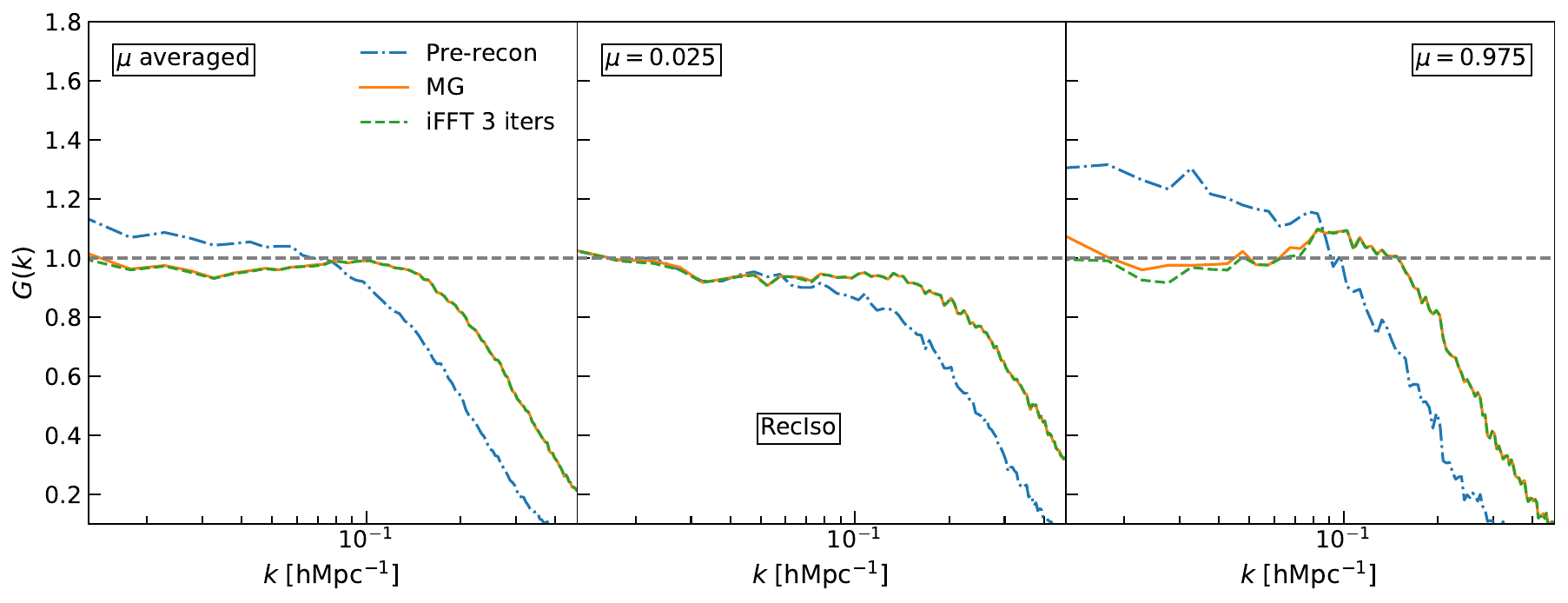}
    \caption{Propagator for the three reconstruction algorithms for one \textbf{BGS} mock with the \textbf{RecIso} convention when $\mu$ is integrated over (left), and in $\mu=0.025$ (middle) and 0.975 (right) bins, in comparison of pre-reconstruction propagator (blue dash dotted). The horizontal dashed lines show the expected amplitude (middle) or the expected amplitude after removing RSDs (left and right). We shift the propagators by 14\% to correct for misestimate of the linear bias. Both reconstruction algorithms return improved propagator, and MG and iFFT (3 iterations) are on top of each other for all three cases. The shapes of the propagator by both algorithms in the $\mu=0.975$ bin show distortions on large scales, suggesting that the estimate of the line-of-sight displacement is challenging for both algorithms.  }
    \label{fig:BGS_Gk_reciso}
\end{figure}

We include four additional figures for BGS that show the power spectrum and propagator comparisons for \textbf{RecIso}. Figures~\ref{fig:BGS_Iso} and~\ref{fig:BGS_2d_iso} show the power spectrum comparison for the \textbf{RecIso} case. In Figure~\ref{fig:BGS_Iso}, it is interesting that BGS shows more discrepancy between iFFT and MG on smaller scales in monopole, whereas for ELG and QSO, there is more discrepancy on large scales. In Figure~\ref{fig:BGS_2d_iso}, large scale discrepancy still presents along the line of sight; however, on BAO scales, the differences between iFFT and MG are within 0.3\% for both along and perpendicular to the line of sight. Figure~\ref{fig:BGS_Gk_reciso} show the results of BGS propagator in \textbf{RecIso}. Along the line of sight, there is again noticeable differences between MG and iFFT (3 iterations) similar to that in the \textbf{RecSym} propagator.

\bibliographystyle{JHEP}
\bibliography{opt_recon, DESI2024_bib}{}

\end{document}